\documentclass[twocolumn]{aastex6}
\usepackage{amsmath}
\usepackage{amssymb}
\usepackage{graphicx}
\usepackage[tight,hang]{subfigure}
\usepackage{natbib}
\usepackage{enumitem}


\newcommand{\msun}{\, \mbox{M}_{\odot}}
\newcommand{\muG}{\, \mu\mbox{G}}
\newcommand{\flash}{{\normalfont\scshape flash }}
\newcommand{\paramesh}{{\normalfont\scshape paramesh }}

\def\mean#1{\left< #1 \right>}

\begin{document}

\title{The Co-Evolution of a Magnetized Intracluster Medium and Hot Galactic Coronae: Magnetic Field Amplification and Turbulence Generation }
\author{Rukmani Vijayaraghavan\altaffilmark{1,2,3}}
\author{Paul M. Ricker\altaffilmark{2}}
\email{rukmani@virginia.edu}

\altaffiltext{1}{Department of Astronomy, University of Illinois at Urbana-Champaign, 1002 W. Green Street, Urbana, IL 61801, USA}
\altaffiltext{2}{Department of Astronomy, University of Virginia, 530 McCormick Rd., Charlottesville, VA 22904, USA}
\altaffiltext{3}{NSF Astronomy \& Astrophysics Postdoctoral Fellow}

\begin{abstract}
We use adaptive-mesh magnetohydrodynamic simulations to study the effect of magnetic fields on ram pressure stripping of galaxies in the intracluster medium (ICM). Although the magnetic pressure in typical clusters is not strong enough to affect the gas mass loss rate from galaxies, magnetic fields can affect the morphology of stripped galaxies. ICM magnetic fields are draped around orbiting galaxies and aligned with their stripped tails. Magnetic fields suppress shear instabilities at the galaxy-ICM interface, and magnetized tails are smoother and narrower than tails in comparable hydrodynamic simulations in \citet{Vijayaraghavan15b}. Orbiting galaxies stretch and amplify ICM magnetic fields, amplifying magnetic power spectra on $10 - 100$ kpc scales. Galaxies inject turbulent kinetic energy into the ICM via their turbulent wakes and $g$-waves. The magnetic energy and kinetic energy in the ICM increase up to $1.5 - 2$ Gyr of evolution, after which galaxies are stripped of most of their gas, and do not have sufficiently large gaseous cross sections to further amplify magnetic fields and inject turbulent kinetic energy. The increase in turbulent pressure due to galaxy stripping and generation of $g$-waves results in an increase in the turbulent volume fraction of the ICM. This turbulent kinetic energy is not a significant contributor to the overall ICM energy budget, but greatly impacts the evolution of the ICM magnetic field. Additionally, the effect of galaxies on magnetic fields can potentially be observed in high resolution Faraday rotation measure (RM) maps as small scale fluctuations in the RM structure.
\end{abstract}

\keywords{Galaxies: clusters: general -- galaxies: groups: general -- methods: numerical}

\maketitle

\section{Introduction}
\label{sec:intro}

Observations suggest that $\mu$G magnetic fields in the intracluster medium (ICM) are ubiquitous, although direct measurements of the magnetic field strength and morphology in clusters are not straightforward (see \citealt{Carilli02}, \citealt{Govoni04}, \citealt{Kronberg05}, \citealt{Ryu12} for reviews). The earliest estimates of galaxy cluster magnetic fields were based on energy equipartition and minimum energy configuration arguments applied to cluster radio halos (e.g.\ \citealt{Willson70}, \citealt{Miley80}, \citealt{Giovannini93}, \citealt{Feretti99}). More recent observations have used Faraday rotation measure (RM) to estimate the strength of the ICM magnetic field. RM is proportional to the magnetic field strength along the line of sight and the electron density; for known electron density (e.g.\ from X-ray observations), the line of sight magnetic field strength can be estimated. \citet{Dreher87} performed the first RM observations of ICM gas surrounding Cygnus A. Further measurements of the RM in clusters agree on $\sim \muG$ strengths for cluster magnetic fields (e.g.\ \citealt{Vallee86, Vallee87}, \citealt{Ge93}, \citealt{Taylor94, Taylor01}, \citealt{Clarke00}, \citealt{Rudnick03}, \citealt{Murgia04}, \citealt{Bonafede10}, \citealt{Govoni10}, \citealt{Bonafede11}, \citealt{Vacca12}, \citealt{Bonafede13}), and indicate the cluster magnetic fields are tangled on $\sim$~kpc scales. From the distribution of RM in a cluster, one can infer the morphology and coherence scales of cluster magnetic fields. \citet{Vogt03, Vogt05}  used a correlation analysis and Bayesian likelihood analysis to determine the magnetic field strengths of three clusters to be $\sim 3 - 13 \muG$, magnetic field autocorrelation lengths to be $0.9 - 4.9$~kpc, and power spectral indices $\alpha$ to range from $1.6 - 2.0$, consistent with being Kolmogorov-like power spectra with $\alpha = 5/3$. Complementary to RM observations, lower limits on ICM magnetic field strengths can be constrained using observations of cluster radio relics and inferred upper limits on inverse Compton emission (e.g.\ $\sim 3 \muG$ magnetic fields in A3667; \citealt{Finoguenov10, Sarazin13}). These measurements, however, cannot constrain magnetic field correlation lengths.

Magnetic fields in the ICM directly affect the evolution of galaxies and their interstellar medium (ISM) gas. Thermal conduction across the ISM-ICM boundary  in typical hot ($kT \sim 0.5 - 1$~keV) galactic coronae should be saturated, since the mean free path of the ICM electrons ($\lambda_{e} \simeq 10$ kpc) is comparable to the sizes of galactic coronae (\citealt{Sarazin86}). Under these conditions, saturated evaporation timescales should be $\sim 10^7$ years (\citealt{Vikhlinin01}), more than two orders of magnitude shorter than ram pressure stripping timescales for galactic coronae in groups and clusters, as quantified in \citet{Vijayaraghavan15b}. However, observations of long-lived coronae in groups and clusters argue against efficient thermal conduction; a combination of radiative cooling and suppression of thermal conduction by draped magnetic fields is likely effective in these environments (\citealt{Lyutikov06}). The draping of ICM magnetic fields over the leading surfaces of moving subclusters has been shown to suppress the formation of hydrodynamic instabilities (\citealt{Dursi07}, \citealt{Dursi08}), thereby suppressing mixing with the ICM. Magnetic field draping effects have only been recently investigated with numerical simulations in the context of galaxy-scale coronae (\citealt{Shin14}), and can potentially play an important role in their survival and longevity. 

Numerical simulations have been used to study the impact of ICM magnetic fields on the cold disk gas of cluster galaxies. \citet{Ruszkowski14} simulated disk galaxies exposed to a uniformly magnetized ICM wind, and showed that ambient magnetic fields result in $100$ kpc long filamentary structures in the stripped tails of galaxies, forming bifurcated structures similar to those in observed ram pressure stripped galaxies. They also found that magnetic pressure can support these tails, and that magnetic field vectors are aligned with the tails. They found that the ICM magnetic field did not significantly affect the removal of gas due to ram pressure stripping. \citet{Shin14} simulated the effects of magnetic fields aligned parallel and perpendicular to the direction of an elliptical galaxy's motion in a uniform ICM, and showed that the morphology of the stripped tail was strongly dependent on the relative alignment of the initial magnetic field -- strongly collimated for aligned magnetic fields, and sheet-like for magnetic fields perpendicular to the direction of motion. They also showed that magnetic fields were amplified in the stripped tail in both cases. \citet{Tonnesen14} investigated the effect of magnetic fields in the disks of galaxies themselves. They showed that while galactic magnetic fields do not significantly affect the amount of gas removed by stripping, they produce unmixed structures in the tail and result in an overall increase in the magnetic energy density in the stripped tail. 

Galaxies themselves affect the overall evolution of magnetic fields in the ICM. Galaxy motions and stripped galaxy wakes can amplify existing cluster magnetic fields and generate turbulence in the ICM. \citet{Jaffe80} and \citet{Roland81} explored the possibility that turbulent galactic wakes are the source of ICM magnetic fields using a dimensional analysis and an analytic model, respectively. \citet{Ruzmaikin89} showed that turbulent motions in galaxy wakes can generate chaotic magnetic fields,  but \citet{DeYoung92} argued using statistical modeling of the magnetic energy spectrum that the turbulence should dissipate before the field can be significantly amplified. Using analytic arguments and simulations, \citet{Subramanian06} showed that major mergers can amplify the existing field via a turbulent dynamo, but that galaxy wakes likely occupy too small a volume filling fraction to be the main source of magnetic field. Nevertheless they argued that the wake filling fraction in projected area should be large, producing random RM values consistent with observations. Using hydrodynamic simulations, \citet{Kim07} found that orbiting galaxies can generate turbulence over larger volumes by resonantly exciting gravity waves. \citet{Ruszkowski10} showed using magnetohydrodynamic (MHD) simulations that galaxy-scale turbulence can randomize tangential magnetic fields, supporting anisotropic thermal conduction between the central cool core and outer regions of a cluster. \citet{Ruszkowski11} extended these simulations to include particles representing massive galaxies, and showed that in addition to turbulent wakes, galaxies also excite larger scale $g$-mode turbulence, in agreement with \citet{Kim07}. \citet{Conroy08} found that dynamical friction from orbiting satellites can contribute to ICM heating. In addition to generating turbulence and amplifying existing ICM magnetic fields, magnetized galaxy winds and stripped gas can be injected into and partly seed ICM magnetic fields (\citealt{Donnert09}, \citealt{Arieli11}).  

Galaxies are by no means the only, or even the most energetic, sources for magnetic field amplification and turbulence generation in the ICM. Cosmological simulations (e.g.\ \citealt{Dolag99}, \citealt{Dolag02}, \citealt{Dubois08}, \citealt{Donnert09}, \citealt{Vazza14}, \citealt{Egan16}) show that initial seed cluster magnetic fields can be amplified by cosmological structure formation and the growth of structure. These simulations do not isolate the effect of galaxies alone.  \citet{Roettiger99} show that during cluster mergers, magnetic fields are initially stretched and compressed by merger shocks and bulk flows, after which turbulent flows amplify magnetic fields on scales comparable to the sizes of cluster cores. \citet{Takizawa08}, using idealized cluster mergers, show that the magnetic field perpendicular to the merger axis is amplified in addition to the generation of an ordered and amplified magnetic field in the merging subcluster's wake. \citet{Iapichino12} showed that in addition to turbulence produced by shocks in cluster mergers, turbulent pressure support upstream of merger shocks can amplify magnetic fields to $\mu$G levels. \citet{Beresnyak16} argue that cluster magnetic fields trace clusters' past turbulent activity. \citet{Xu09, Xu10, Xu11} showed that magnetic fields injected by AGN can be amplified by turbulence generated during cluster formation, and that magnetic field amplification is strongly dependent on the mass and merger history of clusters. Subsequently, \citet{Xu12} showed that the resulting RM measurements and radio halos of these magnetic fields agree with observations. However, \citet{Sutter12} showed that the strength of magnetized outflows from AGN in simulations can vary significantly depending on the mode of injection, accretion strength, and numerical resolution. 

In addition to quantifying the effect of turbulence on magnetic field amplification, various studies have quantified the overall turbulent kinetic energy and non-thermal pressure in clusters.  Numerical simulations have shown that cluster mergers can generate significant turbulence (e.g.\ \citealt{Roettiger93},  \citealt{Roettiger97}, \citealt{Ricker01}, \citealt{Nagai03}, \citealt{Takizawa05}, \citealt{Paul11}, \citealt{Donnert13}). Cosmological simulations of cluster formation and growth also show the overall growth of turbulence from infalling substructure (e.g. \citealt{Dolag05}, \citealt{Maier09}, \citealt{Iapichino08}, \citealt{Vazza09}, \citealt{Vazza11}, \citealt{Miniati14}, \citealt{Miniati15}). Nonthermal pressure due to bulk and turbulent motions in the ICM may significantly bias estimates of cluster masses (e.g.\ \citealt{Nagai07}, \citealt{Nelson14}; though see \citealt{Hitomi16}). In principle, turbulence and bulk flows in the ICM can be indirectly detected via the broadening of spectral lines. However until recently existing X-ray telescopes have not had sufficient spectral resolution to resolve turbulent velocity dispersions of $100 - 200$ km s$^{-1}$. Upcoming X-ray observatories, particularly \textit{ATHENA}, can potentially detect turbulent velocity dispersions at these levels.\footnote{The loss of \textit{Hitomi} (\textit{ASTRO-H}) will be keenly felt in this field. In \S~\ref{sec:obs_diagnostics} we include predictions for what \textit{Hitomi} would have been able to detect in case another similar mission is attempted.}

Previous hydrodynamic cluster galaxy simulation work has generally focused on clusters in cosmological volumes combined with subgrid prescriptions or semi-analytic models (e.g.\ \citealt{Vogelsberger14}), individual galaxies in wind tunnels or isolated cluster potentials (e.g.\ recent work by  \citealt{Tonnesen14}, \citealt{Ruszkowski14}, \citealt{Shin14}, \citealt{Roediger15a,Roediger15b}), or galaxies as point mass perturbers (e.g.\ \citealt{Ruszkowski11}). To isolate the collective effects of galaxies on the ICM apart from infall and major mergers, there is a need for numerical experiments that fit between idealized studies of single galaxies and fully cosmological simulations. In \citet{Vijayaraghavan15b} (hereafter Paper~I), we simulated the hydrodynamic evolution of a distribution of galactic coronae in an isolated group and cluster. We showed that ICM ram pressure on galaxies produces characteristic leading surfaces where the thermal and ram pressure of the ICM balance the internal thermal pressure of the galaxies; weakly bound gas is stripped from galaxies and deposited in tails that trace galaxies' orbits. These stripped tails form shear instabilities and dissipate as they move through the ICM. Galaxies lose $\sim 90\%$ of their gas within a dynamical time due to stripping processes alone, and in the absence of any form of shielding or replenishment.

In this paper, we extend our work to consider MHD simulations that quantify the impact of magnetic fields on the stripping of galactic coronae, the amplification of magnetic fields by orbiting galaxies and their stripped tails and wakes, and the generation of ICM turbulence by galaxies. The idealized clusters in our simulations evolve in isolation and do not undergo any major mergers or accrete material; all galaxies begin to orbit within the group and cluster simultaneously. We describe simulations of an isolated $3.2 \times 10^{13} \, \msun$ group  and an isolated $1.2 \times 10^{14} \, \msun$ cluster with magnetized intracluster media and their galaxies with a cosmologically motivated distribution of initial galaxy masses. Our simulations are controlled experiments particularly designed to quantify the interactions between galaxies and the ICM. These simulations bridge the gap between cosmological simulations and wind tunnel simulations. Cosmological simulations realistically account for the buildup of clusters, but the unique effects of galaxies alone cannot be gleaned in the presence of mergers and accretion. Wind tunnel simulations cannot characterize the global properties of the ICM, and they do not capture the interactions of galaxies with other galaxies and their wakes. With our simulations, we can characterize the co-evolution of the ICM and cluster galaxies in the absence of other cluster-scale processes. Since the ICM is initially in hydrostatic equilibrium, the evolution of the magnetic field and ICM gas is due to physical processes induced solely by galaxies (in addition to numerical effects, which are quantified). Since the total initial mass in galaxies is an upper limit to the mass fraction bound to galaxies, the net impact on the ICM is an upper limit to the impact galaxies can possibly have over many Gyr in real clusters.  The temporal evolution of the ICM in our simulations, while not necessarily correlated with a cosmological temporal evolution of the ICM, quantify the dynamical time over which galaxies' effects persist within the ICM.  Therefore, the major strength of our simulations is in isolating the magnitude and timescale of the unique impacts that orbiting galaxies have on the ICM. 
 
The paper is structured as follows. We summarize our code and methods in \S~\ref{sec:methods}. In \S~\ref{sec:ic_mhd}, we describe our initial conditions. We describe the simulation results in \S~\ref{sec:results_mhd}, observational diagnostics and implications in \S~\ref{sec:obs_diagnostics}, and discuss our results in \S~\ref{sec:disc_mhd}. \S~\ref{sec:conclusions} summarizes our conclusions. Where needed we assume standard cosmological parameter values $H_0 = 71$ km s$^{-1}$ Mpc$^{-1}$, $\Omega_{\rm m} = 0.3$, and $\Omega_{\Lambda} = 0.7$.

\section{Methods}
\label{sec:methods}

The simulations in this paper were performed using \flash 4.2 (\citealt{Fryxell00}, \citealt{Dubey08, Dubey11}), a parallel $N$-body plus adaptive mesh refinement (AMR) Eulerian hydrodynamics code. The isolated group, cluster, and their galaxies consist of `live' dark matter halos represented using particles, and gas halos and coronae are initially in hydrostatic equilibrium with the group, cluster, and galaxy potentials. We refer the reader to Paper~I for details on the initial properties of the collisionless and hydrodynamic components of the group and cluster halos and their galaxies. We use cloud-in-cell (CIC) mapping to generate mesh density fields for particles. AMR is implemented using \paramesh (\citealt{MacNeice00}). We use a direct multigrid solver (\citealt{Ricker08}) to calculate the gravitational potential on the mesh. 

We treat the ICM plasma using the equations of ideal single-fluid MHD. In Gaussian units, these are
\begin{eqnarray}
\frac{\partial\rho}{\partial t} + \nabla\cdot(\rho{\bf u}) &=& 0\\
\frac{\partial \rho \mathbf{u}}{\partial t} + \nabla \cdot (\rho \mathbf{u} \mathbf{u}) + \nabla P &=& - \rho \nabla \Phi + \frac{1}{4 \pi} (\nabla \times \mathbf{B}) \times \mathbf{B} \\
\frac{\partial \rho  E}{\partial t} + \nabla \cdot [(\rho E + P)\mathbf{u}] &=& - \rho \mathbf{u} \cdot \nabla \Phi\ ,
\end{eqnarray}
where $\rho$, $P$, $\mathbf{u}$, $E$, and $\Phi$ are density, thermal pressure, velocity, specific total energy, and gravitational potential. The magnetic field $\mathbf{B}$ satisfies the induction equation,
\begin{equation}
\frac{\partial \mathbf{B}}{\partial t} + \nabla \times (\mathbf{B} \times \mathbf{u}) = 0 .
\end{equation} 

To numerically solve the MHD equations, we use an unsplit staggered mesh (USM) algorithm described in \citet{Lee09} and \citet{Lee13}. The USM algorithm, a finite-volume, second-order Godunov method, uses a directionally unsplit scheme to evolve the MHD equations. The divergence-free constraint on magnetic fields, $\nabla \cdot \mathbf{B} = 0$, is enforced using the constrained transport method of \citet{Evans88}. We use the HLLD Riemann solver (\citealt{Miyoshi05}) in \flash to calculate high-order Godunov fluxes. Magnetic fields are injected from cells on coarser to finer levels of refinement on the AMR grid using the prolongation method in \citet{Balsara01}, preserving the divergence-free character of the magnetic field.

\subsection{Initial Conditions}
\label{sec:ic_mhd}

The group and cluster halo and their galaxies are initialized using the method in Paper~I, with the initial parameters of the group and cluster halos identical to those in Table~1 of that paper. The satellite and central galaxies have the same masses, positions, and velocities. The ICM in the simulations in this paper, in addition to the hydrodynamic component, is threaded by magnetic fields. The primary goal of these simulations is to study the effect that ICM magnetic fields have on galactic coronae and the effect of galaxy motions on the ICM magnetic field; the galaxies themselves do not have magnetic fields separate from that of the ICM.

The initial strength and structure of the ICM magnetic field are determined from observations of relaxed clusters. The strength of the magnetic field is controlled by the plasma $\beta$ parameter, where $\beta \equiv P / P_{\rm magnetic} = 8\pi P/|{\bf B}|^2$. The thermal pressure $P$ as a function of cluster- or group-centric radius is calculated as in Paper~I, assuming hydrostatic equilibrium and a pre-determined cool-core entropy profile. The plasma $\beta$ parameter, hereafter referred to simply as $\beta$, is inversely proportional to the square of the magnetic field: a higher value of $\beta$ implies a weaker magnetic field and vice versa.

Observational evidence based on rotation measure (RM) studies (e.g.\ \citealt{Kim90, Kim91}, \citealt{Taylor93}, \citealt{Clarke00}, \citealt{Carilli02}, \citealt{Vogt05}) indicates that the typical magnetic field strength in the ICM is $\sim 1 - 10 \, \muG$. For typical ICM thermal pressure values, this corresponds to $\beta \simeq 100$. In this work, we adopt $\beta = 100$ as the initial ratio of the total thermal energy to the total magnetic energy in the simulation volume. This value of $\beta$ corresponds to initial magnetic field strengths of $\sim 0.5 - 4 \muG$; subsequent amplification by galaxies results in peak magnetic field strengths of $10 - 20 \muG$.

We assume that the magnetic fields in the cluster and group are random and isotropically oriented with a Kolmogorov-like power spectrum. This assumption is motivated by \citet{Vogt03, Vogt05}, who determined the power spectrum of the cluster magnetic field in  three clusters using RM analyses. In our simulations, stochastic magnetic fields are generated using the procedure outlined in \citet{Ruszkowski07}. A similar approach has been used to generate ICM magnetic fields for \flash simulations in \citet{Ruszkowski10} and \citet{ZuHone11b}. We describe our method and assumptions below.

To ensure the magnetic field initially satisfies $\nabla\cdot{\bf B}=0$, we construct it from the vector potential ${\bf A}$ via ${\bf B} = \nabla\times{\bf A}$. $\mathbf{A}$ is initialized on a uniform grid in $\mathbf{k}$-space. The amplitude of the Fourier transform $\tilde{A}(\mathbf{k})$ is
\begin{equation}
\tilde{A}(k) \propto k^{-1}\tilde{B}(k), 
\end{equation}
where $k = |\mathbf{k}|$. $\tilde{B}(k)$ is assumed to have a Kolmogorov-like spectrum with exponential cutoff terms, and in line with previous studies (\citealt{Ruszkowski07}, \citealt{ZuHone11b}), we adopt
\begin{equation}
\tilde{B}(k) \propto k^{-11/6} \exp[-(k/k_{\rm high})^2] \exp[-k_{\rm low}/k] ,
\end{equation}
where $k_{\rm high} = 2 \pi / \lambda_{\rm min}$ is a high wavenumber cutoff, corresponding to the assumed coherence length of the magnetic field, and $k_{\rm low} = 2 \pi / \lambda_{\rm max}$ is a low wavenumber cutoff, comparable to the size of the group or the cluster. In these simulations, we use $\lambda_{\rm min} = 43$ kpc and $\lambda_{\rm max} = 500$ kpc, consistent with previous ICM simulations by \citet{ZuHone11b}. 

To ensure that the phase of the final magnetic field is uniformly distributed, the three Cartesian components $\tilde{A}_x(\mathbf{k})$, $\tilde{A}_y(\mathbf{k})$, and $\tilde{A}_z(\mathbf{k})$ are treated independently and set to
\begin{equation}
\tilde{A}_x(\mathbf{k}) = \tilde{A}(k)[G(u_{x1}) + iG(u_{x2})],
\end{equation}
\begin{equation}
\tilde{A}_y(\mathbf{k}) = \tilde{A}(k)[G(u_{y1}) + iG(u_{y2})],
\end{equation}
\begin{equation}
\tilde{A}_z(\mathbf{k}) = \tilde{A}(k)[G(u_{z1}) + iG(u_{z2})],
\end{equation}
where $G(u_i)$ returns Gaussian-distributed random values of the uniformly distributed random variables $u_i$.
$\tilde{A}_x(\mathbf{k})$, $\tilde{A}_y(\mathbf{k})$, and $\tilde{A}_z(\mathbf{k})$ are inverse Fourier transformed, and the corresponding amplitudes of $A_x(\mathbf{x})$, $A_y(\mathbf{x})$, and $A_z(\mathbf{x})$ are calculated on a uniform grid and interpolated onto the AMR grid. $B_x(\mathbf{x})$, $B_y(\mathbf{x})$, and $B_z(\mathbf{x})$ are then calculated using second-order finite differences. The final initialization step is to normalize $\mathbf{B}(\mathbf{x})$, to ensure that the ratio of the total thermal energy to the total magnetic energy is equal to the chosen value of $\beta$. To satisfy this criterion, we calculate $\beta_{\rm avg} = \left(\int dV\,P\right)/\left(\int dV\,P_{\rm magnetic}\right)$ and then multiply $\mathbf{B}(\mathbf{x})$ by  $\sqrt{\beta / \beta_{\rm avg}}$ throughout the domain. Note that the magnetic field is initialized after galaxies have been initialized, so $\beta_{\rm avg}$ includes the thermal energy of the galaxies' ISM. No additional galactic fields are created; initially the galaxies are threaded by the group/cluster field.

The simulation boxes in which the group and cluster are evolved are identical to those in Paper~I. The group halo and its galaxies are simulated in a cubic box of side $10^{25}$~cm (3.24~Mpc) and the cluster halo and its galaxies in a cubic box of side $2 \times 10^{25}$~cm (6.48~Mpc). We use a minimum of 4 (5) levels of mesh refinement in the group (cluster), corresponding to a base grid resolution of $25.6$~kpc. The maximum resolution is $1.6$~kpc, corresponding to 8 (9) levels of refinement in the group (cluster). All the galaxies and their stripped tails, in addition to the ICM within the virial radius, are fully resolved at the maximum spatial resolution. The refinement criteria are as used in Paper I. We use periodic boundary conditions for the gravity and magnetohydrodynamics; since the virial radius of both the group and cluster are significantly smaller than the box sizes the periodicity of the boundaries does not affect the evolution of the group and cluster.

\section{Simulation Results}
\label{sec:results_mhd}

\subsection{The evolution of the ICM magnetic field in the absence of galaxies}
\label{sec:isobfield}

The magnetic field initialized in the fashion described in \S~\ref{sec:ic_mhd} is not force-free and therefore not relaxed. In the absence of any other dynamical processes, the magnetic field relaxes over many Gyr and the overall magnetic pressure decreases. For $\beta \gtrsim 100$, the magnetic field is dynamically unimportant. Since the group or cluster starts close to hydrostatic equilibrium it should therefore remain so.

\begin{figure}[!htbp]
  \begin{center}
    {\includegraphics[width=0.5\textwidth]{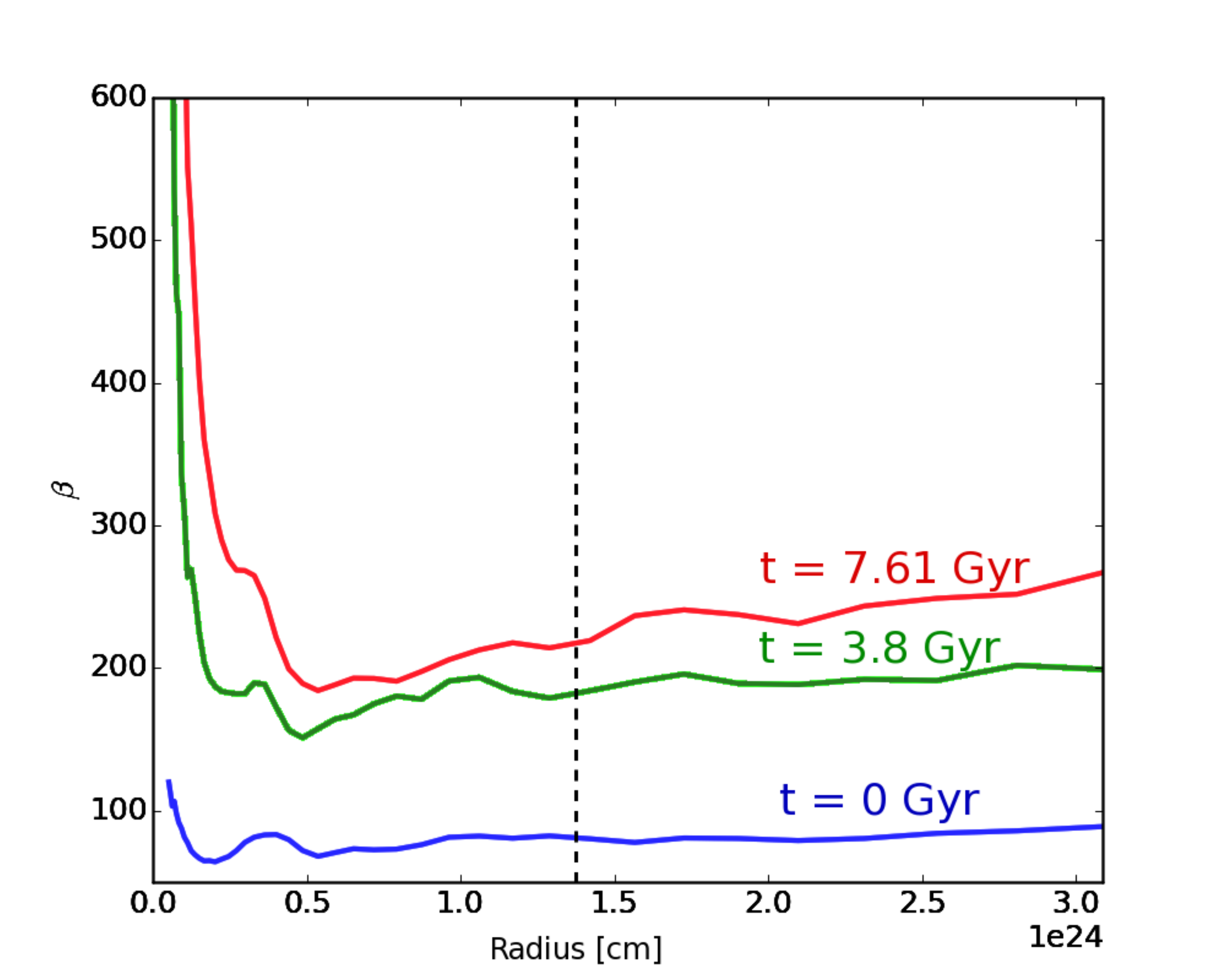}}    
    \caption{Evolution of the azimuthally averaged $\beta$ profile for a $3.2 \times 10^{13} \msun$ group with no galaxies. The black dashed line corresponds to the location of the group's $R_{200}$. Colors correspond to different simulation times.\label{fig:beta_nogal}}
  \end{center}  
\end{figure}

Figure~\ref{fig:beta_nogal} illustrates the evolution of the azimuthally averaged radial profile of $\beta$ for a $3.2 \times 10^{13} \msun$ group. Overall, $\beta$ increases with time as the magnetic field relaxes and the magnetic pressure decreases. This effect has been quantified in the context of galaxy cluster evolution in previous studies by  \citet{Ruszkowski07} and \citet{ZuHone11b}. As seen in the next section, in the presence of galaxy motions, the magnetic field strength \emph{increases}. Relaxation is therefore suppressed during the early phases of cluster galaxies' orbital evolution. 

\subsection{Galaxy stripping in a magnetized ICM}
\label{sec:mhd_stripping}

In the absence of viscosity and thermal conduction, the ICM magnetic field should have two distinct effects on the evolution of galaxies: if the magnetic field is strong enough, the increased ICM pressure on galaxies due to the magnetic pressure term can lead to increased gas loss, but the magnetic field itself can suppress the formation of hydrodynamic instabilities, thereby suppressing gas loss in the tails of stripped galaxies. For $\beta \simeq 100$, the magnetic pressure is too low to significantly affect the ICM pressure on galaxies, but lower values of $\beta$ and the suppression of instabilities can significantly affect the dynamics of galactic gas.

\begin{figure*}[!htbp]
  \begin{center}
  %\vspace{-2em}
    \subfigure[]
    {\includegraphics[width=3.2in]{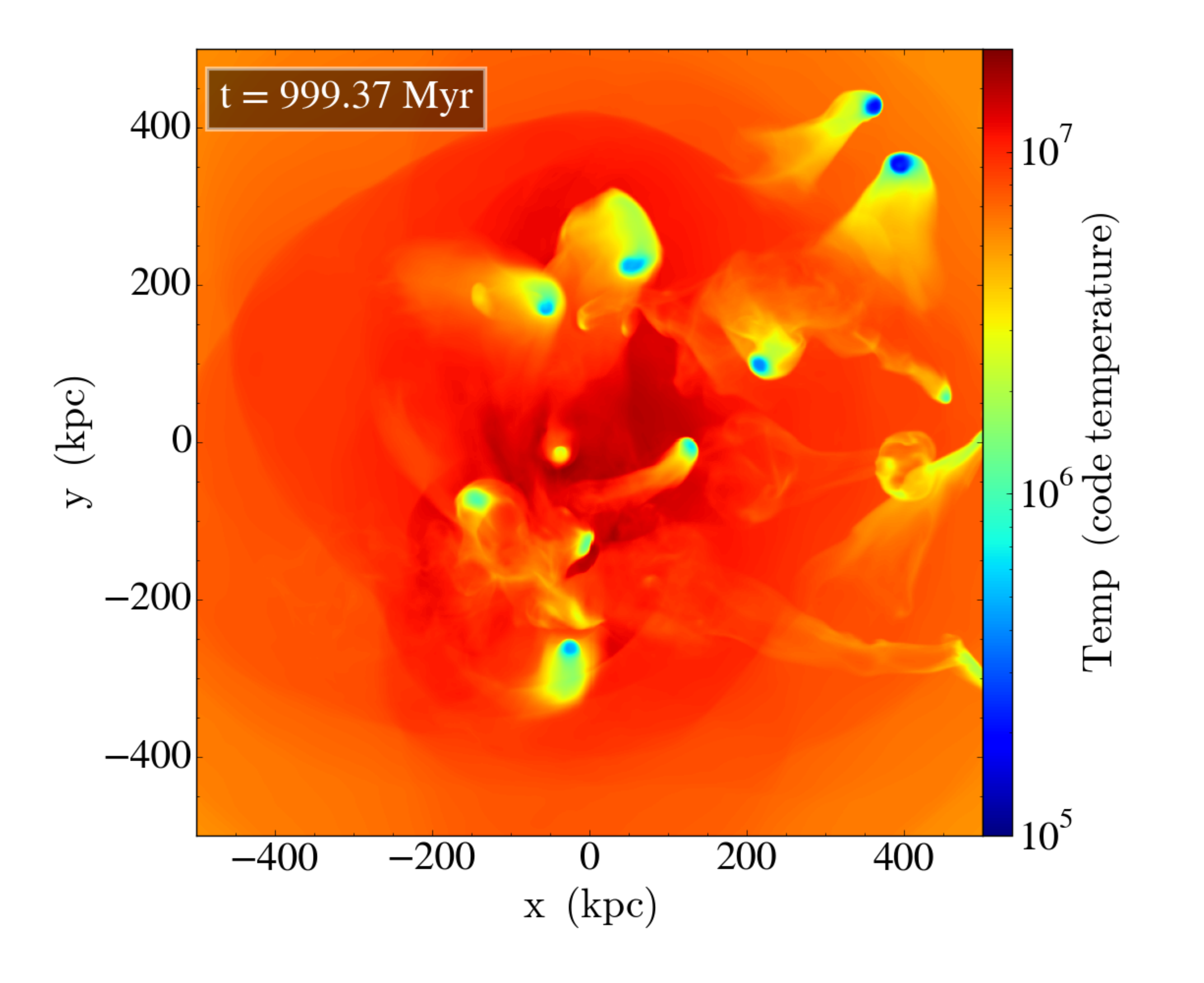}\label{fig:groupmhdTEM80}}
    \subfigure[]
    {\includegraphics[width=3.2in]{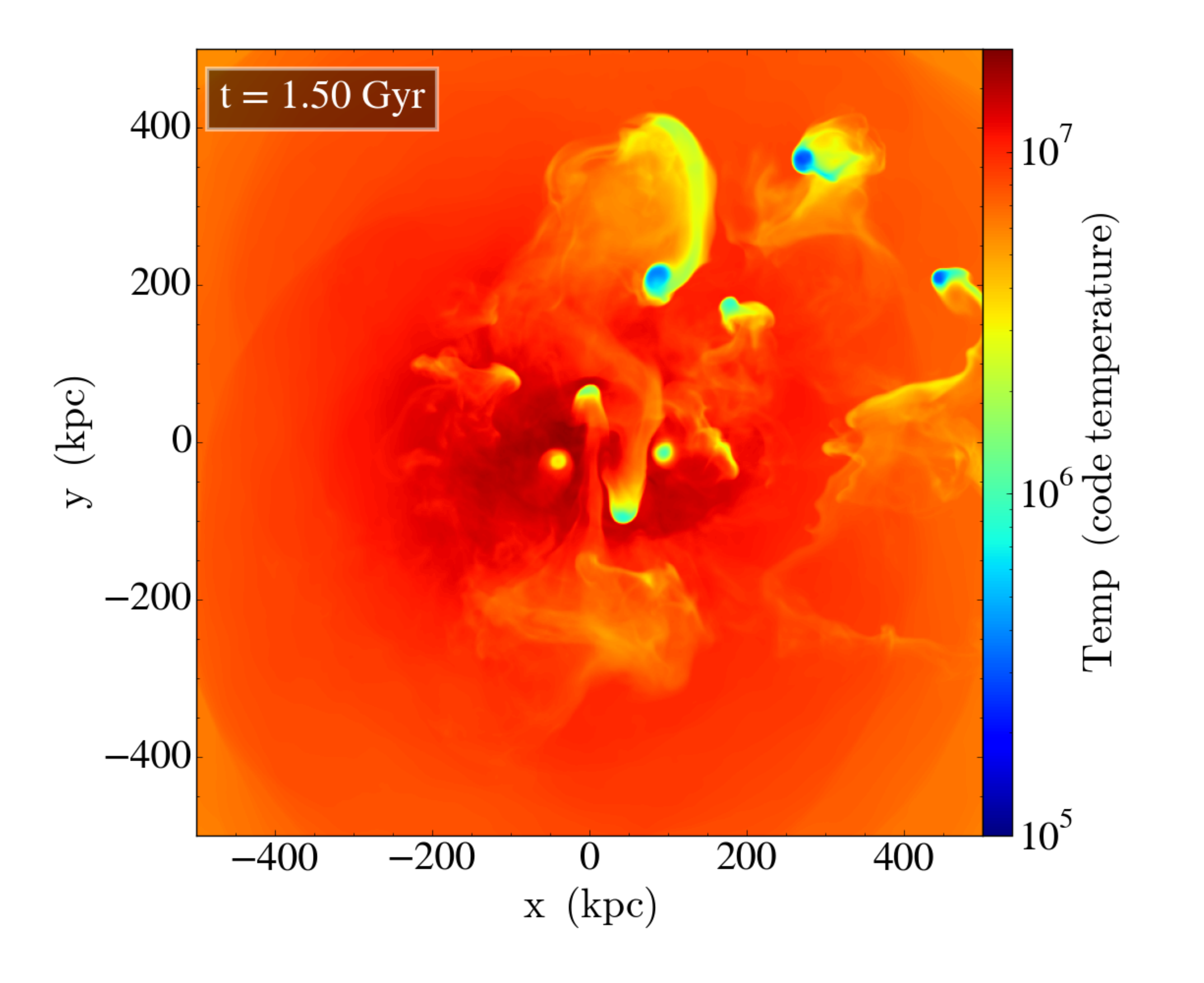}\label{fig:groupmhdTEM120}}
    %\\ \vspace{-2.em}
     \caption{The projected emission measure-weighted temperature (in K) of group galaxies. An animation of this figure showing the evolution from 0 to 3.11 Gyrs is available. \label{fig:mhdT_EM_group}}
  \end{center}  
\end{figure*}

\begin{figure*}[!htbp]
  \begin{center}
  %\vspace{-2em}
    \subfigure[]
    {\includegraphics[width=3.2in]{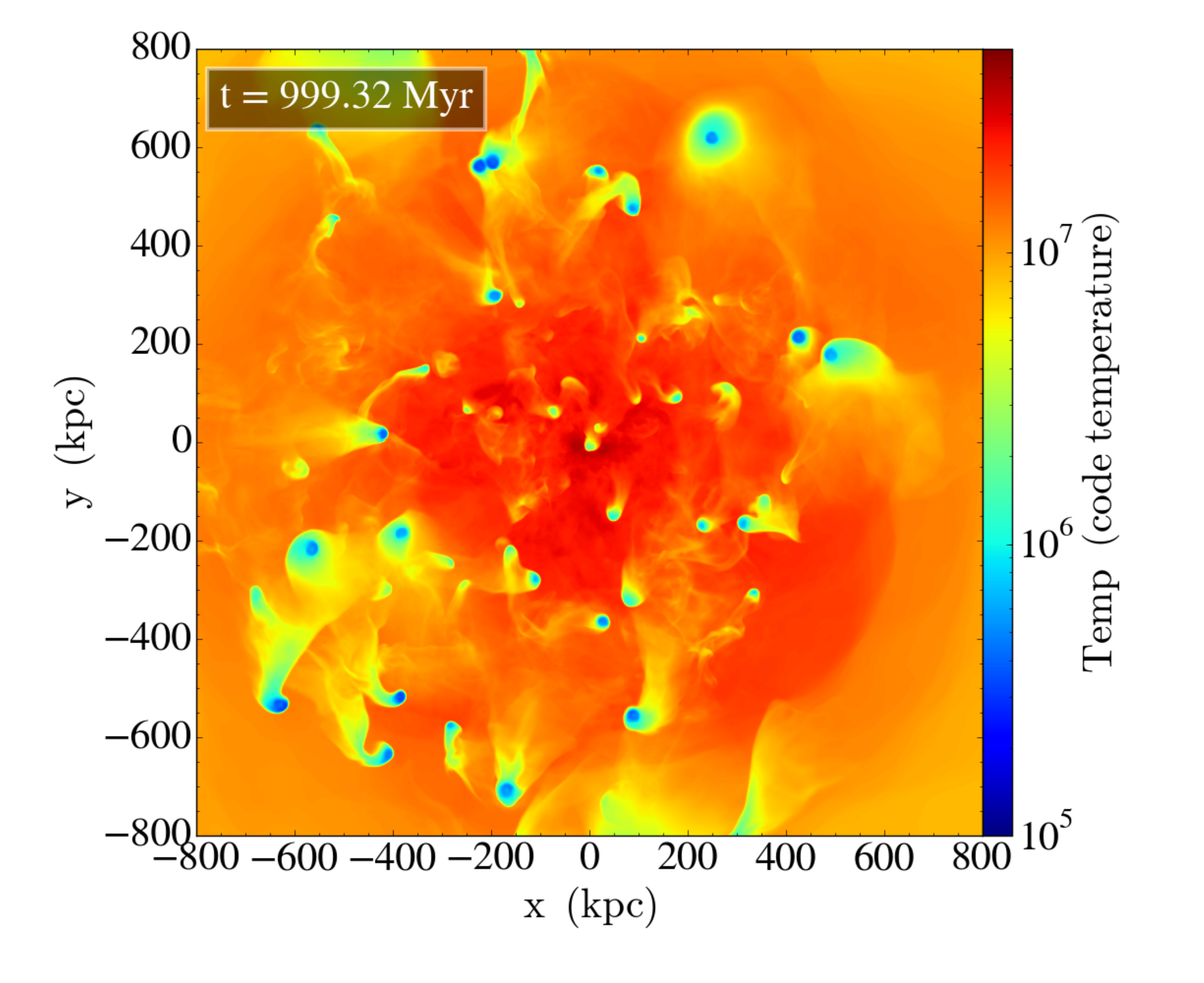}\label{fig:clustermhdTEM80}}
    \subfigure[]
    {\includegraphics[width=3.2in]{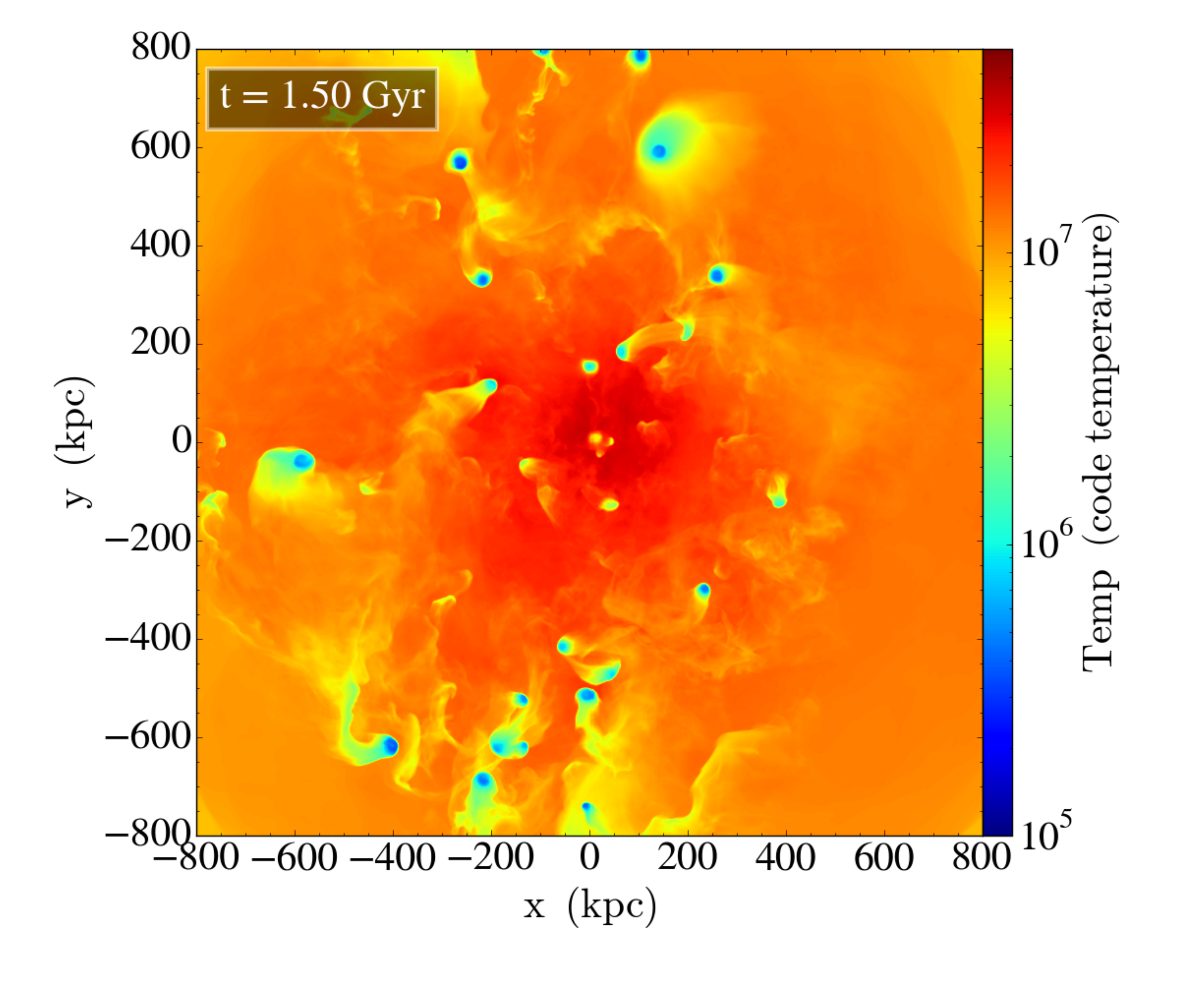}\label{fig:clustermhdTEM120}}
     \caption{The projected emission measure-weighted temperature (in K) of cluster galaxies. An animation of this figure showing the evolution from 0 to 3.11 Gyrs is available. \label{fig:mhdT_EM_cluster}}
  \end{center}
\end{figure*}

\begin{figure*}[!htbp]
  \begin{center}
  %\vspace{-2em}
    \subfigure
    {\includegraphics[width=3.2in]{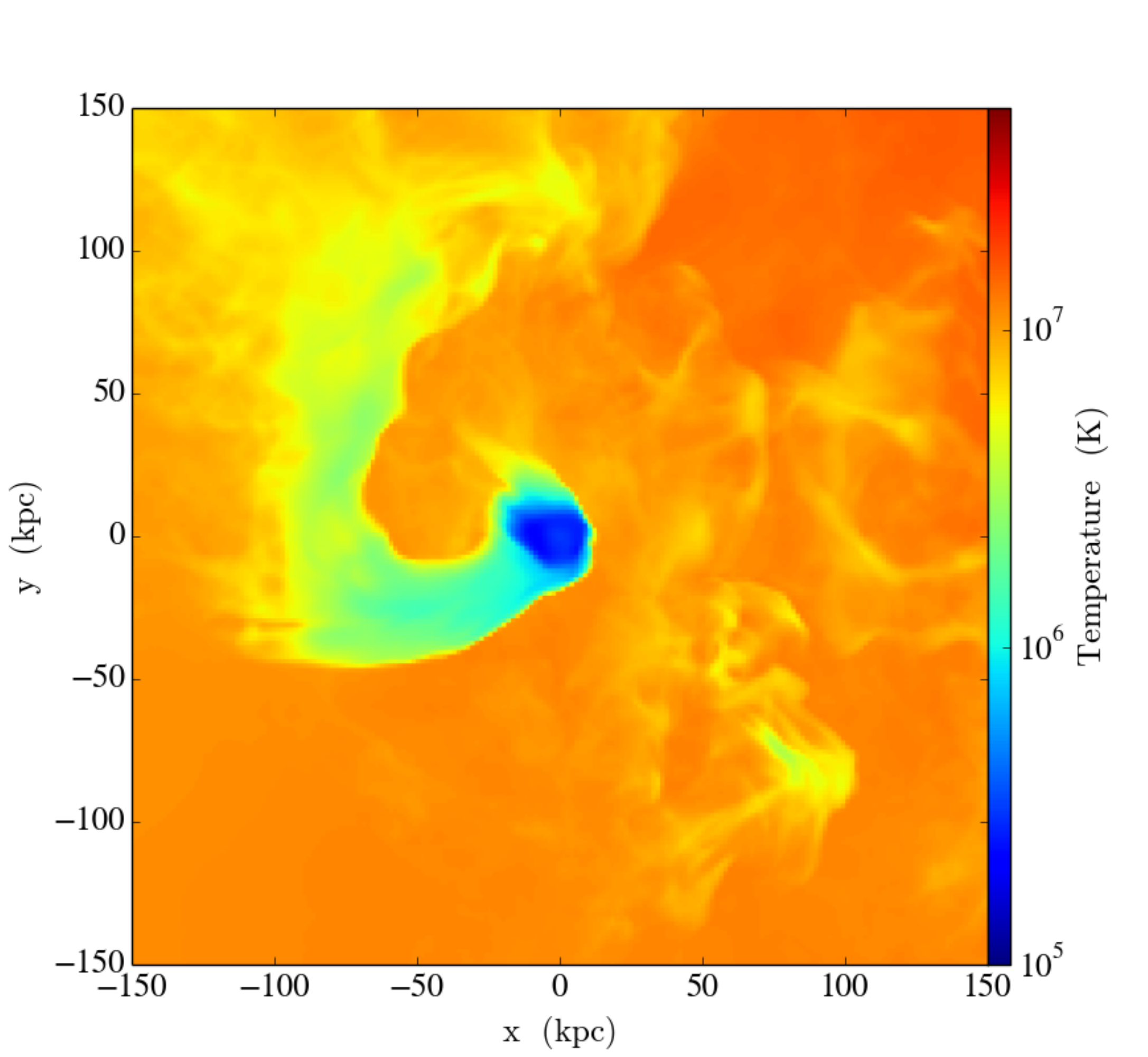}\label{fig:gal21_hydro}}
    \subfigure
    {\includegraphics[width=3.2in]{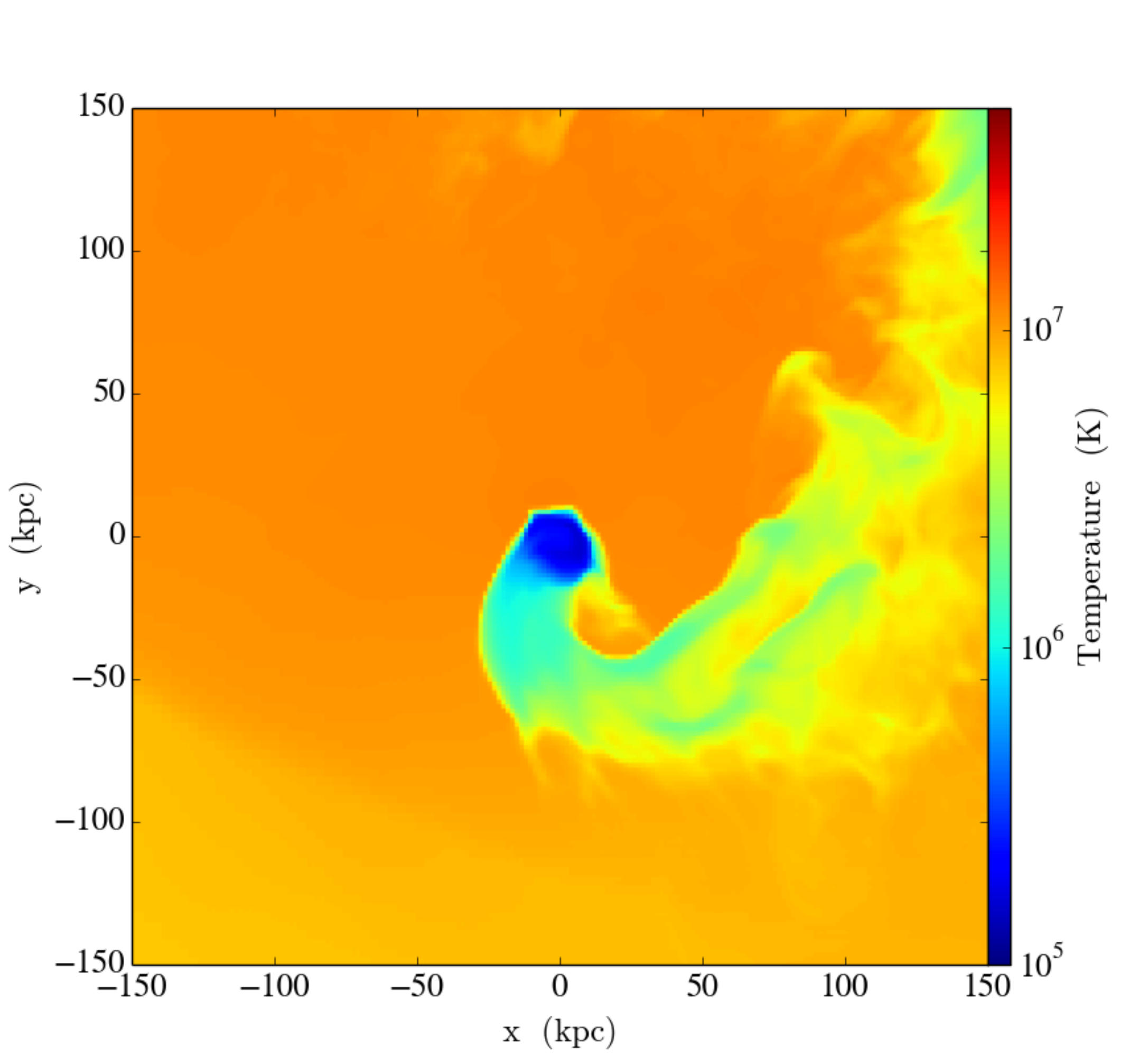}\label{fig:gal94_hydro}}
    %\\ \vspace{-2.em}
    \subfigure
    {\includegraphics[width=3.2in]{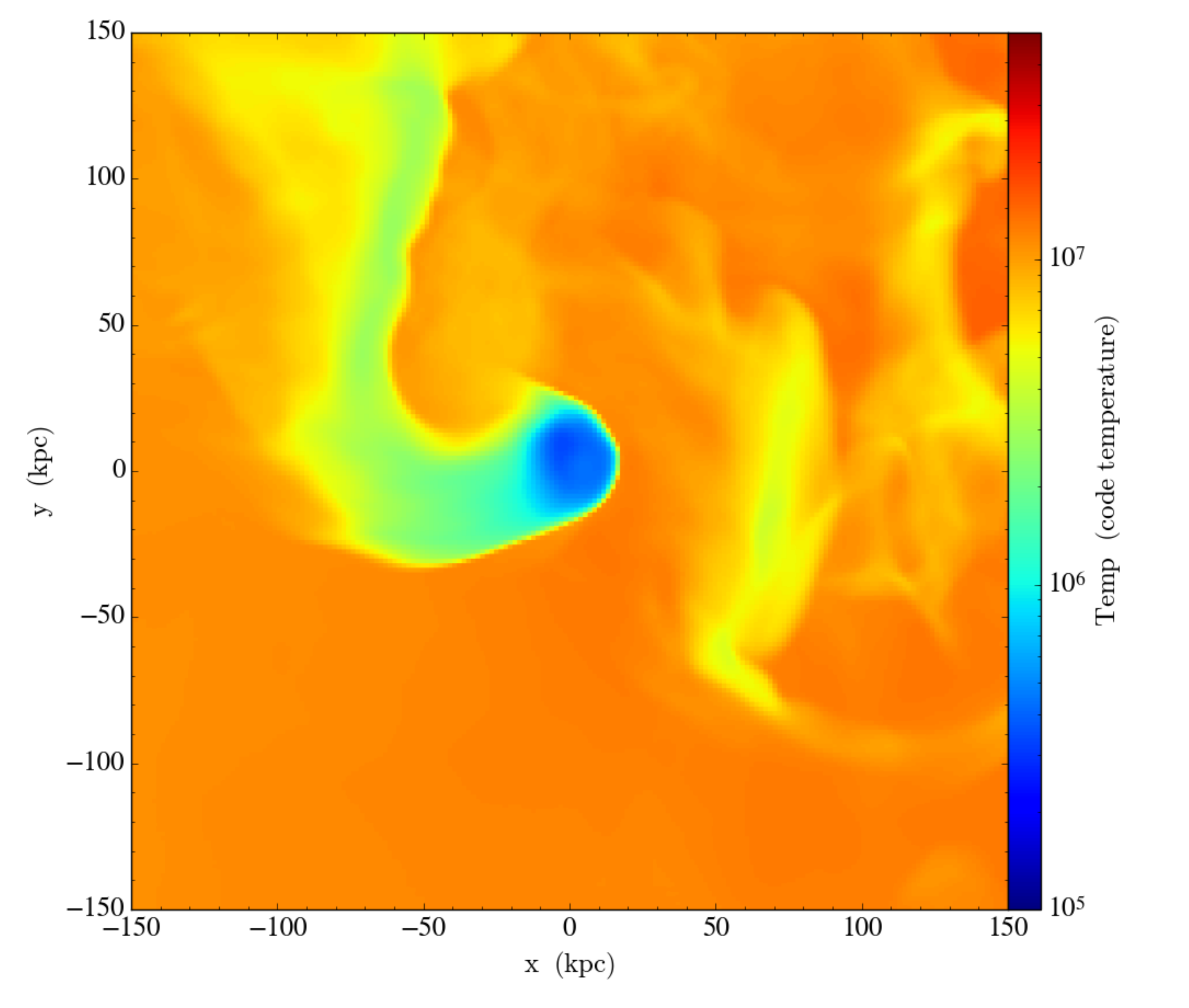}\label{fig:gal21_mhd}}
    \subfigure
    {\includegraphics[width=3.2in]{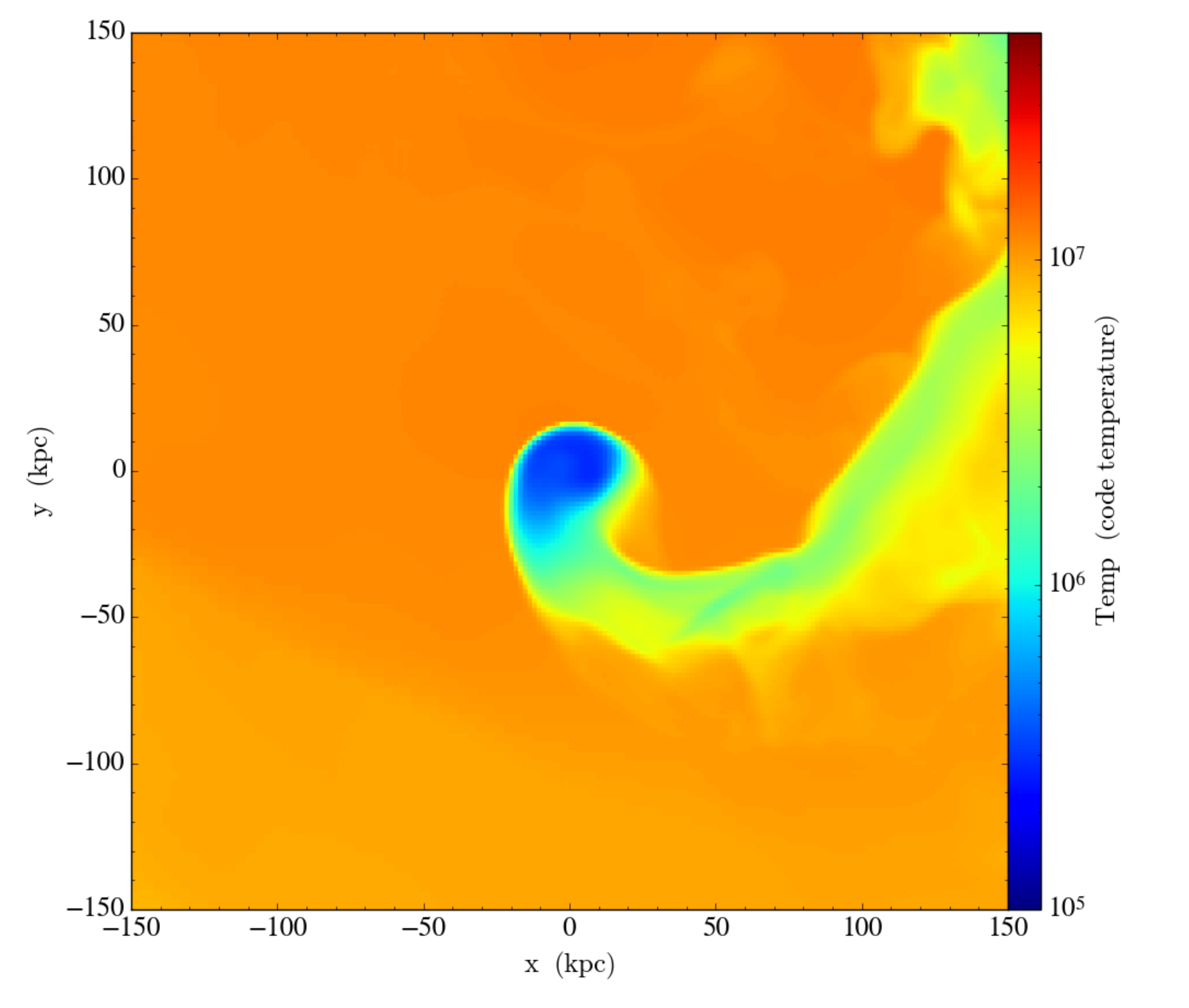}\label{fig:gal94_mhd}}
     \caption{Zoomed in stripped galaxies at $t = 1.46$ Gyr in the cluster in the pure hydrodynamic simulation (top row, from Figure 19 of Paper I), and the same galaxies in the simulation with magnetic fields  (bottom row). \label{fig:gal_hydroMHD}}
  \end{center}  
\end{figure*}

Figures~\ref{fig:mhdT_EM_group} and ~\ref{fig:mhdT_EM_cluster} show snapshots of the emission measure weighted temperature  of galaxies and the ICM in the group and cluster. These snapshots can be compared to those in Figure 3 of Paper I. At $t = 1$~Gyr, the temperature snapshots in both magnetized and unmagnetized simulations are qualitatively almost identical at first glance, but closer inspection shows that the prominent Kelvin-Helmholtz rolls associated with galaxies in the central regions of the group are absent in galaxies in the MHD simulations, particularly in Figures~\ref{fig:groupmhdTEM80} and ~\ref{fig:clustermhdTEM80}. Both the narrow galaxy tails in the group core and the wide galaxy wakes in the outer regions of the group are noticeably smoother and nearly featureless. At $t = 1.5$~Gyr, this difference still persists: supported by magnetic fields aligned with the stripped tails, galaxy tails are noticeably narrower and smoother compared to the wide, diffuse stripped tails in the hydrodynamic simulations. By $t = 2$~Gyr, the appearance of the stripped tails in projection is markedly different: tails in the MHD simulations are smaller, narrower, and less disrupted (Figure~\ref{fig:gal_hydroMHD}). There are no wide galactic tails in the MHD simulation, but more galaxies have narrow tails attached to them than in the hydrodynamic simulation. Qualitatively, one can therefore conclude that the overall effect of ICM magnetic fields on stripped galactic tails is to  become aligned with the tails as galaxies move through the ICM, and prevent their dissipation through shear instabilities in the galaxy tail -- ICM interface. We further discuss the reason for this phenomenon -- the suppression of KH instabilities by magnetic fields -- in \S~\ref{sec:disc_mhd_coronae}.

\begin{figure}[!htbp]
  \begin{center}
  \subfigure[No magnetic fields]
    {\includegraphics[width=0.5\textwidth]{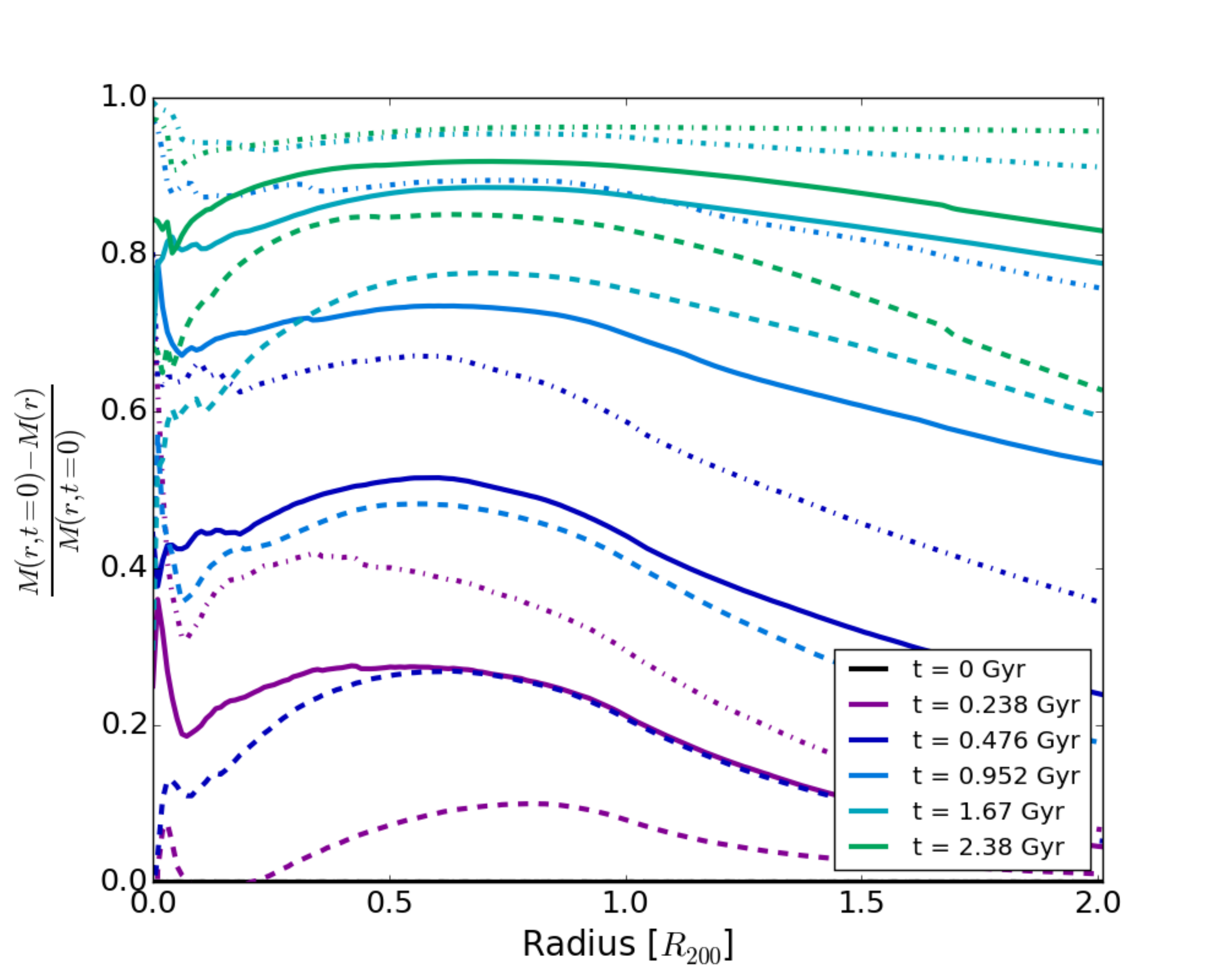}}  
   \subfigure[With magnetic fields]
    {\includegraphics[width=0.5\textwidth]{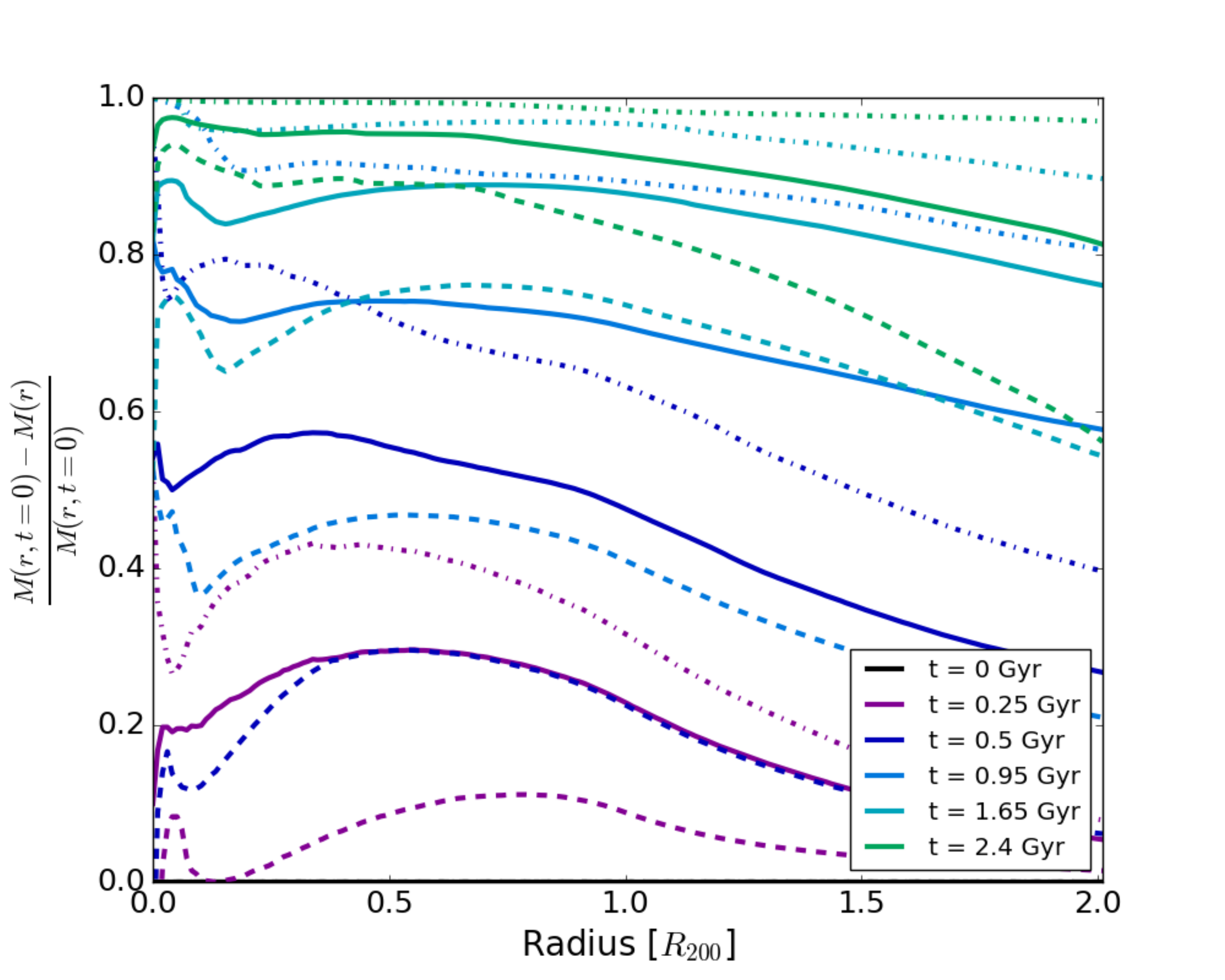}}    
    \caption{Stacked differential mass profiles as a function of time for group  galaxies. The solid lines correspond to all group galaxies. The dashed lines are for galaxies that have initial masses $M > 10^{11}\ \mbox{M}_{\odot}$, and the dotted lines are for galaxies with initial masses $M < 10^{11}\ \mbox{M}_{\odot}$. Top: From hydrodynamic simulation (Paper I). Bottom: From simulation with magnetic fields. While the timesteps at which differential mass loss profiles are calculated are not identical in both simulations, they are close enough that they do not affect our overall interpretation of the results, i.e.\ that a magnetized intracluster medium with $\beta = 100$ does not affect the overall gas mass loss rate from galaxies. \label{fig:gaslossratemhd} } 
  \end{center}  
\end{figure}

Although the appearance of stripped tails is markedly different in simulations with and without magnetic fields, the amount of gas retained in the cores of galaxies is not significantly affected. Figure~\ref{fig:gaslossratemhd} shows radial profiles of the stacked differential mass loss rate for group galaxies up to $t = 2.4$~Gyr in simulations with and without magnetic fields. At comparable timesteps, the amount of mass lost at any given radius is comparable in hydrodynamical and MHD simulations. This makes sense since gas mass loss is primarily driven by the net amount of pressure that a galaxy is subject to, and for $\beta \gg 1$, magnetic pressure does not significantly contribute to this component. At late times (discussed in the following section), $\beta$ decreases on average from $\sim 100 $ to $\sim 50$ at $t = 0.5$~Gyr and to $\sim 20$ at $t = 1 - 2$~Gyr. The corresponding increase in magnetic pressure is still not effective in significantly modifying the overall mass loss rate for at least two reasons: (i) even for $\beta \simeq 20$, the magnetic pressure is 20 times lower than the thermal pressure, and (ii) galaxies have on average lost $50\%$ of the gas within $R_{200}$ by 0.5~Gyr, i.e., before the magnetic pressure has sufficiently increased to modify galaxy mass loss rates. 

\subsection{The evolution of the ICM magnetic field in the presence of galaxies}
\label{sec:galbfield}

Orbiting galaxies, particularly massive, gas-rich galaxies that interact with the magnetized ICM, can strengthen ICM magnetic fields and drive turbulence. In this section, we illustrate this process for an isolated group and cluster from $t = 0$~Gyr to $t = 3$~Gyr. Galaxies are stripped of $80 - 90\%$ of their gas by $t = 2$~Gyr (see Figure~\ref{fig:gaslossratemhd}; detailed analysis without magnetic fields is in Paper I). Due to the decrease in the force due to galaxies' ram pressure, the ICM magnetic field is not amplified  after  $t \simeq 2$ Gyr. Here we briefly illustrate the decay in the field strength in the case of the isolated group.

\begin{figure*}[!htbp]
  %\begin{center}
  \hspace{-1.2cm}
    {\includegraphics[width=1.1\textwidth]{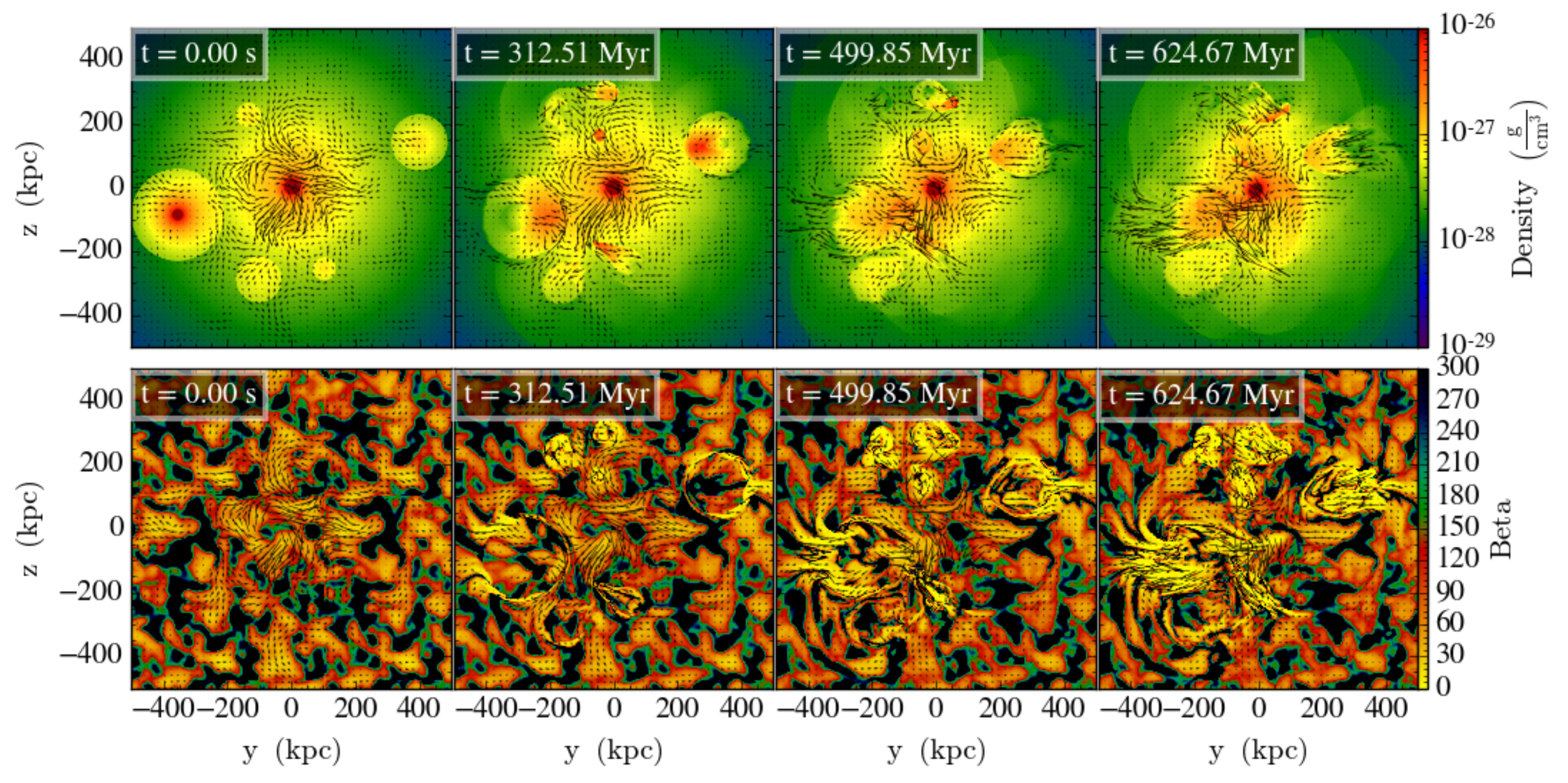}}
     \caption{Slices of density and $\beta$ from $t = 0 - 0.62$~Gyr in the $x = 0$ plane of the isolated group, annotated with magnetic field lines. An animation of this figure showing the evolution from 0 to 3.22 Gyrs is available. \label{fig:groupdensbeta1}}
  %\end{center}  
\end{figure*}

\begin{figure*}[!htbp]
  %\begin{center}
  \hspace{-1.2cm}
    {\includegraphics[width=1.1\textwidth]{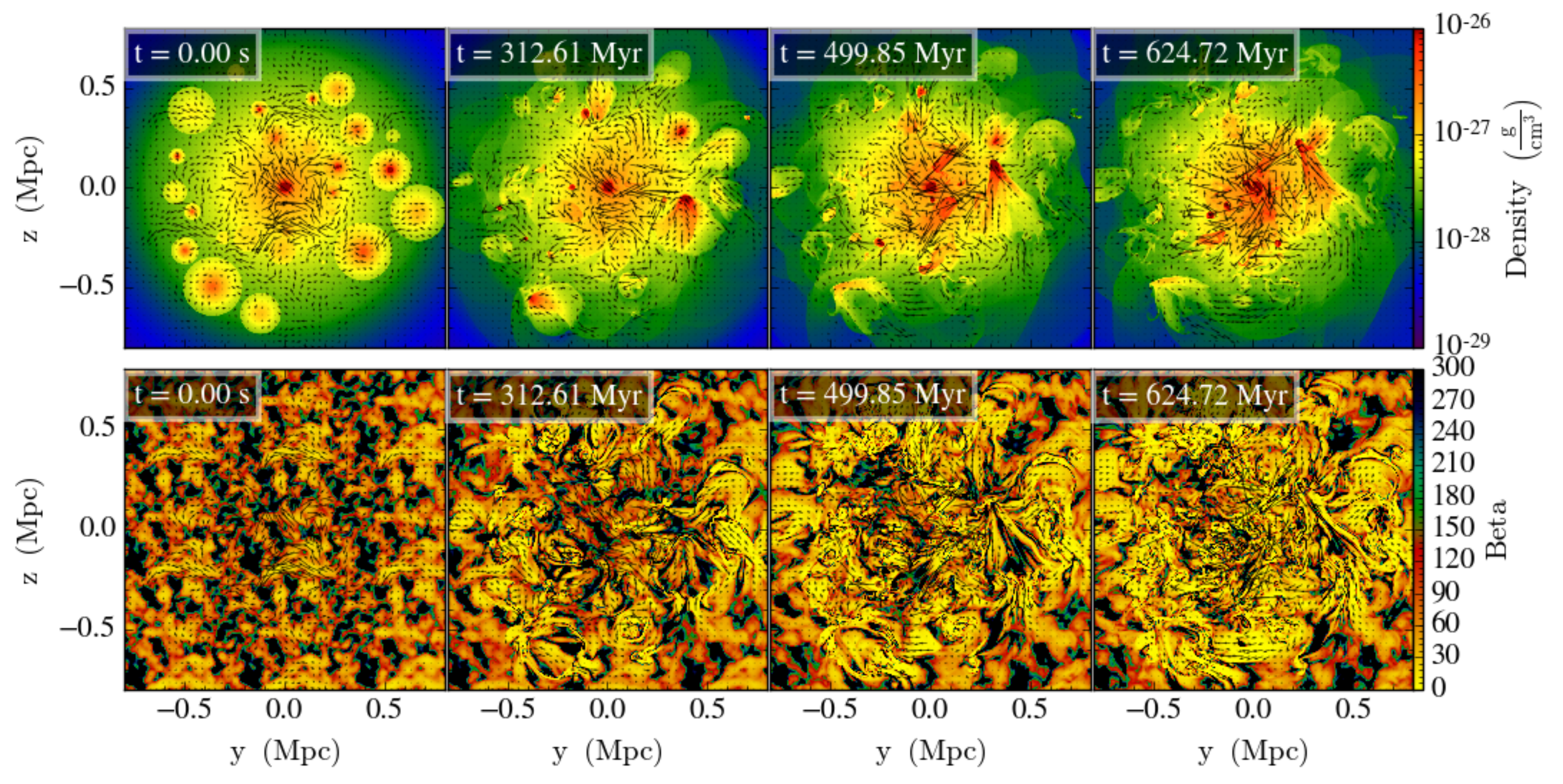}}
     \caption{Slices of density and $\beta$ from $t = 0 - 0.62$~Gyr in the $x = 0$ plane of the isolated cluster, annotated with magnetic field lines. An animation of this figure showing the evolution from 0 to 3.22 Gyrs is available. \label{fig:clusterdensbeta1}}
  %\end{center}  
\end{figure*}

\begin{figure*}[!htbp]
  %\begin{center}
  \hspace{-1.2cm}
    {\includegraphics[width=1.1\textwidth]{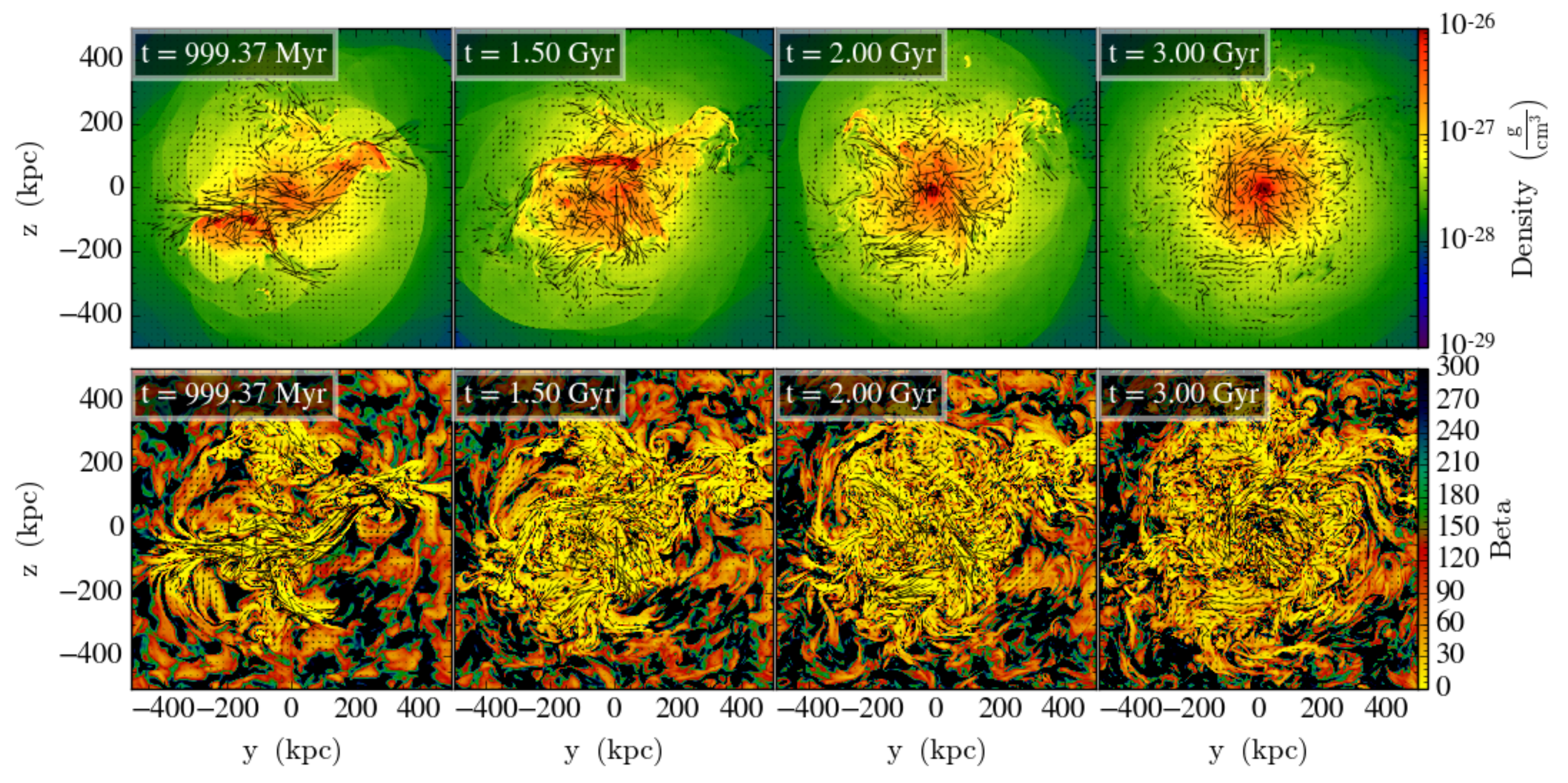}}
     \caption{Slices of density and $\beta$ from $t = 1 - 3$~Gyr in the $x = 0$ plane of the isolated group, annotated with magnetic field lines. An animation of this figure showing the evolution from 0 to 3.22 Gyrs is available. \label{fig:groupdensbeta2}}
  %\end{center}  
\end{figure*}

\begin{figure*}[!htbp]
  %\begin{center}
  \hspace{-1.2cm}
    {\includegraphics[width=1.1\textwidth]{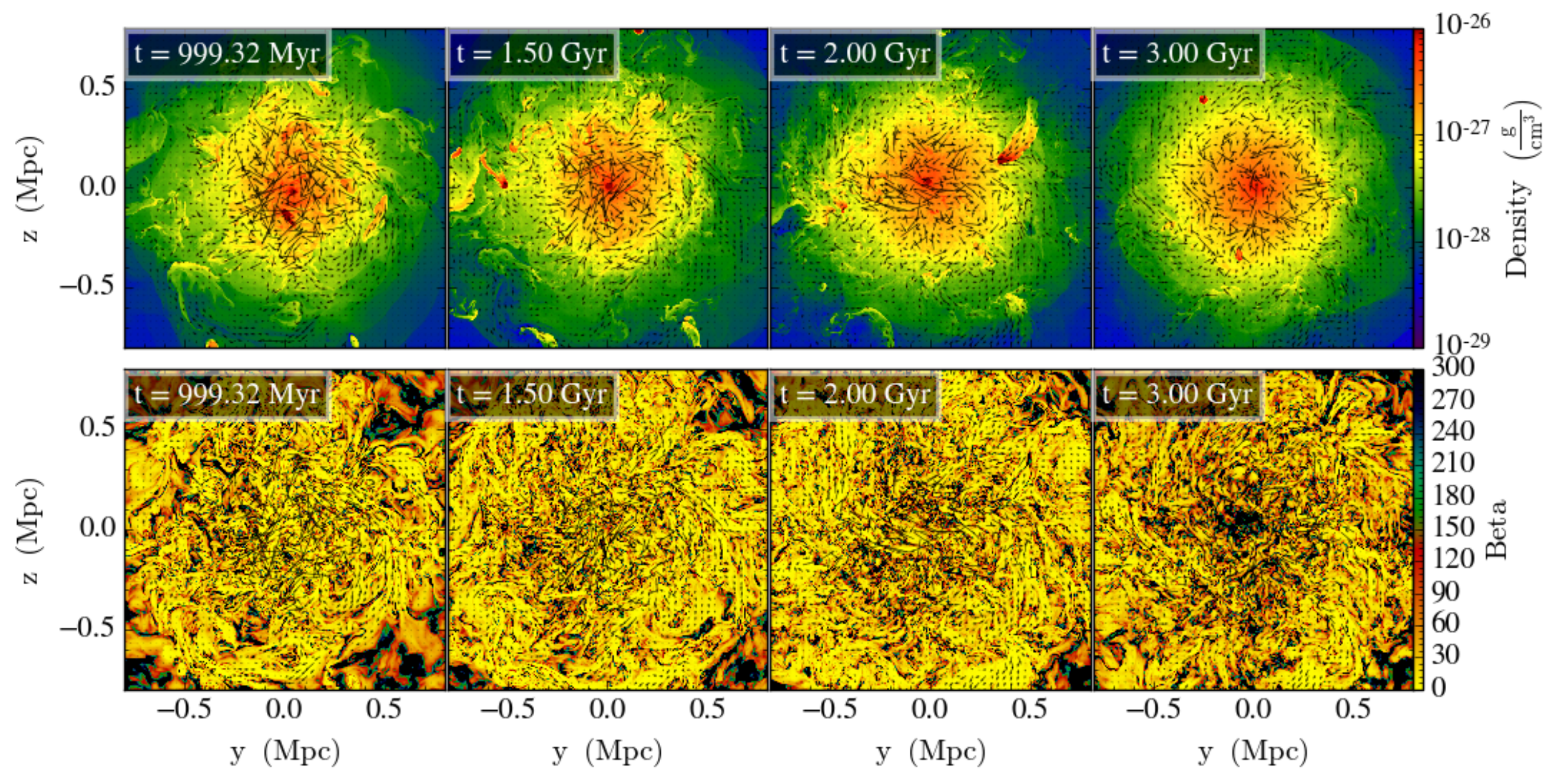}}
     \caption{Slices of density and $\beta$ from $t = 1 - 3$~Gyr in the $x = 0$ plane of the isolated cluster, annotated with magnetic field lines. An animation of this figure showing the evolution from 0 to 3.22 Gyrs is available. \label{fig:clusterdensbeta2}}
  %\end{center}  
\end{figure*}

Figures~\ref{fig:groupdensbeta1} and \ref{fig:groupdensbeta2} show slices of density (in the upper rows) and the plasma $\beta$ parameter (lower rows) in the $x = 0$ plane of the isolated group and its galaxies. These slices are annotated with magnetic field vectors.  Correspondingly, Figures~\ref{fig:clusterdensbeta1} and \ref{fig:clusterdensbeta2} show slices of density and $\beta$ in the $x = 0$ plane for the isolated cluster and its galaxies. In these snapshots, $\beta$ is used as a measure of the magnetic field strength. At $t = 0$~Gyr, the distribution of $\beta$ is random and isotropic. The galaxies do not have distinct magnetic fields themselves. The distribution of galaxies in the density slice at $t = 0$~Gyr is uncorrelated with the magnetic field structure. 

By $t = 0.5$~Gyr (third columns in Figures~\ref{fig:groupdensbeta1} and \ref{fig:clusterdensbeta1}), prominent structures driven by galaxy motions appear in the magnetic field. Most distinct is the increase in $\beta$ \emph{behind} galaxies, along their direction of motion, well before their gas is stripped and forms tails. This is due to field lines being dragged and stretched by galaxies moving through the ICM. In addition, field strength is enhanced along the outer boundaries of galaxy coronae, at the ISM-ICM interface, where the ICM is compressed as it moves past the galaxies' surfaces. As galaxies are further stripped, these structures become more pronounced at  $t \simeq 0.6 - 0.8$~Gyr. $\beta$ increases outside-in in the tails and edges of galaxies being stripped, and the fraction of stripped galactic gas that is magnetized increases significantly. The interiors of the more massive galaxies are yet to be significantly affected by the magnetic field. 

By $t = 0.63$~Gyr (right columns, Figures~\ref{fig:groupdensbeta1} and \ref{fig:clusterdensbeta1}), the magnetic field strength continues to be amplified along low density wakes of ICM gas that trail galaxies. Distinct galaxy tails are not visible for all galaxies, since these are slices rather than projections, but magnetic field lines associated with galaxy tails in regions of low $\beta$ are clearly seen. At $t = 1$ Gyr in the group (first column, Figure~\ref{fig:groupdensbeta2}), the two massive galaxies from the previous snapshots have been stripped to the characteristic central corona plus stripped tail structure. Similarly, corona-plus-tail galaxies are visible in the cluster at $t = 0.63$ Gyr. Stripped and elongated tails are partially supported by magnetic pressure, and the alignment of magnetic field lines along these tails suppresses the formation of shear instabilities at the interface between these tails and wakes and the ICM. Even at $t = 1$ Gyr, while the tail of the most distinctive galaxy in the group snapshot (with the center at $[y, z] \simeq [200, 100]$ kpc) is magnetized, the central coronal region is largely unmagnetized and shielded. Galaxies with less prominent tails  also have coronae with significantly weaker magnetic fields than the surrounding stripped gas. Although the overall structure of this unstripped gas has been subject to compression and tidal stretching, it has yet to actually mix with the ICM. 

At $t \gtrsim 1$ Gyr (Figures~\ref{fig:groupdensbeta2} and \ref{fig:clusterdensbeta2}), distinct unmagnetized coronae are no longer visible except for the most massive galaxies. Magnetic field lines trace the orbits of stripped tails; although some of the tails themselves are no longer visible as overdense regions in the density slice, their associated $\beta$ decrements persist. Areas through which galaxies have passed are clearly visible in the $\beta$ slice and from the aligned magnetic field vectors in the density slice. Additionally, shock waves driven by galaxies are clearly seen at all timesteps in the density slices, but there are no corresponding features in the $\beta$ slices. These weak shocks do not significantly affect the magnetic field. After $t \gtrsim 1.5$ Gyr, stripped tails widen and become more diffuse,  but their associated magnetic field enhancements are not affected. $\beta$ continues to decrease in previously quiescent regions as the orbits of galaxies and their tails sweep increasingly larger fractions of the group volume. 

The tail and magnetic field of the group galaxy centered at $(y, z) \simeq (100, 100)$ kpc at  $t = 1.5$ Gyr (second column, Figure~\ref{fig:groupdensbeta2}, left column are particularly interesting. This galaxy's orbit bends close to the $x = 0$ plane, as a result of which its stripped tail has a bent, almost $90^{\circ}$ shape between $t = 1.15$ and $t = 1.5$ Gyr. The magnetic field lines aligned with this tail also bend correspondingly, showing that dramatic orbital turns can drag along field lines, in addition to gentler bending of field lines seen in other galaxies. 

After $t \simeq 1.5$ Gyr (the two right hand columns of Figures~\ref{fig:groupdensbeta2}, \ref{fig:clusterdensbeta2}),  the ICM magnetic field becomes increasingly chaotic and complex. The magnetic field structure at this time is a result of stretching and alignment of field lines by initially gas-rich galaxies' and their tails' orbital evolution, followed by further stirring by other galaxies on \emph{their} orbits. The tails and wakes of multiple galaxies are superimposed and the collective effect of their motion is felt by the ICM magnetic field. By $t = 2$ Gyr (third columns of Figures~\ref{fig:groupdensbeta2}, \ref{fig:clusterdensbeta2}), only a few galaxies have distinctly visible tails. The magnetic field structure remains disturbed and turbulent, although there are very few coherent structures by this time. 

Orbiting galaxies therefore clearly have a dramatic effect on the morphology of the ICM magnetic field. The magnetic field strength is initially enhanced along the edges of galactic coronae and field vectors are aligned with stripped tails and wakes. These aligned fields suppress shear instabilities at ISM-ICM boundaries. Stripped tails become diffuse and dissipate with time, but the enhanced field strength in the ICM is retained, and the magnetic field is further disturbed by persistent galaxy motions. A relaxed and relatively quiescent ICM magnetic field therefore increases in strength and becomes significantly more turbulent as a result of orbiting, stripped galaxies. 

\begin{figure}[!htbp]
  \begin{center}
    {\includegraphics[width=3.5in]{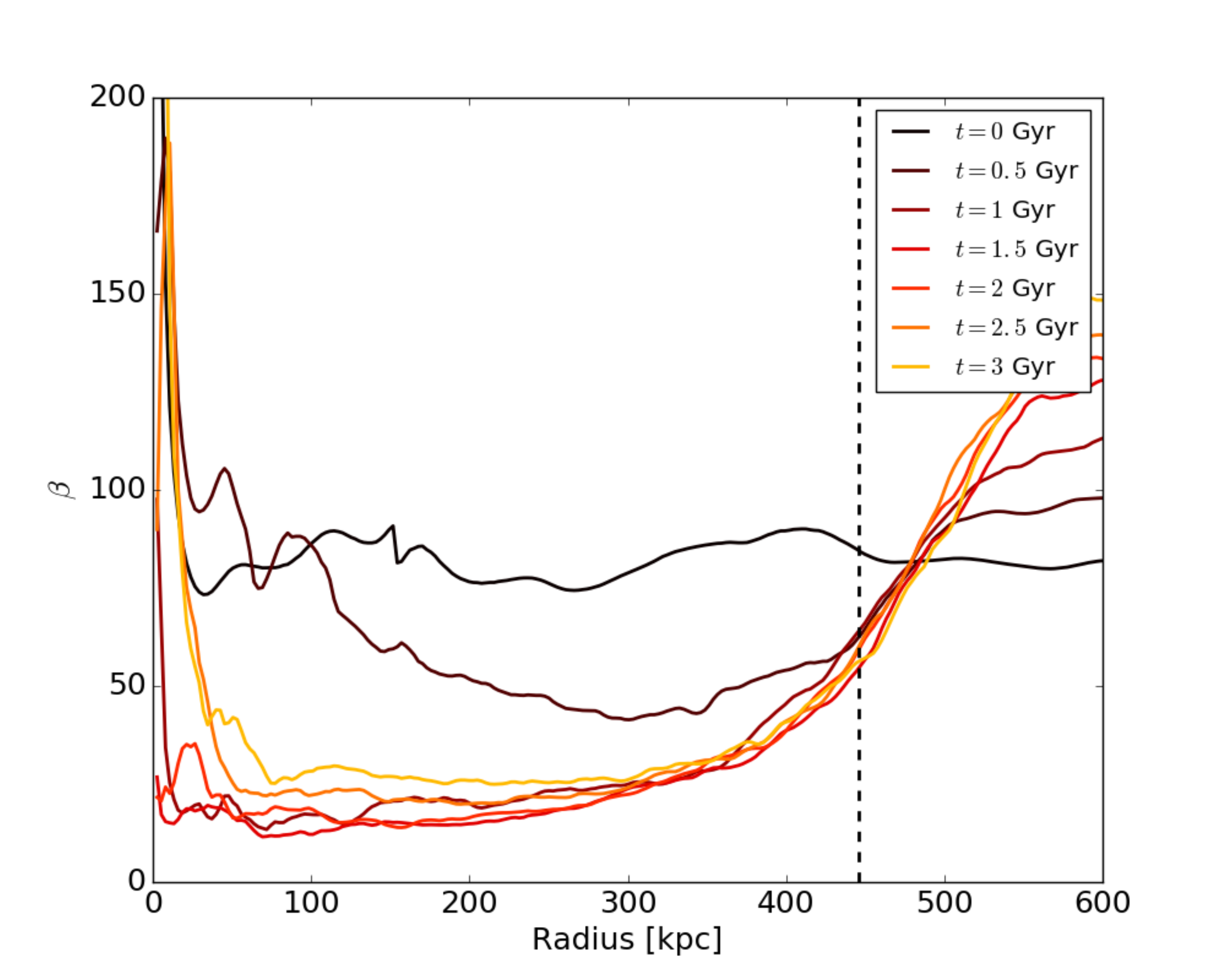}}    
    \caption{The evolution of the azimuthally averaged $\beta$ profile in the presence of galaxies in a $3.2 \times 10^{13} \msun$ group. The black dashed line corresponds to the location of the group's $R_{200}$. Colors correspond to simulation timesteps. \label{fig:beta_gal}}
  \end{center}  
\end{figure}

\begin{figure}[!htbp]
  \begin{center}
    {\includegraphics[width=3.5in]{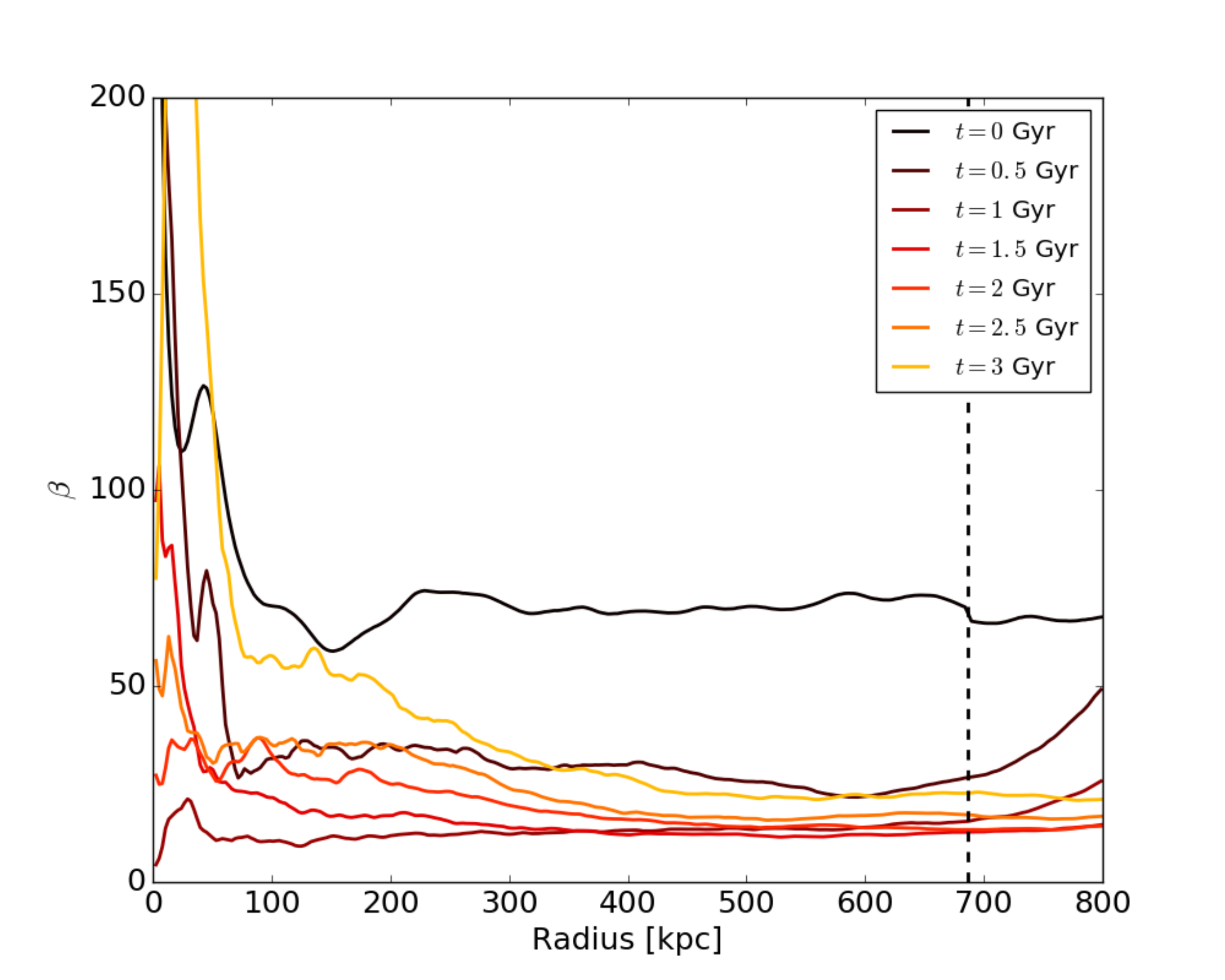}}    
    \caption{The evolution of the azimuthally averaged $\beta$ profile in the presence of galaxies in a $1.2 \times 10^{14} \msun$ cluster. The black dashed line corresponds to the location of the cluster's $R_{200}$. Colors correspond to simulation timesteps. \label{fig:beta_gal_cluster}}
  \end{center}  
\end{figure}

The overall evolution of the magnetic field, for at least 2 Gyr, is therefore not to relax and decay from the initial configuration, but to increase in strength with time as a result of galactic motions and gas flows. These effects are quantitatively seen in the evolution of azimuthally averaged radial $\beta$ profiles (Figures~\ref{fig:beta_gal} and \ref{fig:beta_gal_cluster}). Overall, $\beta$ decreases with time within $R_{200}$, the opposite of the behavior of $\beta$ in the relaxed group (Figure~\ref{fig:beta_nogal}). The rate at which $\beta$ decreases is greatest for $t = 0 - 1$ Gyr, when galaxies are massive and prominent tails are being formed and supported. For $t = 1 - 2$ Gyr, $\beta$ does not change significantly within $R_{200}$. During this period, new galactic tails are not being formed and the tails of less massive galaxies begin to dissipate. Interestingly, while the field does not relax and the overall magnetic field strength within $R_{200}$ increases with time, at large group-centric radii ($R > R_{200}$), this is not the case. These regions are not affected by galaxies. In the absence of galactic motions, the field relaxes unimpeded and $\beta$ increases with time.

At late times, $t \gtrsim 1$ Gyr in the cluster and $t \gtrsim 1.5$ Gyr in the group, galaxies have been stripped of more than $70\%$ of their gas, and no longer exert enough force due to ram pressure to continue amplifying the magnetic field. Correspondingly, the azimuthally averaged value of $\beta$ starts to increase in the group and in the cluster, particularly at smaller group and cluster-centric radii. This occurs partly because galaxies in these denser regions are stripped more rapidly than those in the outskirts. 

\subsection{ICM turbulence and the evolution of magnetic and kinetic energy power spectra}
\label{sec:powerspectrum}

\begin{figure}[!htbp]
  \begin{center}
    {\includegraphics[width=3.5in]{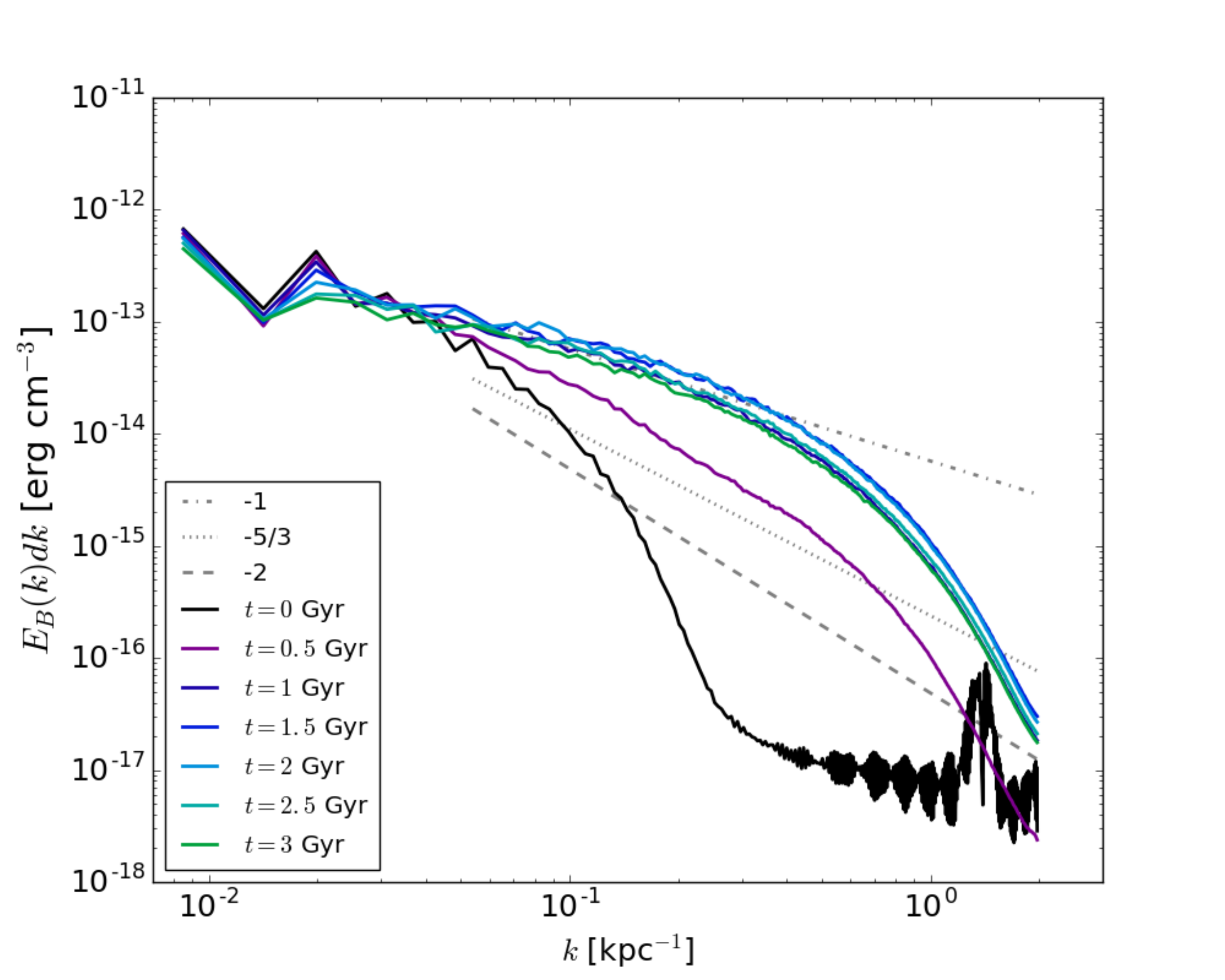}}    
    \caption{The spectrum of magnetic energy density in the box enclosing the isolated group and its galaxies from $t = 0$ Gyr to $t = 3$ Gyr. The black line corresponds to the initial input power spectrum. The grey dashed and dotted lines are reference lines of constant slope, $E (k) \propto k^{-\alpha}$, with the legend denoting the value of $\alpha$. \label{fig:powerspectrum_b_group}}
  \end{center}  
%\end{figure}

%\begin{figure}[!htbp]
  \begin{center}
    {\includegraphics[width=3.5in]{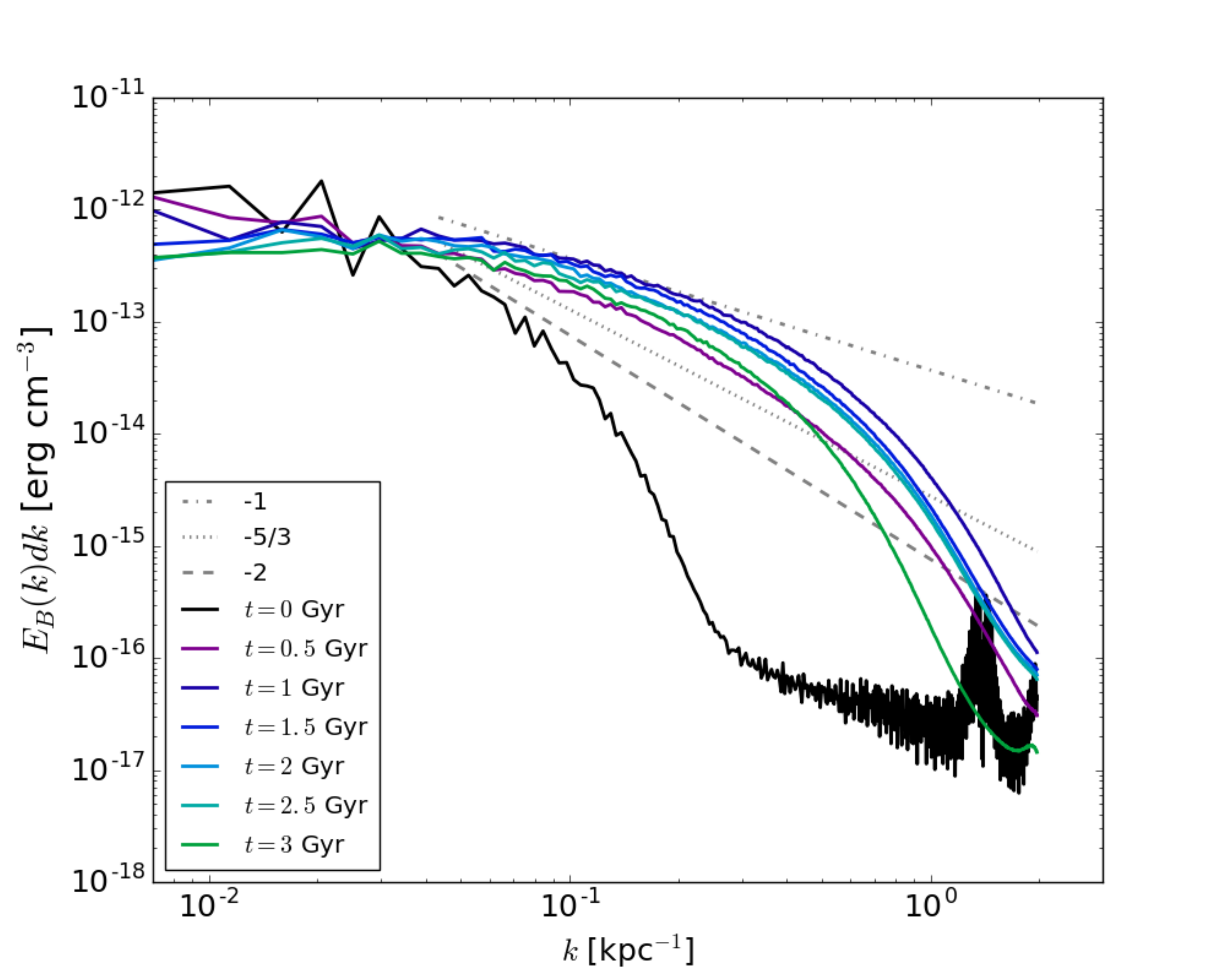}}    
    \caption{The spectrum of magnetic energy density in the box enclosing the isolated cluster and its galaxies from $t = 0$ Gyr to $t = 2$ Gyr. The black line corresponds to the initial input power spectrum. The grey dashed and dotted lines are reference lines of constant slope, $E (k) \propto k^{-\alpha}$, with the legend denoting the value of $\alpha$. \label{fig:powerspectrum_b_cluster}}
  \end{center}  
\end{figure}

\begin{figure}[!htbp]
  \begin{center}
    {\includegraphics[width=3.5in]{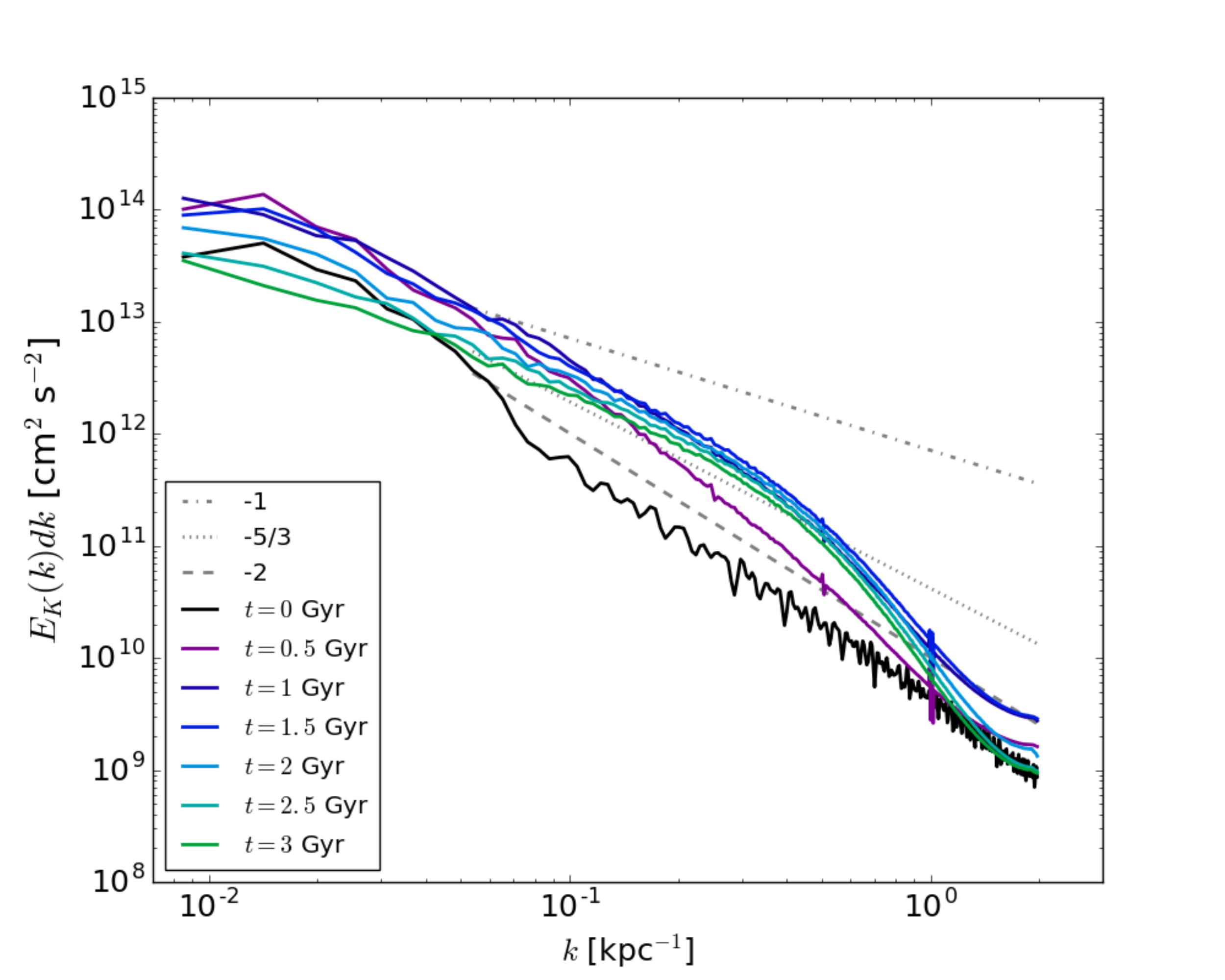}}    
    \caption{The spectrum of $\dfrac{1}{2} v^2$ in the box enclosing the isolated group and its galaxies from $t = 0$ Gyr to $t = 3$ Gyr. The black line corresponds to the initial input power spectrum. The grey dashed and dotted lines are reference lines of constant slope, $E (k) \propto k^{-\alpha}$, with the legend denoting the value of $\alpha$. \label{fig:powerspectrum_v_group}}
  \end{center}  
%\end{figure}

%\begin{figure}[!htbp]
  \begin{center}
u   {\includegraphics[width=3.5in]{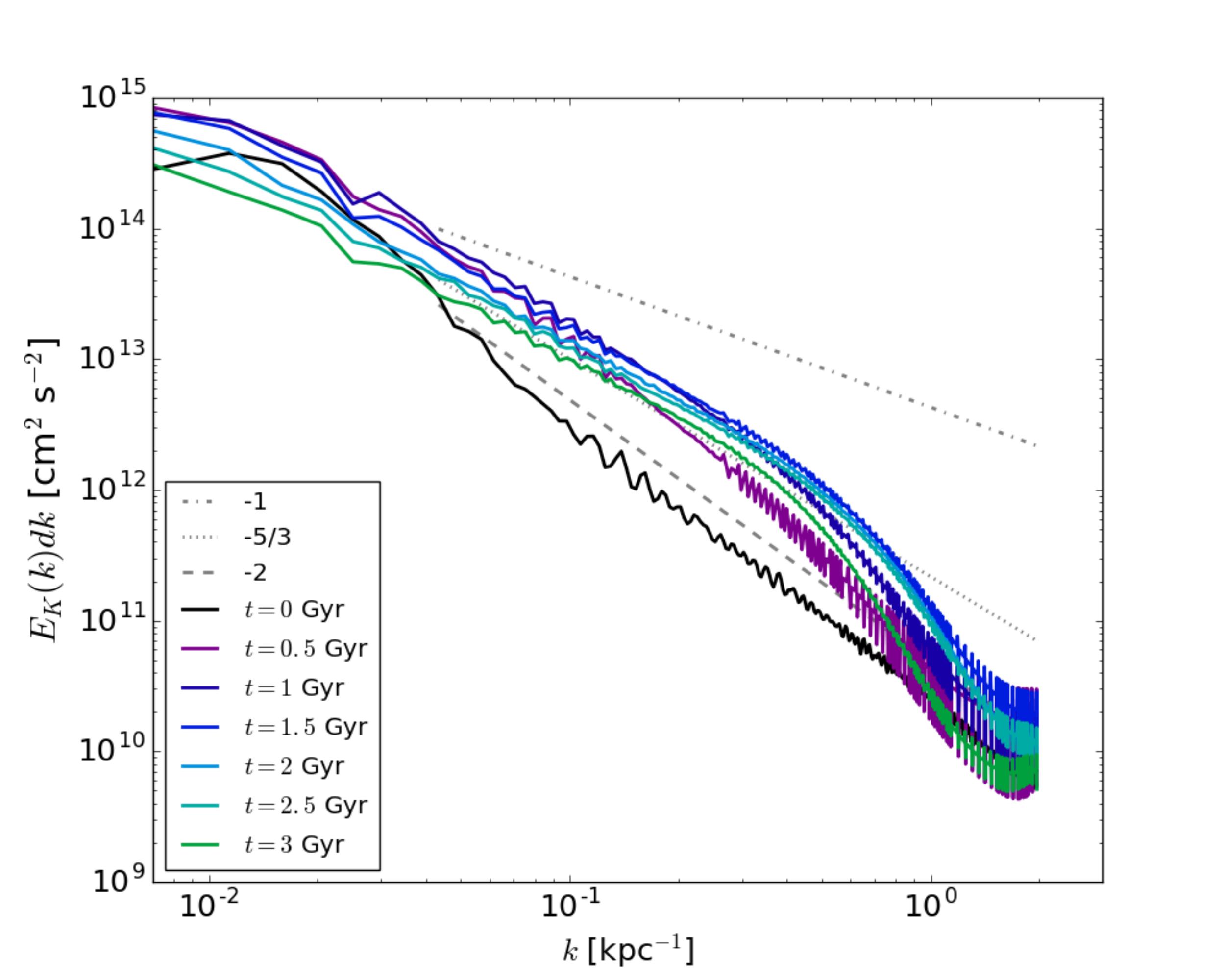}}    
    \caption{The spectrum of $\dfrac{1}{2} v^2$ in the box enclosing the isolated cluster and its galaxies from $t = 0$ Gyr to $t = 2$ Gyr. The black line corresponds to the initial input power spectrum. The grey dashed and dotted lines are reference lines of constant slope, $E (k) \propto k^{-\alpha}$, with the legend denoting the value of $\alpha$. \label{fig:powerspectrum_v_cluster}}
  \end{center}  
\end{figure}

Orbiting galaxies drive turbulence in the ICM. The injection of turbulent kinetic and magnetic energy is reflected in the time evolution of the spectra of these energies. The magnetic energy spectrum is defined as $E_B(k) = |{\bf B}(k)|^2$ (Figures~\ref{fig:powerspectrum_b_group} and \ref{fig:powerspectrum_b_cluster}), and the kinetic energy (per unit mass) spectrum as $E_K(k) = \frac{1}{2}|{\bf v}(k)|^2$ (Figures~\ref{fig:powerspectrum_v_group} and \ref{fig:powerspectrum_v_cluster}) .    

Figure~\ref{fig:powerspectrum_b_group} shows the evolution of the power spectrum of magnetic field fluctuations from $t = 0$ to $t = 3$ Gyr in the isolated group, and Figure~\ref{fig:powerspectrum_b_cluster} from $t = 0$ to $t = 3$ Gyr for the isolated cluster. At $t = 0$ Gyr, the long wavelength, low wavenumber cutoff is $500$ kpc for the input magnetic field power spectrum, corresponding to $k = 1.26 \times 10^{-2}$ kpc$^{-1}$. Below this scale, as reflected in Figures~\ref{fig:powerspectrum_b_group} and \ref{fig:powerspectrum_b_cluster}, the magnetic energy density drops exponentially. There is no significant evolution in the magnetic power spectrum at low wavenumbers corresponding to spatial scales $\gtrsim 250 $ kpc. The periodicity of the $k$-space grid on which the magnetic field is initialized results in oscillations in the power spectrum at $t = 0$; these oscillations are smoothed over time. This is because, in the absence of major mergers or other cluster scale processes, one does not expect any major injection of energy at these scales. Since the galaxy velocity field in our simulations is not symmetric, presumably the galaxies could drive a large-scale dynamo, but if so the timescale for magnetic field growth on large scales appears to be longer than the ram pressure stripping timescale. In real clusters undergoing mergers and accretion this may not be a barrier to development of a large-scale dynamo.

Most of the magnetic power injection occurs at short wavelengths (high wavenumbers). Energy is injected at these small scales by the `fluctuation dynamo' mechanism (\citealt{Brandenburg05}, \citealt{Subramanian06}), wherein the magnetic field is stretched and twisted by velocity shear on scales smaller than the velocity correlation length. In our simulations this length is initially about the size of the galaxies and their wakes ($\sim 10-100$~kpc). Most of the field amplification occurs from $t = 0$ to $t = 1$ Gyr, when the total magnetic energy at 10~kpc scales increases by about an order of magnitude. During this period, $\beta$ also decreases dramatically, as seen in Figures~\ref{fig:beta_gal} and \ref{fig:beta_gal_cluster}. From $t = 1.5$ to $t = 2$ Gyr, there is little change in the power spectrum, consistent with constant $\beta$. At late times ($t \gtrsim 2$ Gyr for the group and $t \gtrsim 1$ Gyr in the cluster), when galaxies have been mostly stripped of their gas, the magnetic field on small scales decays as there is no longer any significant driver of turbulence. The magnetic field decays at a much slower rate than the initial amplification, and decays on small spatial scales where the power spectrum falls off exponentially.

In contrast to the magnetic spectra, $E_K$ grows on all scales during the first Gyr, then decays most rapidly on large scales (Figures~\ref{fig:powerspectrum_v_group} and \ref{fig:powerspectrum_v_cluster}). This occurs because the galaxy wakes drive a turbulent cascade that ceases as the galaxies lose their gas. After about 1.5~Gyr, the galaxies contain insufficient gas to continue to drive the cascade and kinetic energy decays on scales above 10~kpc. The slope of $E_K (k)$ varies between $-5/3$ (Kolmogorov) and $-2$ across two orders of magnitude, consistent with previous results on ICM turbulence by \citet{Vazza09}, \citet{Gaspari13}, \citet{Vazza14} and others. 

\begin{figure}[!htbp]
  \begin{center}
 	 {\includegraphics[width=3.5in]{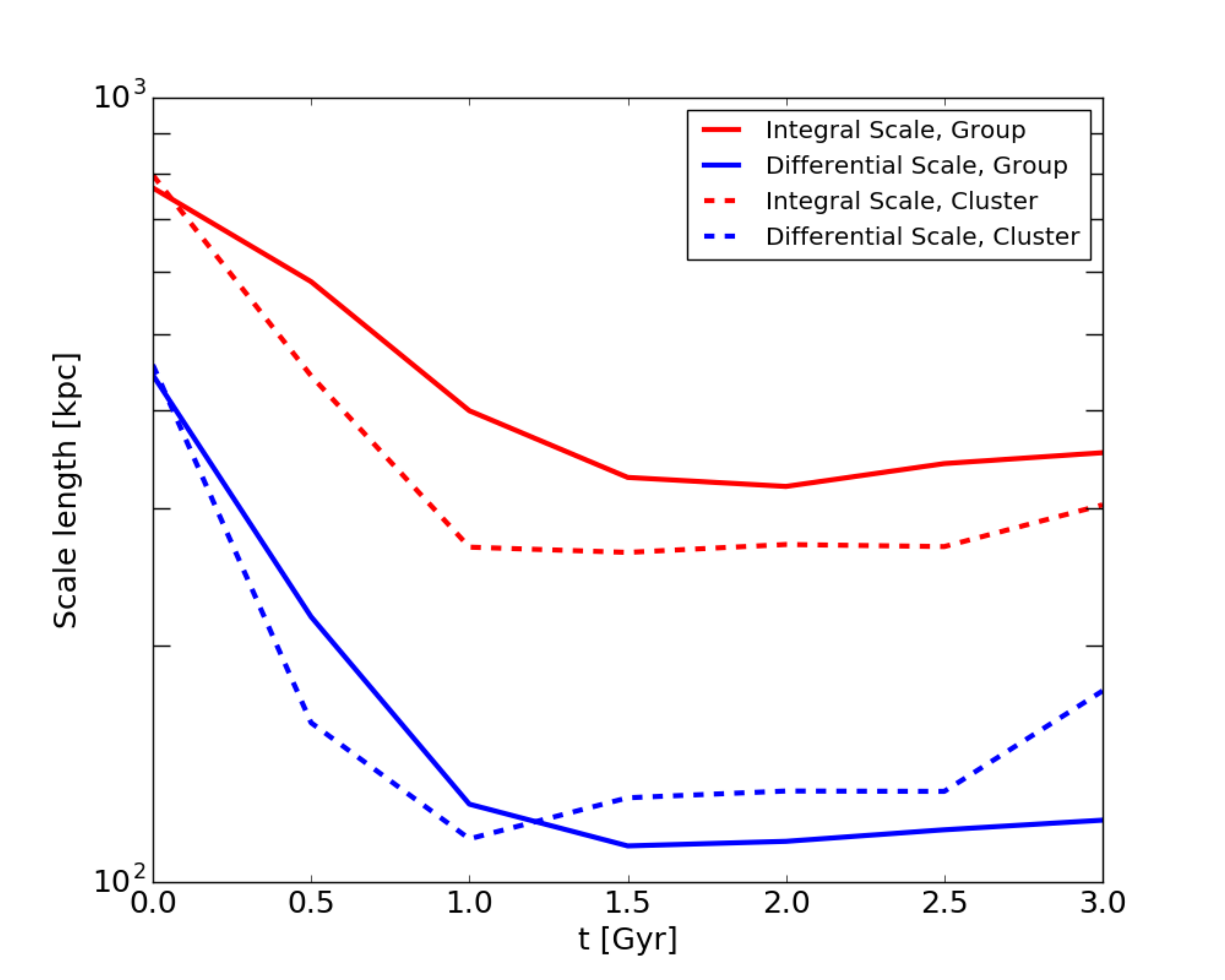}}    
    \caption{Evolution of the integral and differential scales of the magnetic field in the group and cluster. \label{fig:intdiffscale}}
  \end{center}  
\end{figure}

We further characterize the evolution of turbulence in the magnetic and velocity fields by means of the integral and differential length scales. The integral scale of a field corresponds to the largest scale over which a field is correlated. These quantities are defined in \citet{Monin75}, where the integral scale is defined as
\begin{equation}
L \equiv K_i \frac{\int k^{-1} E(k) dk}{\int E(k) dk}\ ,
\end{equation}
and the differential length scale is defined as
\begin{equation}
\lambda \equiv K_d \left(\frac{\int E(k) dk}{\int k^2 E(k) dk}\right)^{0.5}\ .%
\end{equation}
$K_i$ and $K_d$ are constants of order unity. Their precise values depend on the scalar or vector nature of a given field and its homogeneity, isotropy, and overall morphology. For a homogeneous, isotropic, solenoidal field, the longitudinal integral scale has $K_i = 3\pi/4$ and $K_d = 5$. The magnetic field is by definition solenoidal, although not necessarily homogeneous in these simulations since there is a radial dependence to the field strength. The velocity field is not solenoidal.

The evolution of the integral and differential length scales for the group and cluster magnetic field are shown in Figure~\ref{fig:intdiffscale}. These length scales decrease significantly up to $t \sim 1.5$ Gyr in the group and $t \simeq 1$ Gyr in the cluster, when galaxies maximally drive turbulence and amplify ICM magnetic fields. During this period, the plasma $\beta$ parameter also declines and the magnetic power spectrum increases significantly. Later, when turbulence decays, these scales grow.

\begin{figure*}[!htbp]
  	\hspace{-1.2cm}
    {\includegraphics[width=1.1\textwidth]{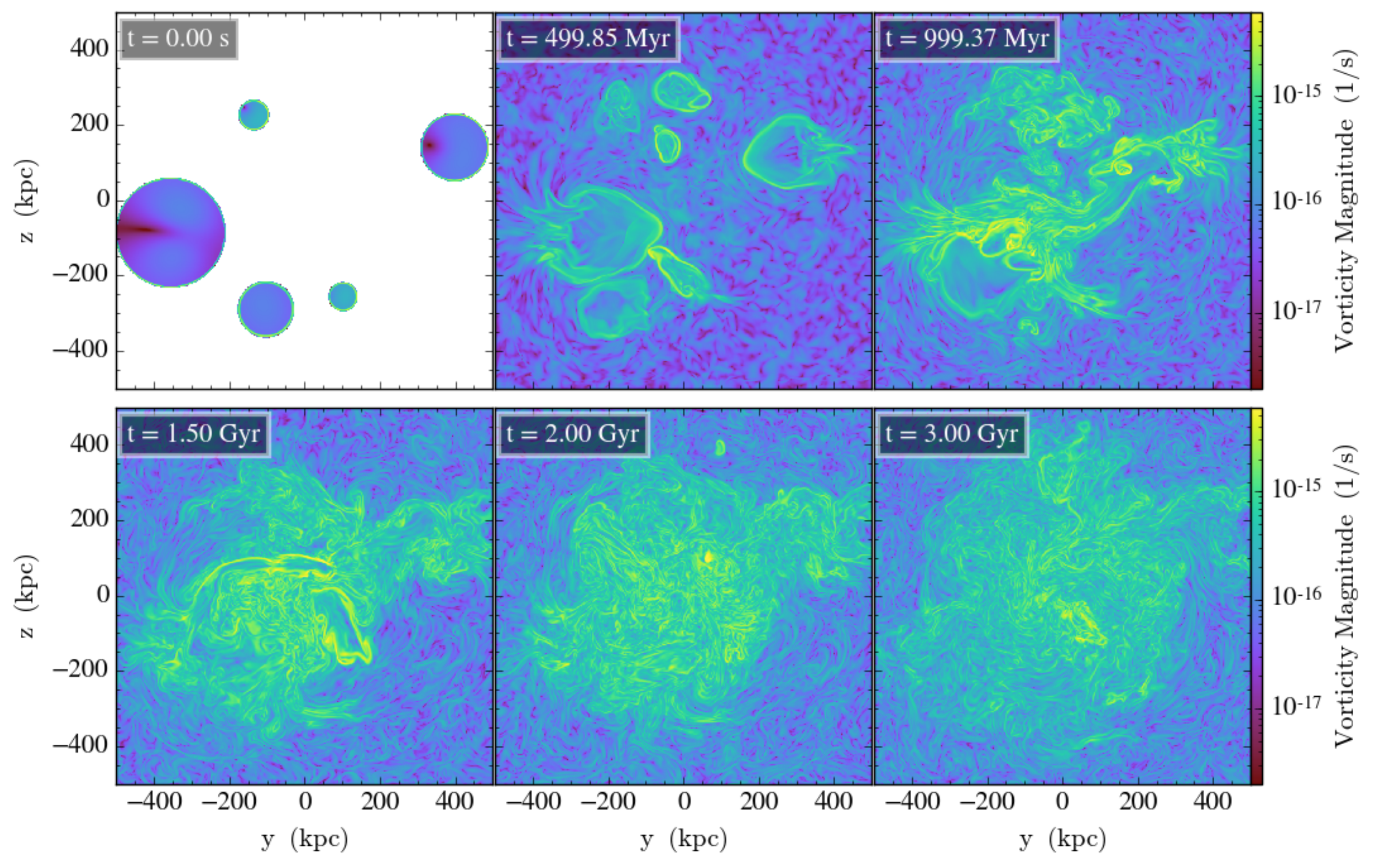}}
	\caption{Absolute magnitude of vorticity, $\mathbf{\Omega} = \nabla \times \mathbf{v}$ in the $x = 0$ plane of the group.  \label{fig:vorticity_group}}
\end{figure*}

In addition to the turbulence generated in the stripped galaxy tails and wakes, orbiting, massive galaxies can also generate turbulence in the ICM through the excitation of $g$-modes. Previous analytic and numerical calculations by \citet{Balbus90}, \citet{Lufkin95}, and \citet{Ruszkowski11} show that when a galaxy's orbital frequency $\omega < \omega_{\rm BV}$ (where $\omega_{\rm BV}^2 = \omega_{\rm circ}^2 \tfrac{3}{5} \tfrac{\mathrm{d} \ln S}{\mathrm{d} \ln r}$ is the Brunt-V\"ais\"al\"a frequency; we refer the reader to \citealt{Balbus90} and \citealt{Lufkin95} for detailed calculations), the galaxy can resonantly excite $g$-mode oscillations. The cluster-centric radius within which $\omega < \omega_{\rm BV}$ (which, for the power-law entropy profile used in our simulations, is proportional to $G M_{\rm cluster} / v_{\rm galaxy}^2$) defines a resonant cavity within which $g$-waves are trapped. These $g$-waves perturb the velocity field. The vorticity $\Omega \equiv \nabla \times  \mathbf{v}$ is a good tracer of $g$-waves (\citealt{Lufkin95}). \citet{Ruszkowski11} also argue that given the similarity between the vorticity and magnetic field equations, the growth in vorticity due to $g$-mode oscillations can drive dynamo action, subsequently amplifying ICM magnetic fields.

The generation of vorticity in our simulations is illustrated in Figure~\ref{fig:vorticity_group} in a slice through the group. At $t = 0$ Gyr, the ICM is at rest and in hydrostatic equilibrium; consequently there is no vorticity in the ICM. Galaxies have a non-zero initial velocity with respect to the ICM; as they orbit within the group and drive $g$-waves, the vorticity in the ICM increases. At $t = 0.5$ Gyr, the shearing motions between the galaxies and the ICM result in an increase in vorticity along the surfaces and tails of galaxies (compare the enhancements in vorticity in Figure~\ref{fig:vorticity_group} to the enhancements in $\beta$ in Figures~\ref{fig:groupdensbeta1} and \ref{fig:groupdensbeta2}. Additionally, the vorticity significantly increases from zero in the background ICM, even in regions which are \emph{not yet impacted by galaxy wakes}, indicating that the vorticity and subsequent turbulence in these regions is not generated by galaxy tails and wakes, but $g$-waves. Further, we see that the overall vorticity increases up to $t \simeq 1.5 - 2$ Gyr and begins to decrease by $t = 3$ Gyr, similar to the evolution of the magnetic and kinetic power spectra and the overall velocity dispersion. This makes sense since dynamically the generation of galaxy wakes and $g$-waves depends on galaxy masses; once galaxies are stripped of most of their mass by tidal effects and ram pressure stripping, they are less effective in driving both types of turbulence.

\begin{figure}[!htbp]
  \begin{center}
 	 {\includegraphics[width=3.5in]{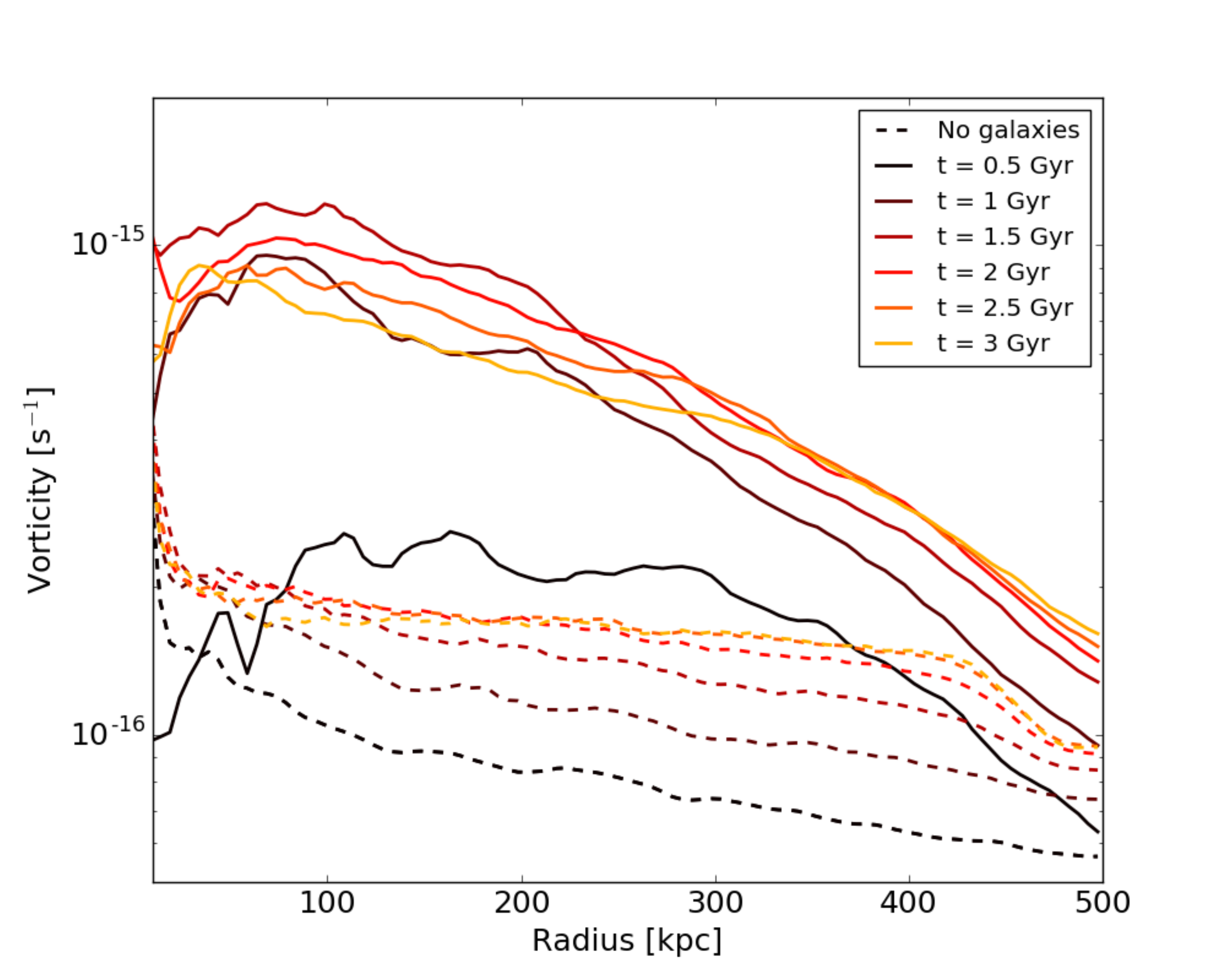}}    
    \caption{Radial profiles of the magnitude of vorticity in the group simulation without orbiting galaxies (dashed lines) and with galaxies (solid lines). \label{fig:vorticityprofile_timeseries}}
  \end{center}  
\end{figure}

We have verified that the increase in vorticity is indeed due to the galaxies and not merely spurious numerical effects. Using a simulation of a group with the same collisionless dark matter distribution, hydrodynamic properties, and identical magnetic field structure, but without any galaxies and initially in hydrostatic equilibrium, we measure the amount of vorticity generated only due to numerical effects. This simulation is at the same spatial resolution as our other simulations. Figure~\ref{fig:vorticityprofile_timeseries} shows azimuthally averaged radial profiles of the magnitude of vorticity in the simulations with and without galaxies. We see that there is a higher level of vorticity generated due to orbiting galaxies: up to a factor of 10 in the central regions, and a factor of 2 or more in the outskirts. Furthermore, we see that when there are no galaxies, the amount of numerical vorticity generated saturates by $t = 2$ Gyr, and does not significantly decline later. In the presence of orbiting galaxies, however, the temporal behavior of the vorticity is similar to that of the overall turbulence and magnetic pressure: increasing up to $t = 1.5 - 2$ Gyr, followed by a decline.

\section{Observational Diagnostics and Implications}
\label{sec:obs_diagnostics}

\subsection{Turbulence and velocity dispersion: limits for \textit{ATHENA} or a \textit{Hitomi}-like mission}

\begin{figure*}[!htbp]
  \begin{center}
  \hspace{-0.5in}
 	\subfigure[$v_x$ field in the $z = 0$ plane]
    {\includegraphics[width=2.3in]{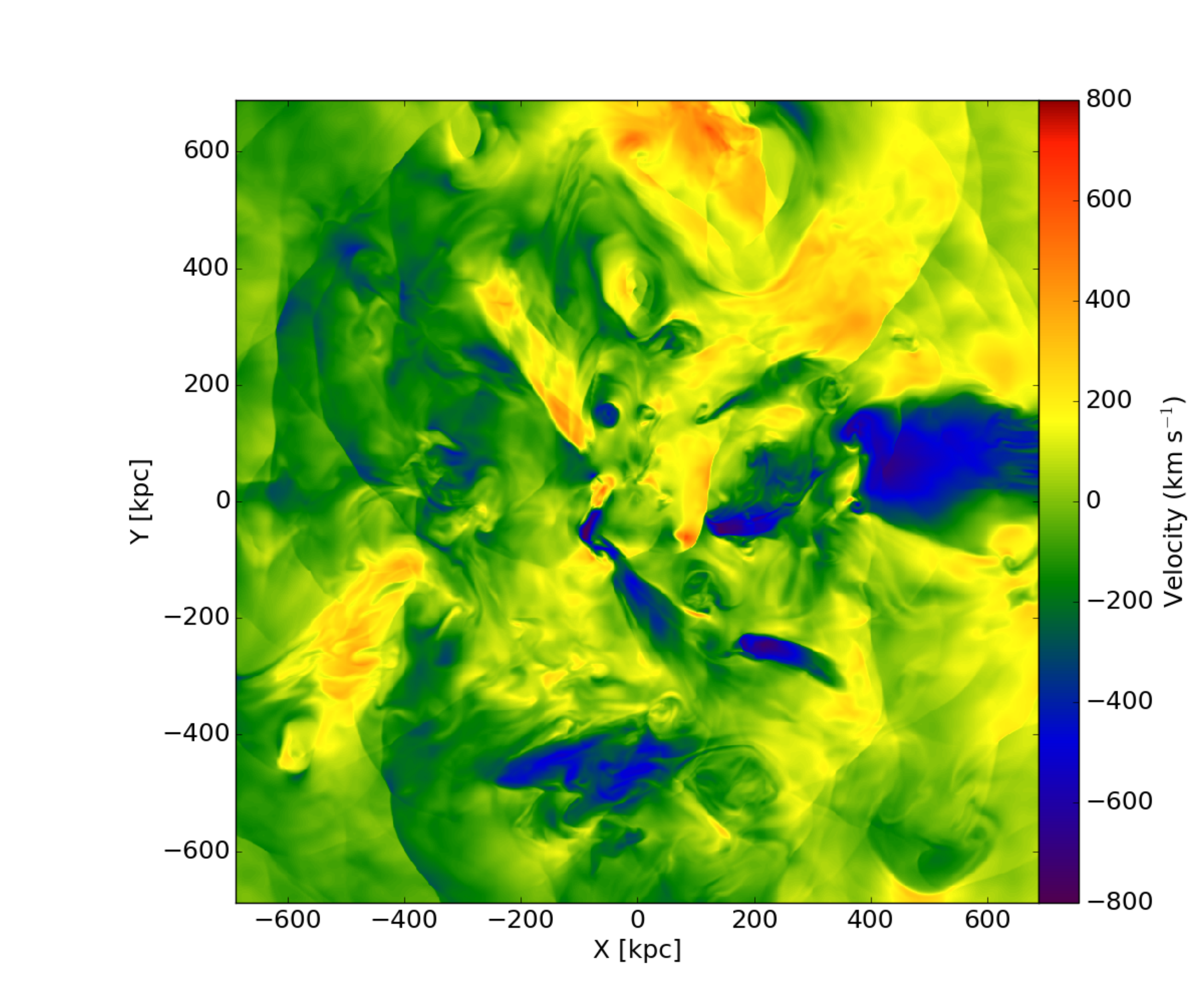}\label{fig:vx_all}}  
    \subfigure[$v_x$, smoothed with a Gaussian kernel of width 50 kpc]
    {\includegraphics[width=2.3in]{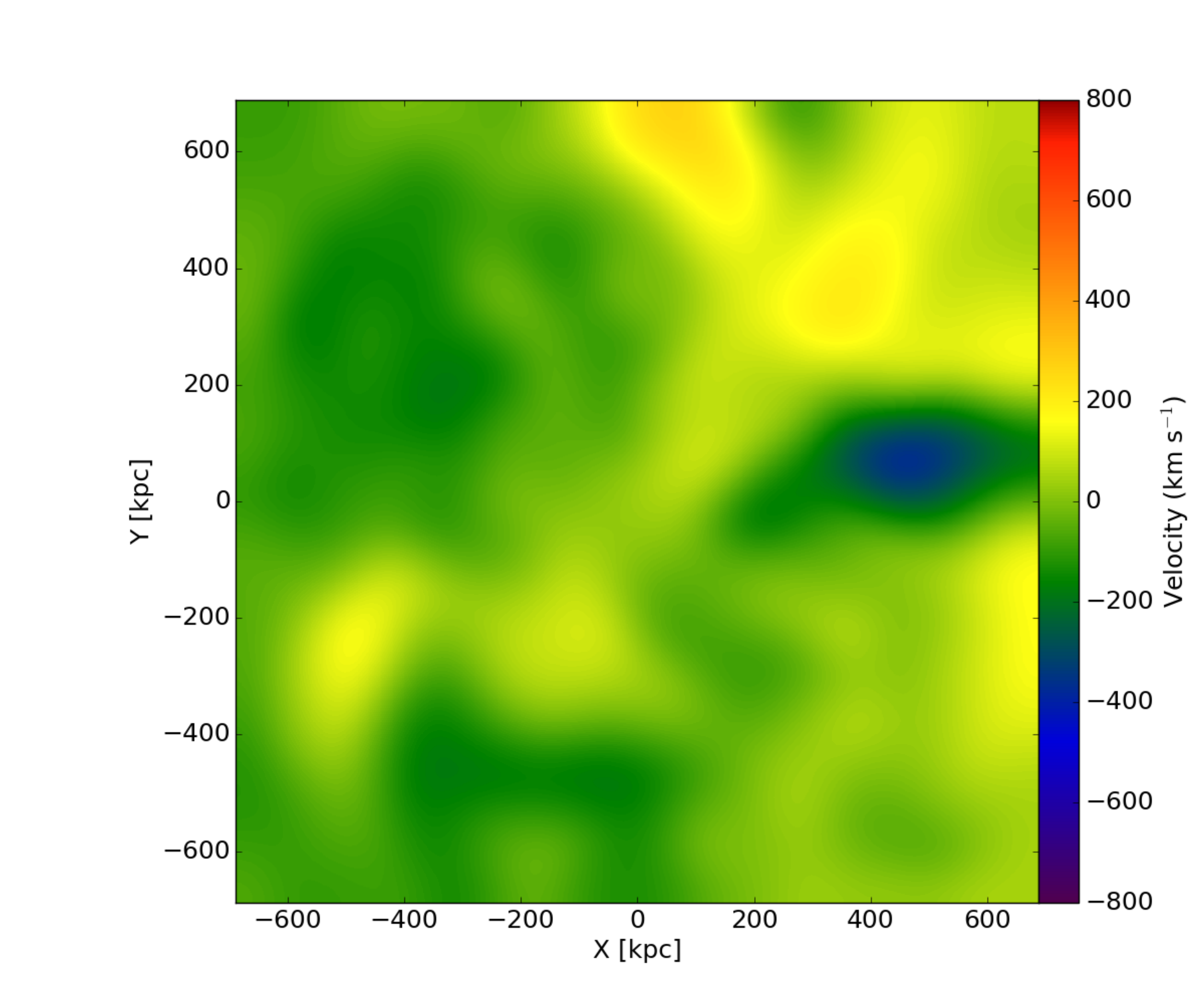}\label{fig:vx_gsmoothed}}    
    \subfigure[$v_{x, {\rm pec}} = v_x - v_{x, {\rm smoothed}}$]
    {\includegraphics[width=2.3in]{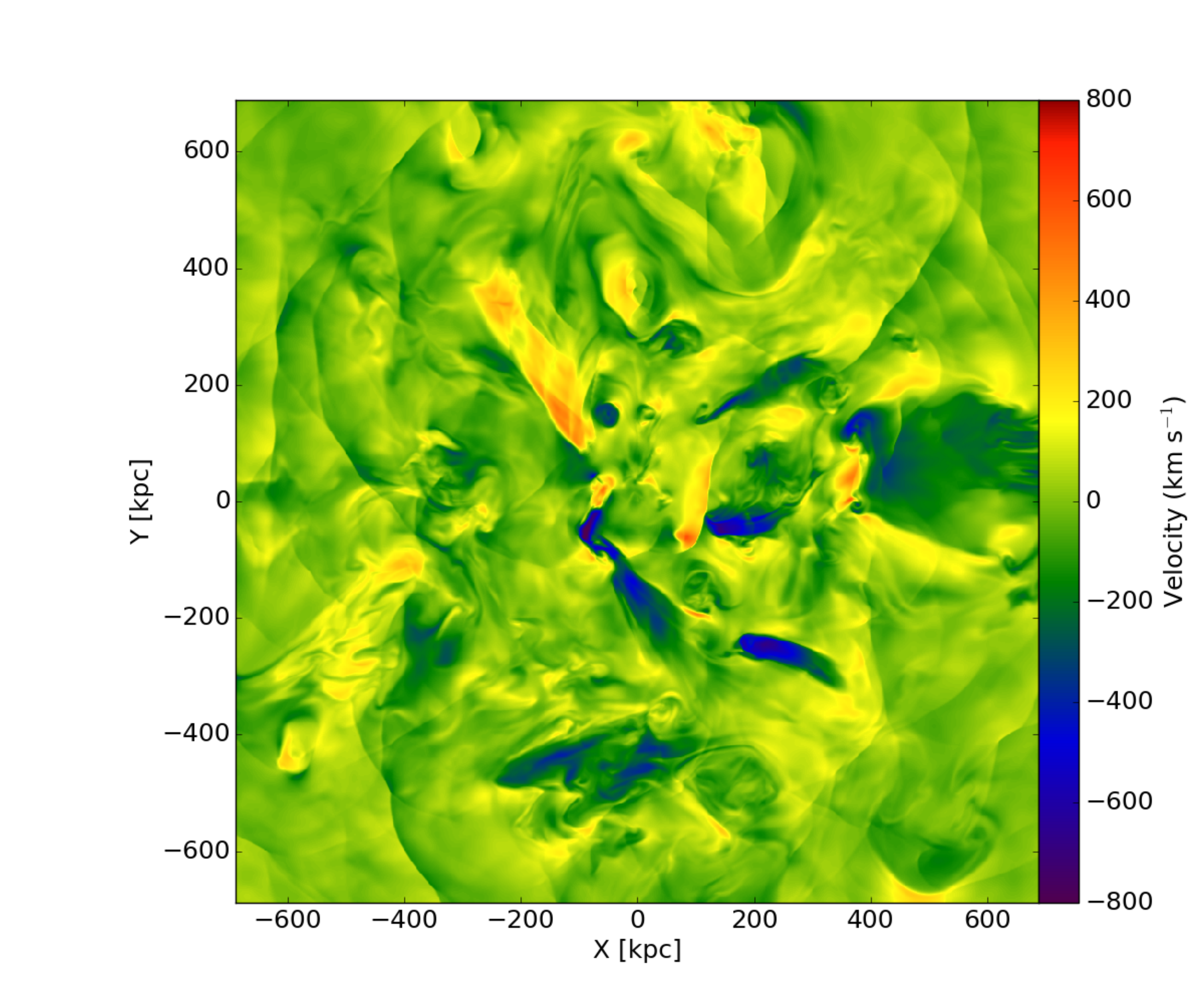}\label{fig:vx_turb}} 
    \caption{These figures illustrate the calculation of the peculiar velocity field in the cluster at $t = 0.5$~Gyr in the $z = 0$ plane. The figure on the left is the $x$ component of the velocity field on the grid, including the bulk motion of galaxies and their tails. The figure in the center is $v_x$ smoothed with a Gaussian kernel of width 50 kpc, and the figure on the right is the velocity field after subtracting the smoothed component, or the bulk motion --- the $x$ component of the turbulent velocity field in our calculation. \label{fig:vx_smoothed}}
  \end{center}  
\end{figure*}

We showed in the previous section that orbiting galaxies generate $g$-mode turbulence in the ICM in addition to the turbulence generated in their stripped tails and ICM wakes. Future X-ray missions, particularly \textit{ATHENA} (or a \textit{Hitomi}-like mission with an X-ray microcalorimeter) can measure the turbulent broadening of X-ray spectral lines.  \textit{ATHENA} (currently scheduled to launch in 2028) is planned to have an X-ray Integral Field Unit (X-IFU), which will consist of an array of cooled microcalorimeters with a spectral resolution of 2.5 eV, energy range of $0.2 - 12$ keV, field of view of 5\arcmin , and spatial resolutions of 5\arcsec $-$ 10\arcsec\ (\citealt{Barret15}). At these limits. the X-IFU can measure turbulent velocities of $100 - 1000$ km s$^{-1}$ with $10 - 20$ km s$^{-1}$ precision (\citealt{Ettori14}). The Soft X-ray Spectrometer (SXS) on \textit{Hitomi} using a micro-calorimeter was designed to be capable of producing spectra with an energy resolution $< 7$ eV in the $0.3 - 12$ keV energy band, with an angular resolution of 1.3\arcmin and a 3\arcmin $\times$ 3\arcmin\ field of view (\citealt{Takahashi14}). Correspondingly, \textit{Hitomi} was capable of measuring RMS velocity dispersions with precisions of $\sim 100 - 150$ km s$^{-1}$ and mean line-of-sight velocities with precisions of $\sim 100$ km s$^{-1}$ for clusters at redshifts $0.03 < z < 0.1$  (\citealt{Kitayama14}). Below we discuss limits applicable to both a \textit{Hitomi}-like mission and \textit{ATHENA}.

In addition to orbiting galaxies and their stripped tails and wakes, turbulence can be generated in the ICM via multiple potential mechanisms, including major and minor mergers with clusters and subclusters, accretion of gas from the surrounding large scale structure, and AGN activity, as described in \S~\ref{sec:intro}. In our idealized experiment, we do not include these effects, and can therefore uniquely constrain the amount of turbulence generated \emph{only} due to galaxy-driven mechanisms. The ICM in our simulations is initially in hydrostatic equilibrium, with no bulk or turbulent velocity component apart from that of orbiting galaxies. The ICM velocity field therefore consists of galaxies' bulk velocities, velocities of stripped tails, turbulent  velocities in ICM wakes, as well as $g$-mode turbulence driven into the diffuse background ICM. 

We analyze the evolution of both the overall velocity field, $\mathbf{v_{\rm total}}$ and the turbulent velocity field, $\mathbf{v_{\rm turbulent}}$, where $\mathbf{v_{\rm turbulent}} = \mathbf{v_{\rm total}} - \mathbf{v_{\rm bulk}}$. To calculate the bulk velocity, $\mathbf{v_{\rm bulk}}$, we apply a Gaussian filter with a smoothing scale of $r_{\rm smooth} = 50$ kpc on the $\mathbf{v_{\rm total}}$ field. The choice of $r_{\rm smooth}$ corresponds to the typical size of galaxies and their wakes; modifying this scale by a factor of 2 does not significantly affect our overall results. Figure~\ref{fig:vx_smoothed} illustrates our calculation of $\mathbf{v_{\rm turbulent}}$. Figure~\ref{fig:vx_all} shows a slice of $v_{{\rm total}, x}$ at $t = 0.5$ Gyr, Figure~\ref{fig:vx_gsmoothed} shows  $v_{{\rm total}, x}$ after Gaussian smoothing, and Figure~\ref{fig:vx_turb} shows the residual turbulent component of $v_x$ after subtracting the smoothed velocity field.

\begin{figure*}[!htbp]
  \begin{center}
	\subfigure[Velocity dispersion, $\sigma_v = \sqrt{\mathbf{v}^2 - \mean{\mathbf{v}}^2}$]
    {\includegraphics[width=3in]{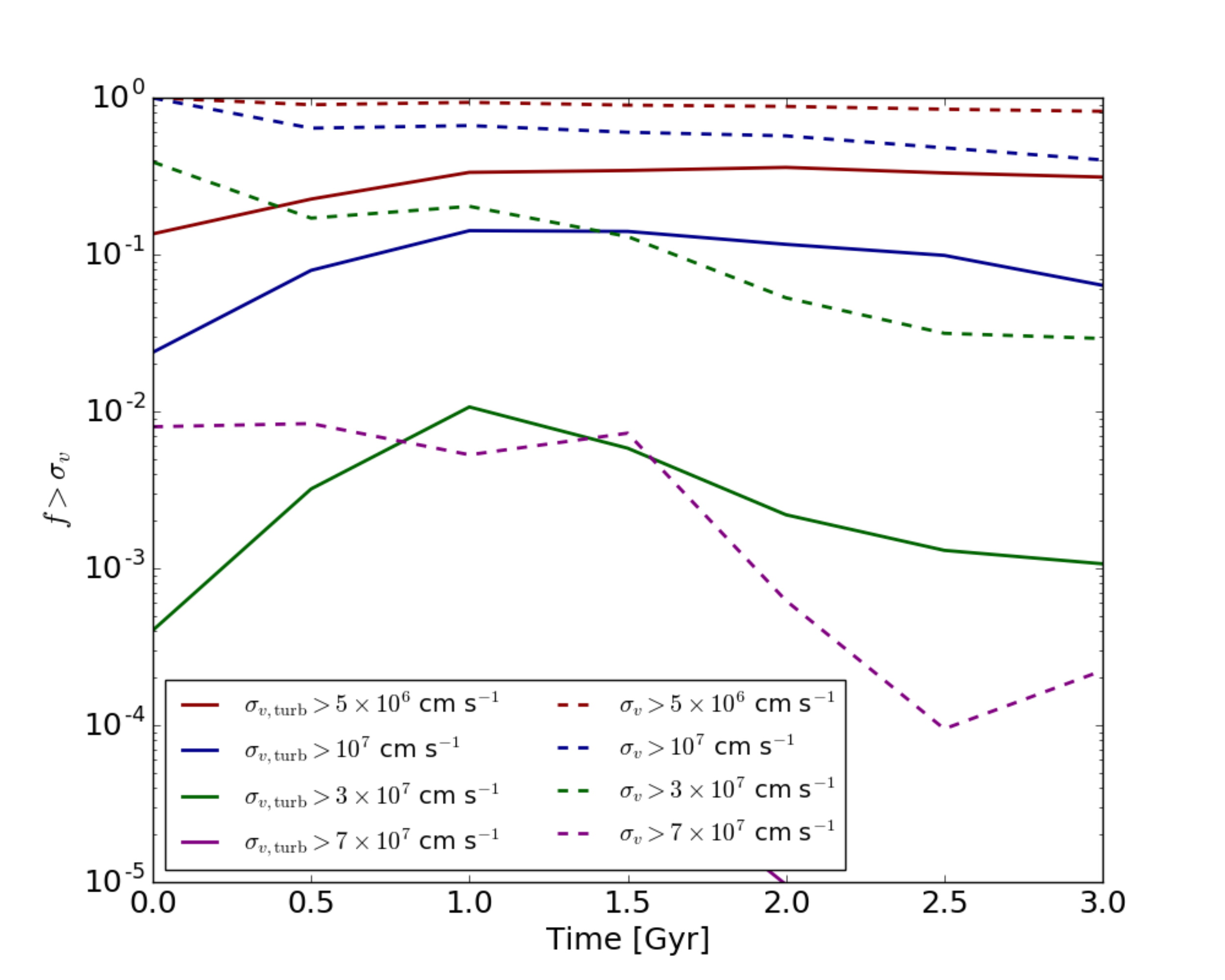}\label{fig:vdisp_tot_group}}
    \subfigure[Turbulent pressure, $\rho \sigma_{v, {\rm turbulent}}^2$]
    {\includegraphics[width=3in]{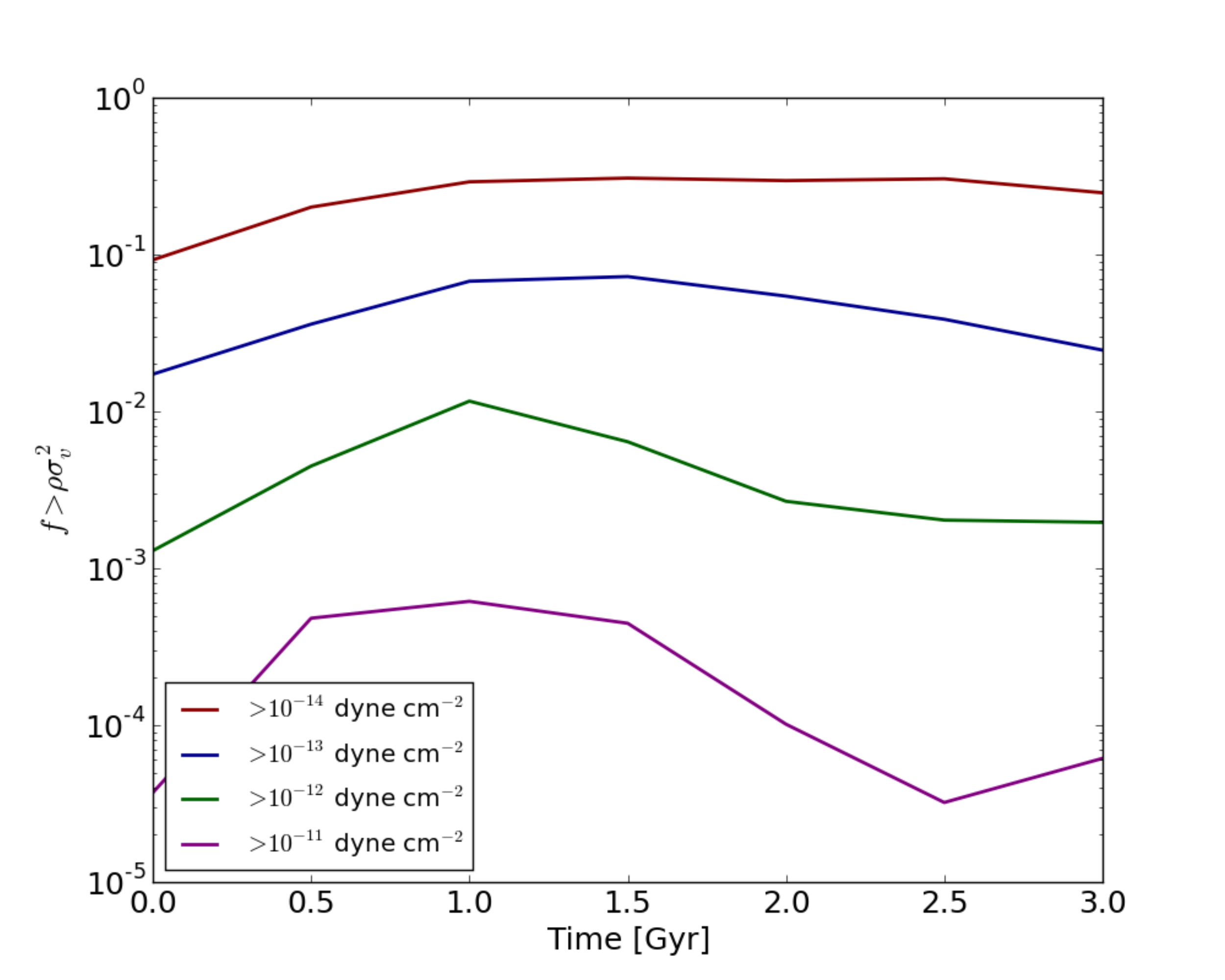}\label{fig:pturb_group}} 
    \caption{The evolution of the volume fraction of the isolated group where the velocity dispersion, turbulent velocity dispersion, and turbulent pressure exceed specified values. $\sigma_v$ is the velocity dispersion before subtracting the smoothed component and $\sigma_{v, {\rm turb}}$ is the velocity dispersion after subtracting the smoothed component. \label{fig:group_vturb_frac}}
  \end{center}  
\end{figure*}

\begin{figure*}[!htbp]
  \begin{center}
	\subfigure[Velocity dispersion, $\sigma_v = \sqrt{\mathbf{v}^2 - \mean{\mathbf{v}}^2}$]
    {\includegraphics[width=3in]{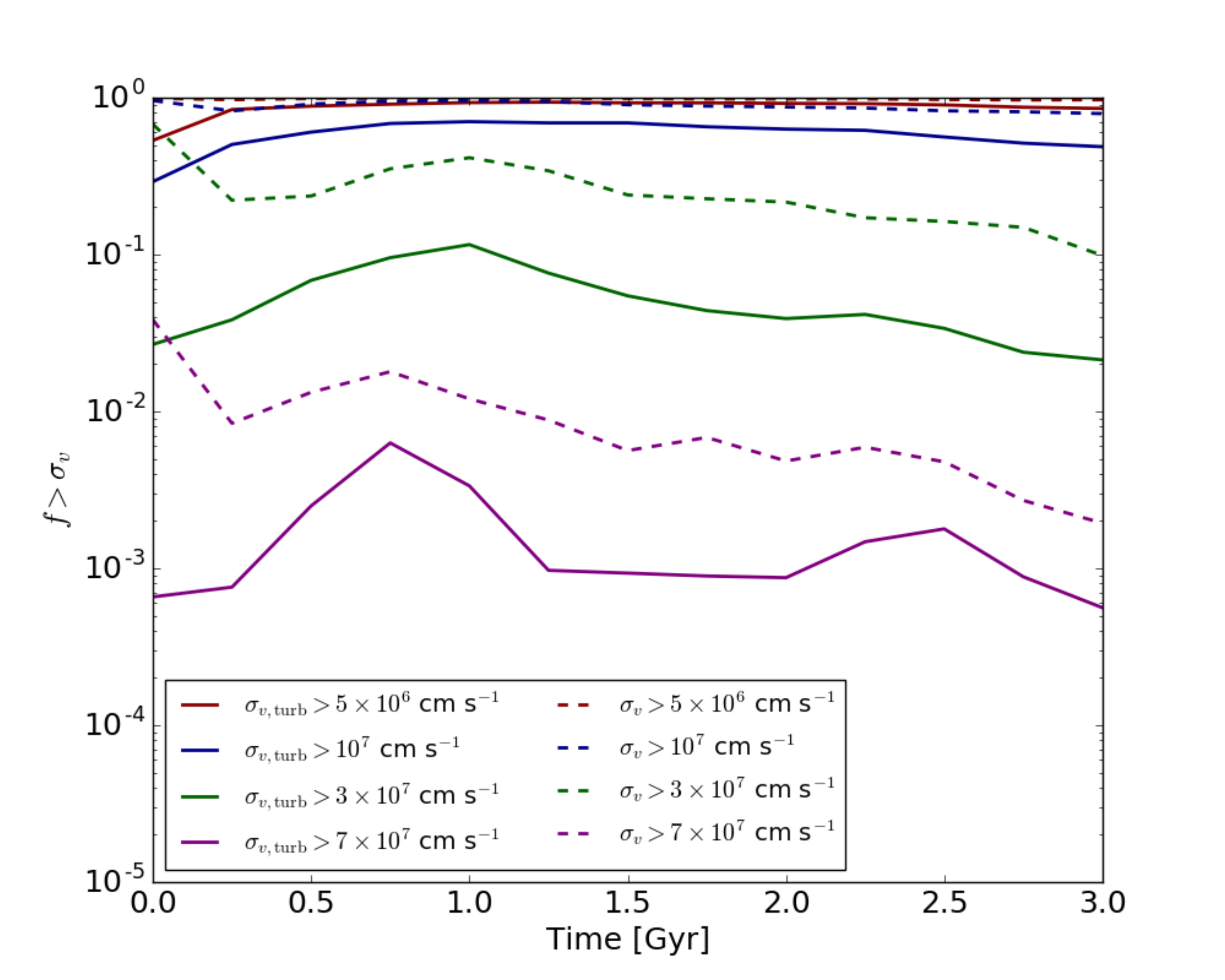}\label{fig:vdisp_tot_cluster}}  
    \subfigure[Turbulent pressure, $\rho \sigma_{v, {\rm turbulent}}^2$]
    {\includegraphics[width=3in]{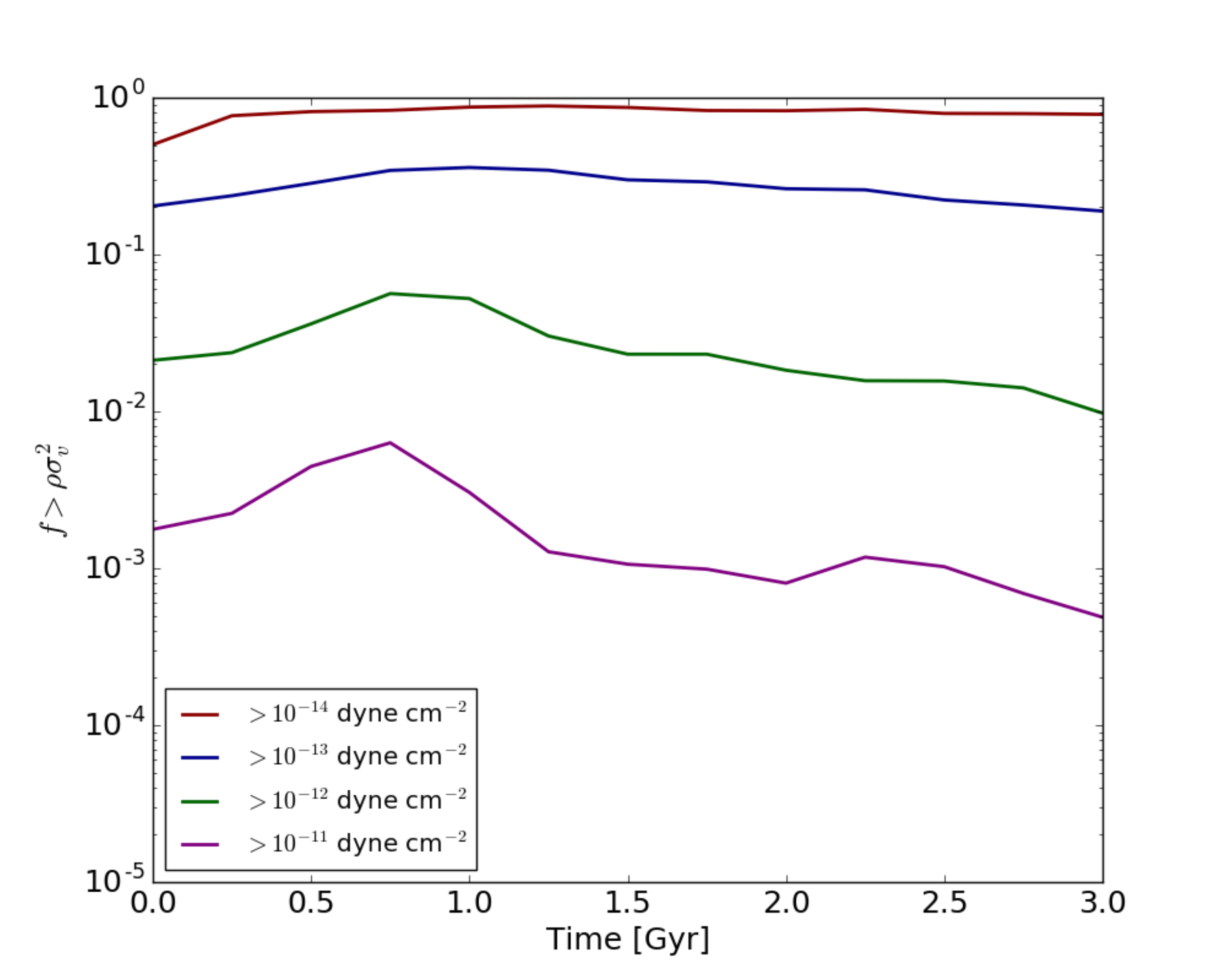}\label{fig:pturb_cluster}} 
    \caption{The evolution of the volume fraction of the isolated cluster where the velocity dispersion, turbulent velocity dispersion, and turbulent pressure exceed specified values. $\sigma_v$ is the velocity dispersion before subtracting the smoothed component and $\sigma_{v, {\rm turb}}$ is the velocity dispersion after subtracting the smoothed component.   \label{fig:cluster_vturb_frac}}
  \end{center}  
\end{figure*}

Figures~\ref{fig:group_vturb_frac} and ~\ref{fig:cluster_vturb_frac} show the evolution of the total velocity dispersion field, the turbulent velocity dispersion, and turbulent pressure from $t = 0$ to $t = 3$ Gyr. In these figures, we show the overall evolution of the fractional volume of the group and cluster (within their virial radius, $R_{200}$) where the velocity dispersion exceeds 50, 100, 300, and 700 km s$^{-1}$ and the turbulent pressure exceeds $10^{-11}$, $10^{-12}$, $10^{-13}$, and $10^{-14}$ dyne cm$^{-2}$. These values are chosen to represent typical detection limits of \textit{ATHENA} (and a \textit{Hitomi}-like mission). The evolution of the total velocity field is primarily driven by the initial bulk velocities of galaxies. As galaxies are stripped and their dark matter and gas are eventually virialized within the cluster potential, their bulk kinetic energy is converted to turbulent kinetic energy and then heat, and the overall velocity dispersion gradually declines. Correspondingly, the fractional volume of gas where the velocity dispersion exceeds the above typical detection limits declines monotonically in Figures~\ref{fig:vdisp_tot_group} and \ref{fig:vdisp_tot_cluster}.

Although the overall velocity dispersion continually declines, the volume fraction of gas where the turbulent velocity is higher than the specified detection limits peaks between  $t = 0.5 - 1.5$  Gyr. This corresponds to the period when turbulent kinetic energy is injected, seen in Figures~\ref{fig:powerspectrum_v_group} and \ref{fig:powerspectrum_v_cluster}. As the turbulent kinetic energy decays, the volume of the ICM with increased turbulent velocity dispersion also declines, as seen in Figures~\ref{fig:vdisp_tot_group} and \ref{fig:vdisp_tot_cluster}. The more massive cluster has a higher intrinsic velocity dispersion; therefore a larger fraction of its gas exceeds the above velocity dispersion detection limits. The evolution of the turbulent pressure is similar to that of the turbulent velocity. The highest fractional volume of gas exceeding the previous detection limits occurs between  $t = 0.5 - 1.5$  Gyr, followed by a gradual decline as turbulence initially driven by galaxies decays (Figures~\ref{fig:pturb_group} and \ref{fig:pturb_cluster}).

\subsection{Faraday rotation measure}

An important observational diagnostic of magnetic fields in the ICM is the Faraday rotation measure, first described in \citet{Burn66}. Since then, the Faraday RM has been a key diagnostic of astrophysical magnetic fields, particularly in the diffuse intracluster medium (see \citealt{Dreher87} and other references in the introduction for more recent results). Magnetic fields in an ionized plasma, like the ICM, affect the direction along which electrons gyrate, which in turn rotates the plane of polarization for electromagnetic radiation (reviewed in \citealt{Carilli02}) by an angle $\Delta \chi = \mbox{RM} \lambda^2$. The Faraday rotation measure (RM) depends on the local electron number density, magnetic field, and the net integrated path length through the medium via
\begin{equation}
\mbox{RM} = K \int n_e \mathbf{B} \cdot d\mathbf{l},
\end{equation} 
where RM is in units of radians m$^{-2}$ and $K$ = 0.81 rad m$^{-2}$ cm$^{3}$ $\mu {\rm G}^{-1}$ pc$^{-1}$. At a given wavelength, the RM depends on the \emph{integrated} value of the magnetic field through the ambient ICM. Therefore, for a random, isotropic, magnetic field, the average value of RM along any individual line of sight is 0. 

\begin{figure*}[!htbp]
  \begin{center}
    \subfigure[RM map projected along $x$ at $t = 0.5$ Gyr]
    {\includegraphics[width=3in]{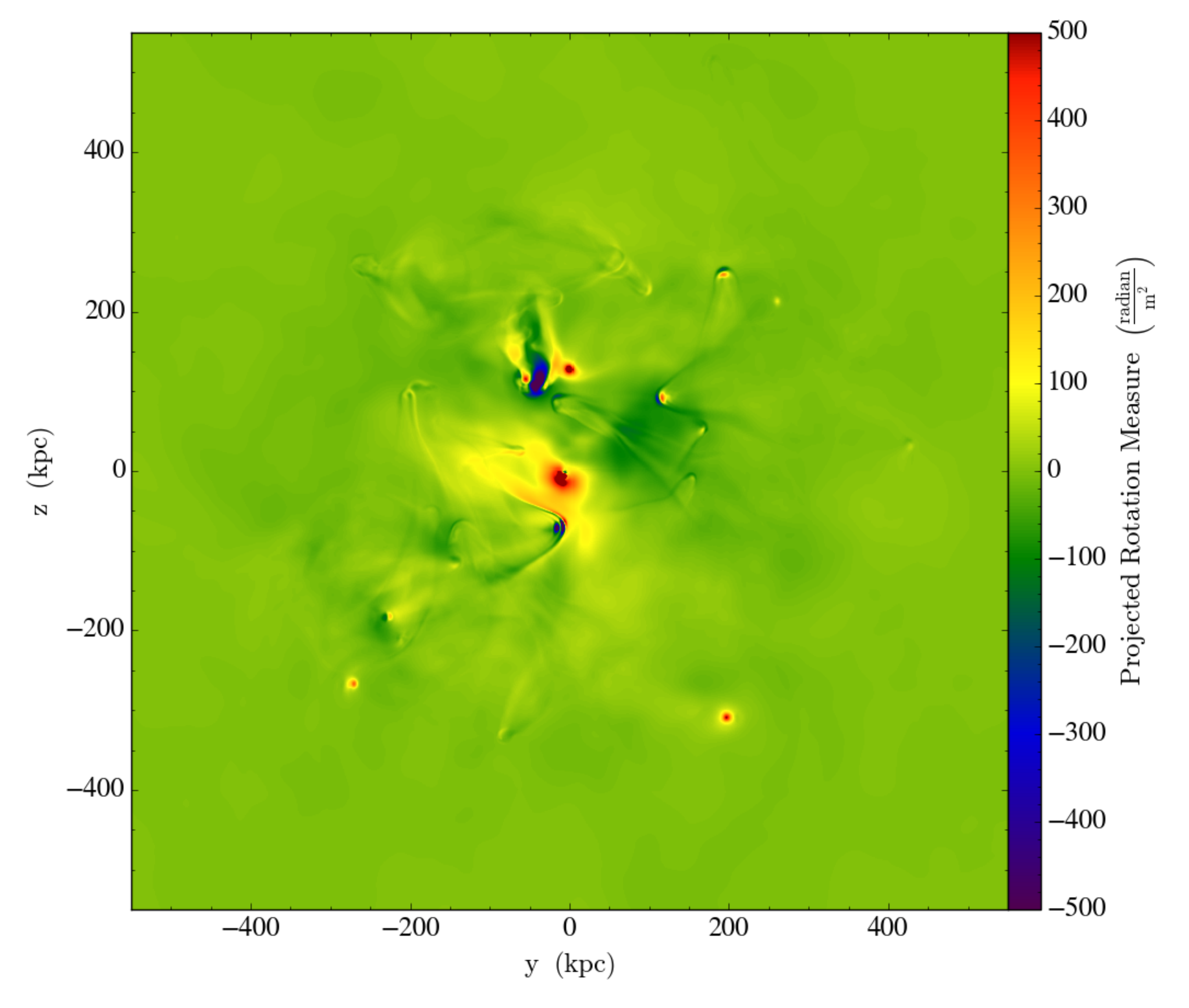}\label{fig:projection_RM_group_t05}}    
    \subfigure[RM map projected along $x$ at $t = 1$ Gyr]
    {\includegraphics[width=3in]{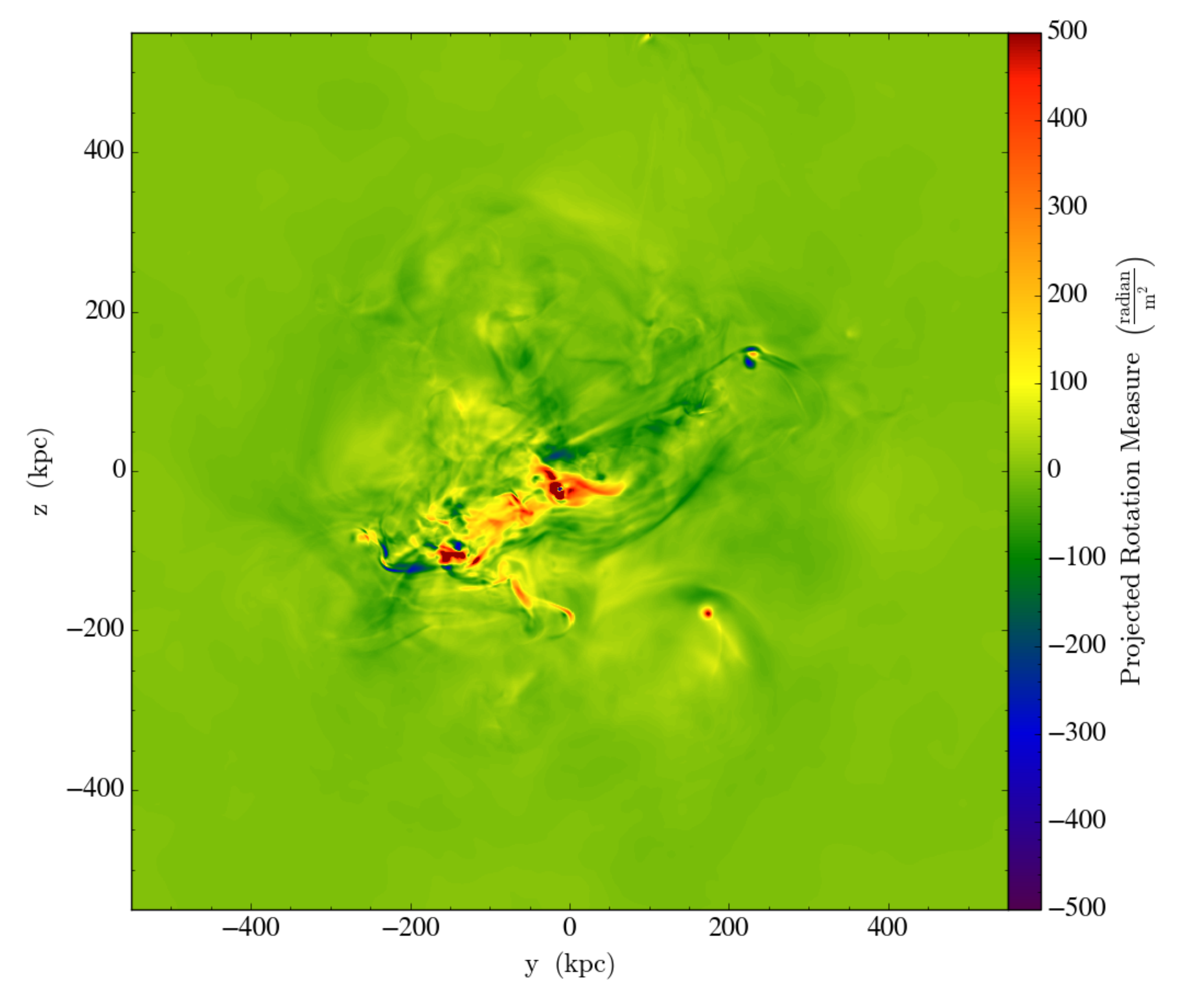}\label{fig:projection_RM_group_t1}}   
    \\
    \subfigure[Radial profile of RM at $t = 0.5$ Gyr]
    {\includegraphics[width=3in]{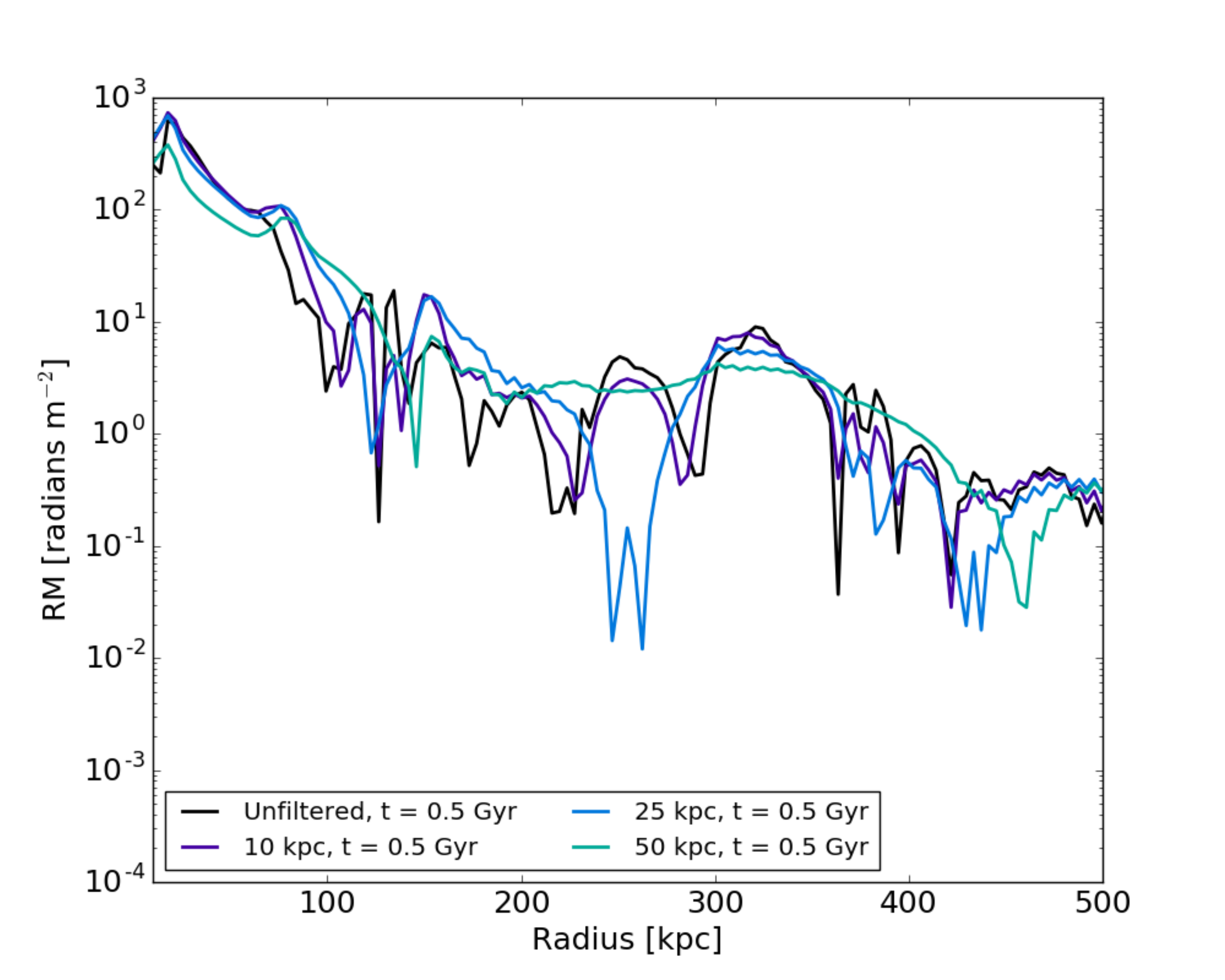}\label{fig:profile_RM_t05}} 
    \subfigure[Radial profile of RM at $t = 1$ Gyr]
    {\includegraphics[width=3in]{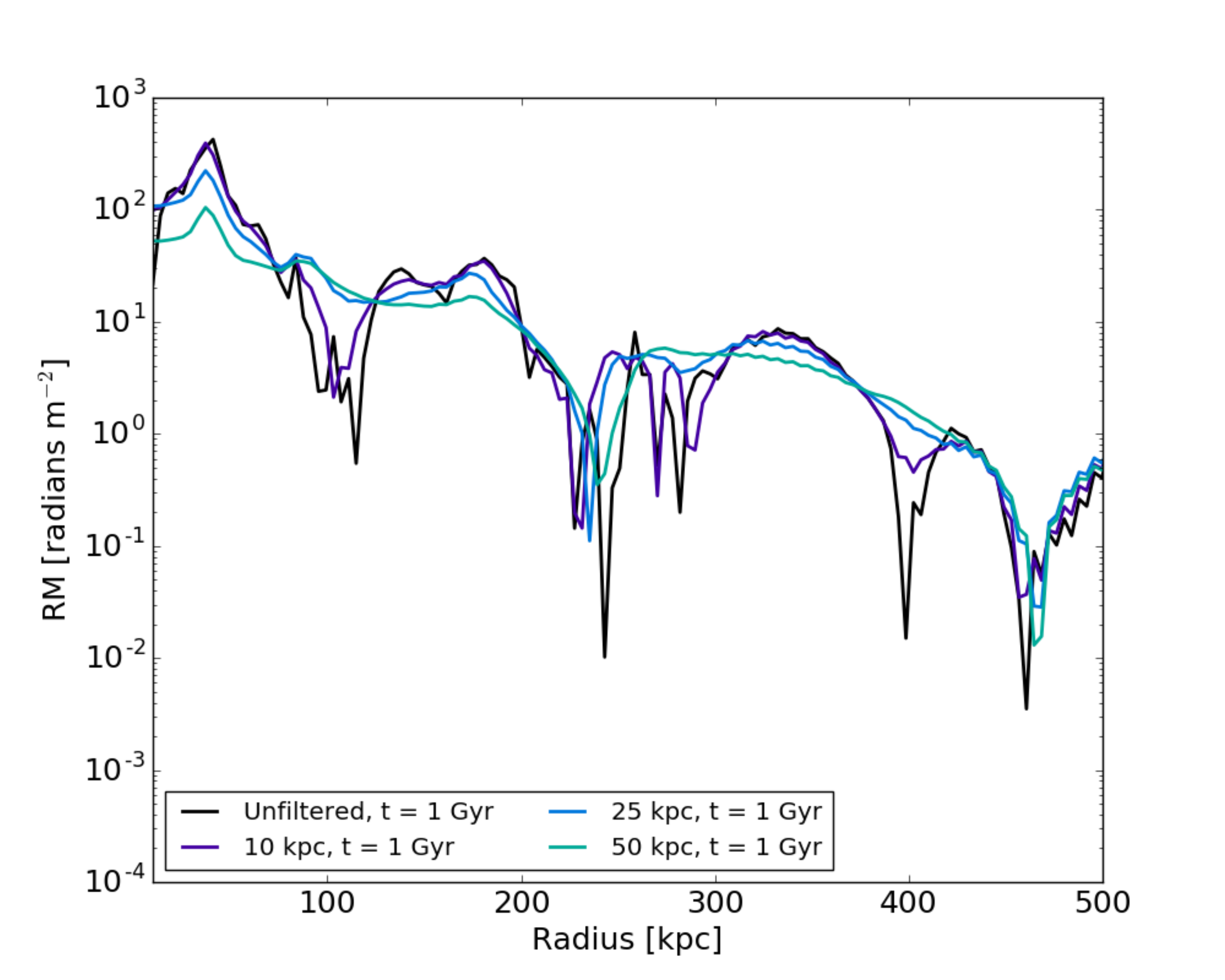}\label{fig:profile_RM_t1}} 
    \caption{Top row: Projected rotation measure map of the isolated group along the $x$ direction, magnetic fields amplified over the leading surfaces of galaxies and along stripped galaxy tails are clearly visible. Bottom row: azimuthally averaged radial profiles of the absolute value of RM in the isolated group along the $x$ direction, after filtering on spatial scales from $10 - 50$ kpc, which effectively filters out small scale fluctuations, at $t = 0.5$ Gyr and $t = 1.0$ Gyr.\label{fig:group_RM}}
  \end{center}  
\end{figure*}

As galaxies move through the ICM, magnetic fields are draped around the leading surfaces of galaxies, stretched, and amplified. Magnetic field lines in the stripped tails of galaxies are also stretched and amplified. These changes introduce local anisotropy into the field. A line of sight through these regions will therefore have an enhanced RM. Figures~\ref{fig:projection_RM_group_t05} and \ref{fig:projection_RM_group_t1} show the RM projected along the $x$-axis of our simulated group at $t = 0.5$ and $1$ Gyr. Outside the core of the group, the RM is close to zero everywhere except in the vicinity of galaxies, whereas in the core the RM is large because the field experiences fewer reversals in a region with high electron density.

Magnetic draping, clearly visible as enhancements in the RM maps, is harder to disentangle in azimuthally averaged radial profiles of the RM. Figures~\ref{fig:profile_RM_t05} and \ref{fig:profile_RM_t1} show profiles of the absolute value of the RM in the group at $t = 0.5$ Gyr and $1$ Gyr. In these profiles the radial bin sizes are 4.7 kpc, significantly smaller than the magnetic field correlation length in our simulations. At each timestep, we also filter the RM on spatial scales from 10 kpc $-$ 50 kpc, corresponding to low spatial resolution mapping and profiling of the RM. With no filtering, the variations in the RM due to magnetic field amplification manifest as fluctuations about a smooth RM profile. With decreasing spatial resolution, corresponding to increasing filtering scale, the overall effect is for the magnitude of these fluctuations to decrease and the profile to become increasingly smooth.

\begin{figure*}[!htbp]
  \begin{center}
 	\subfigure[Radial profile of the absolute value of RM]
    {\includegraphics[width=3in]{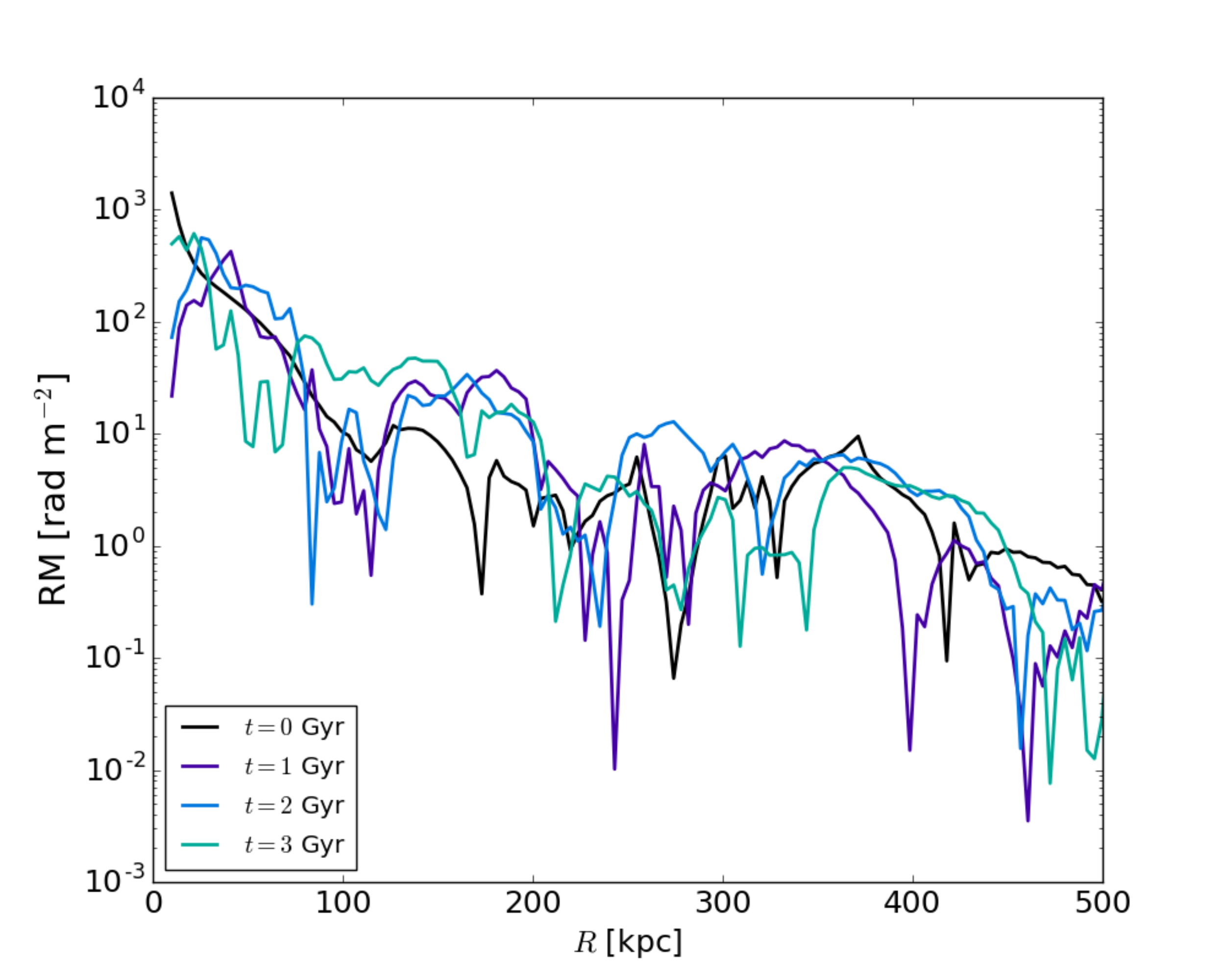}\label{fig:RM_profile_group}}  
    \subfigure[Radial profile of $\sigma_{\rm RM}$]
    {\includegraphics[width=3in]{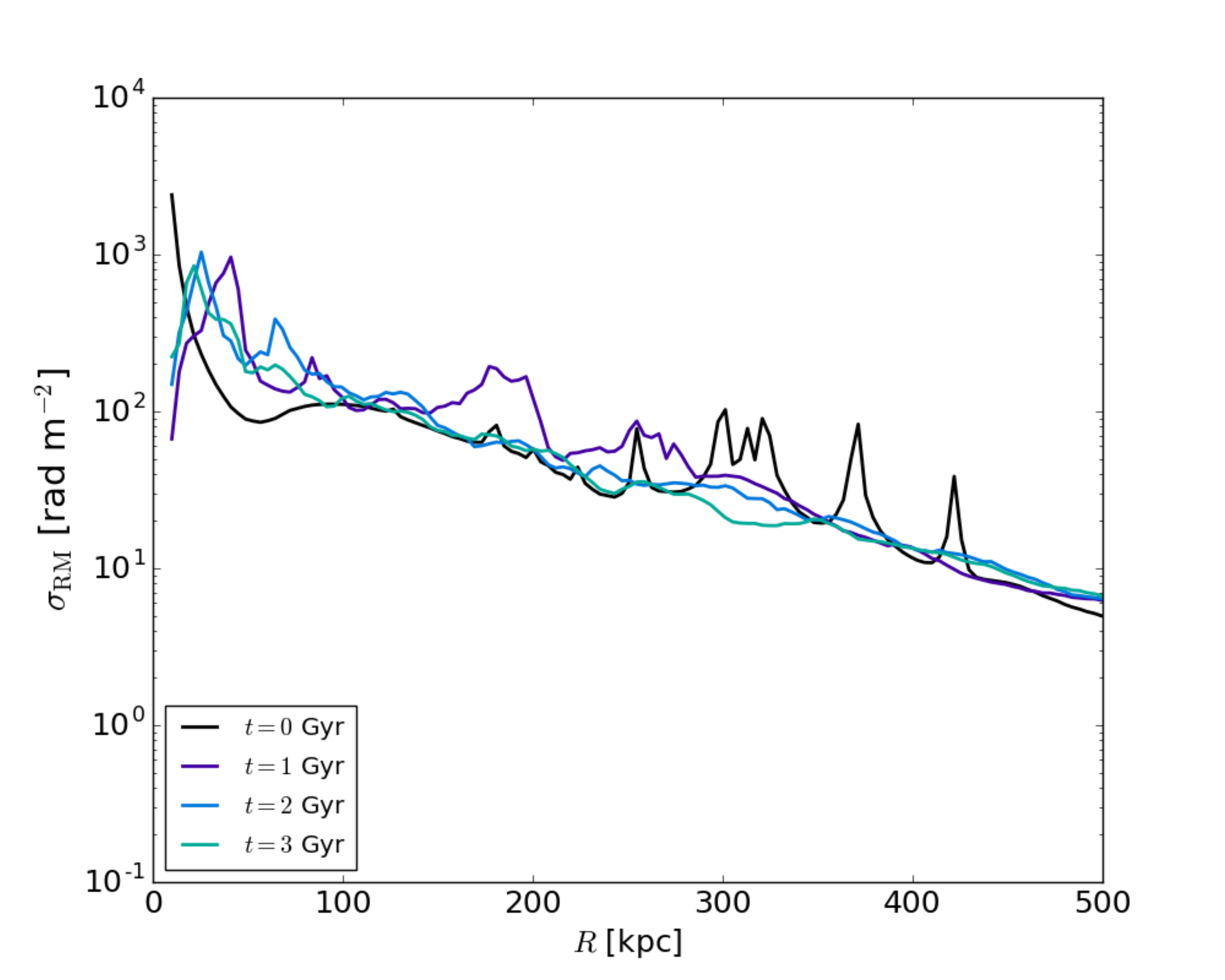}\label{fig:sigmaRM_profile_group}}    
%    \subfigure[Radial profile of RM  / $\sigma_{\rm RM}$]
%    {\includegraphics[width=2.3in]{RM_sigma_RM_all_profile.pdf}\label{fig:RM_sigmaRM_profile_group}} 
    \caption{Evolution in the azimuthally averaged profiles of the absolute value of the RM and dispersion in RM for the isolated group.  \label{fig:group_RM_profile}}
  \end{center}  
%\end{figure*}

%\begin{figure*}[!htbp]
  \begin{center}
 	\subfigure[Radial profile of the absolute value of RM]
    {\includegraphics[width=3in]{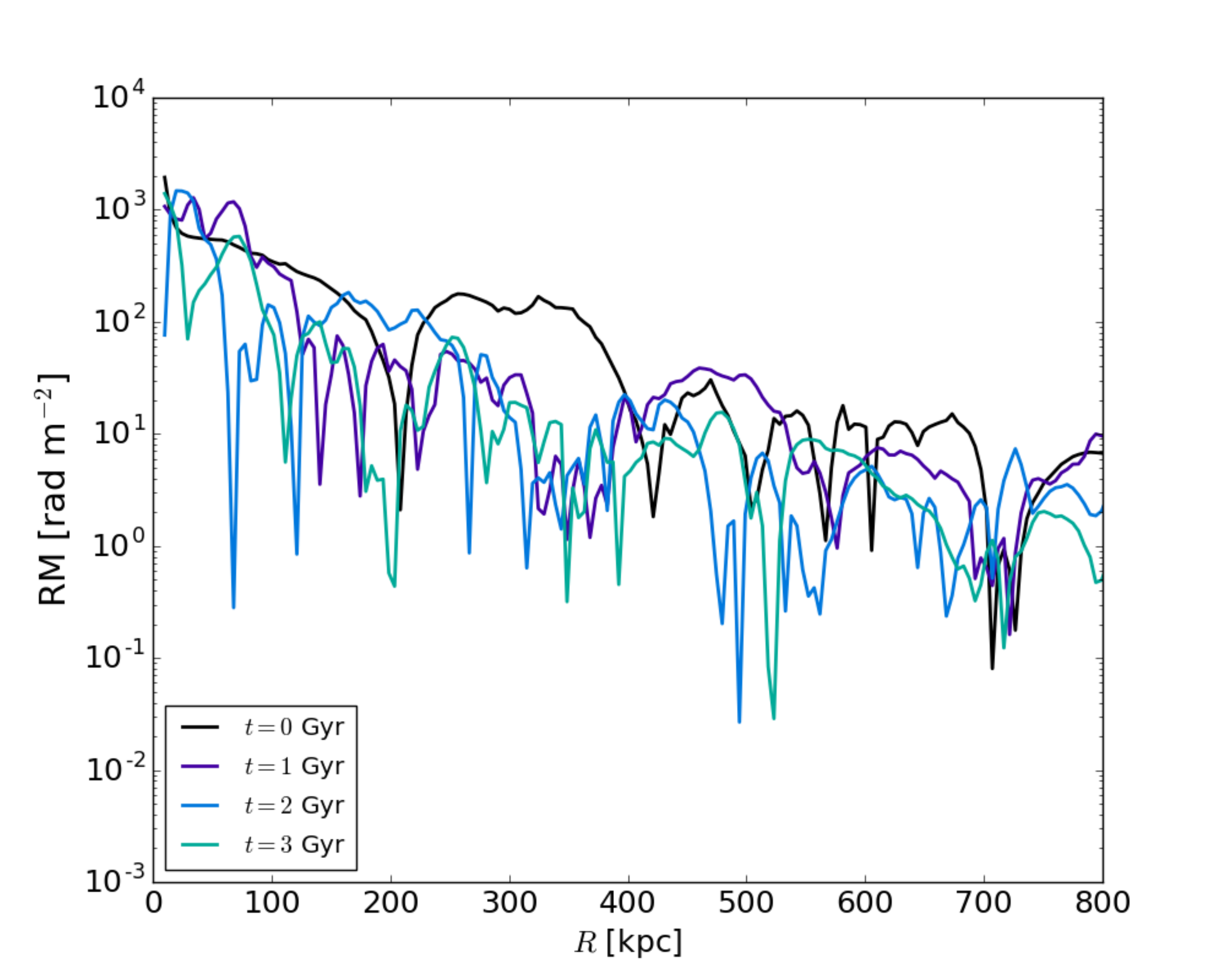}\label{fig:RM_profile_cluster}}  
    \subfigure[Radial profile of $\sigma_{\rm RM}$]
    {\includegraphics[width=3in]{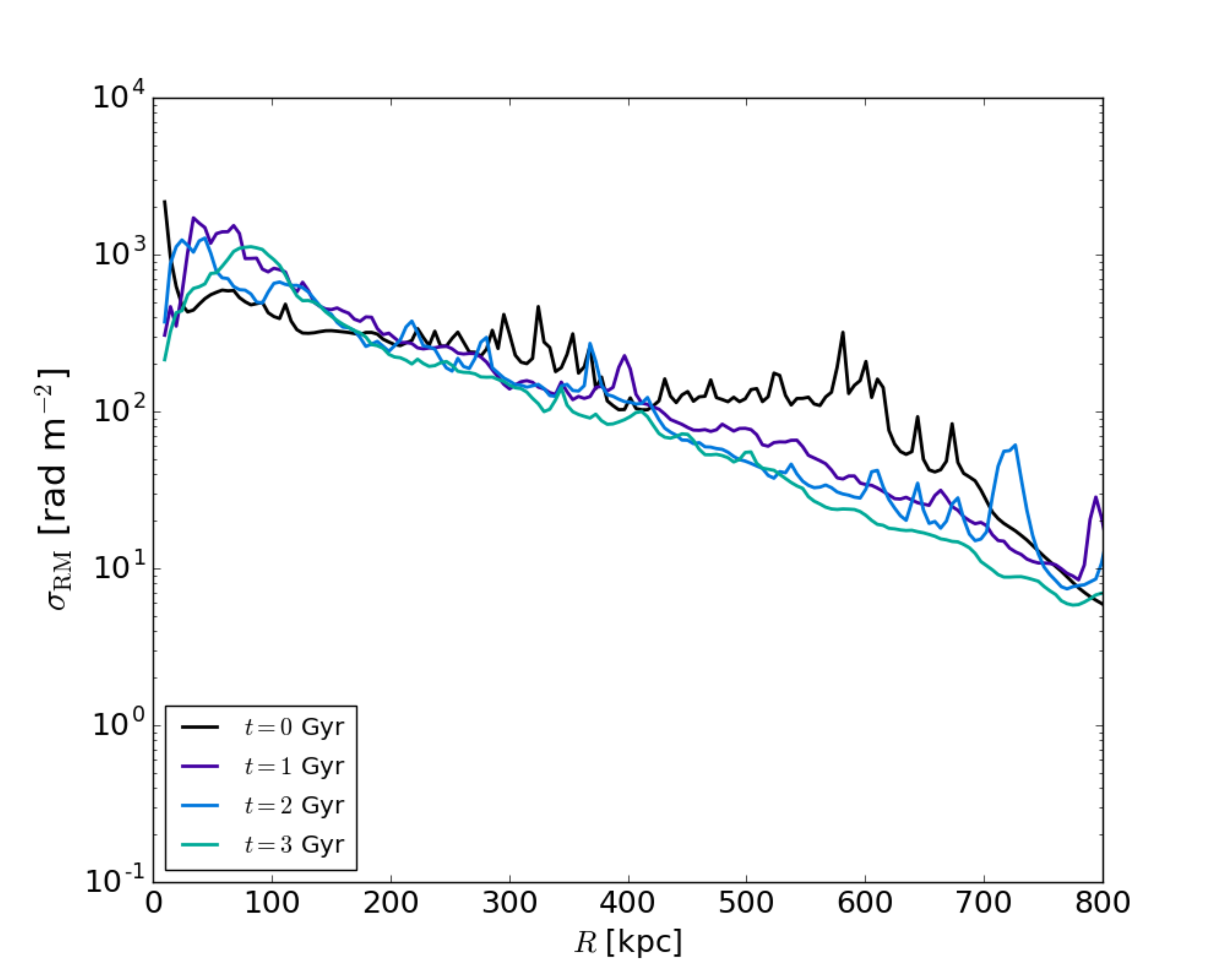}\label{fig:sigmaRM_profile_cluster}}    
    \caption{Evolution in the azimuthally averaged profiles of the absolute value of the RM and dispersion in RM for the isolated cluster.  \label{fig:cluster_RM_profile}}
  \end{center}  
\end{figure*}

We compare the unfiltered radial profiles of the absolute value of RM and its dispersion, $\sigma_{\rm{RM}}$, at different timesteps in Figures~\ref{fig:group_RM_profile} and \ref{fig:cluster_RM_profile}.  RM and $\sigma_{\rm{RM}}$ decline with radius as both density and the absolute value of the magnetic field decrease with increasing radius. Figures~\ref{fig:RM_profile_group} and \ref{fig:RM_profile_cluster} show that there is no significant evolution in RM from $t = 0 - 3$ Gyr. Although the ICM magnetic field is amplified by galaxies up to $t = 2$ Gyr, this is not immediately apparent from the azimuthally averaged profiles alone given the extent of spatial fluctuations. Figures~\ref{fig:sigmaRM_profile_group} and \ref{fig:sigmaRM_profile_cluster} show the dispersion in RM as a function of radius. Here, the fluctuations in the magnetic field due to galaxies is more apparent.  At earlier times, when there is an enhanced magnetic field due to the increased magnetic field strengths and densities in the centers of galaxies ($t = 0$ Gyr), or magnetic fields are enhanced along stripped galaxy tails ($t = 1$ Gyr), there is a larger corresponding deviation in $\sigma_{\rm RM}$. At later times, these enhancements in $\sigma_{\rm RM}$ disappear as the magnetic field isotropizes.

\section{Discussion}
\label{sec:disc_mhd}

\subsection{Interpretation of results}

As previously discussed, our simulations sit in terms of complexity between idealized wind tunnel experiments and cosmological simulations. They include some of the properties of the real global ICM but neglect mergers and accretion. A natural extension of this work would be an idealized cluster merger simulation including galaxy models, which we have explored before using particle sampling \citep{Vijayaraghavan13} and plan to address with model subhalos in a future paper.

Because of our idealized initial conditions, some of the early evolution can be attributed to artificial transients. In Figures~\ref{fig:groupdensbeta1} and \ref{fig:clusterdensbeta1}, we see that the initially round shapes of galaxies, as well as the correlation of galaxy size with starting radius implied by our nonoverlapping criterion, has largely disappeared by $t=500$~Myr. Mass stripping has not stopped by this point in either case. As seen in Figure~\ref{fig:gaslossratemhd}, group galaxies lose $\sim 40 - 50\%$ of their gas inside $R_{200}$ within $t = 500$~Myr; this is primarily the diffuse outer gas. Between $t = 500$~Myr to $t = 2.4$~Gyr, galaxies lose another $50\%$ of their gas; this is the denser bound core gas. Cluster galaxies are stripped at a somewhat faster rate: they lose $\sim 60\% $ of their gas within $t = 500$~Myr (Paper I). In each case the mass loss rate slows down only after about 1~Gyr, well after transient effects appear to have dissipated. Thus our simulations resolve much of the initial rapid mass-removal phase as well as the late-time slower removal of gas from the dense galaxy cores.

Insofar as our simulations can be compared directly to real clusters, they should be most similar to relatively isolated clusters with low accretion rates. These are not straightforward to identify observationally. Recent theoretical work on splashback features in cluster density profiles \citep{Diemer2014,Adhikari2014} has shown that mass accretion rate correlates with the radius at which the density profile steepens significantly from the asymptotic \cite{Navarro97} behavior, as well as the steepened slope there and the central concentration. These theoretical predictions have not yet translated into an observational proxy that can be cleanly separated from projection effects \citep{Zu2016}, though there is some hope for methods based on caustics \citep{DeBoni2016}. Thus at present we must look at proxies for recent merger activity as a way to separate out relatively relaxed and isolated clusters. Examples of relaxed clusters include the CLASH sample \citep{Postman12,Donahue16} and the \citet{Mantz14} sample. Merging cluster samples include the Chandra-Planck Legacy Program for Massive Clusters of Galaxies (\url{http://hea-www.cfa.harvard.edu/CHANDRA_PLANCK_CLUSTERS}). Systematic comparisons of the RM, $\sigma_{\rm RM}$, and therefore the magnetic field strength and distribution in the ICM of relaxed and unrelaxed clusters along the lines of \citet{Bonafede11} can allow us to understand the magnetic field evolution in these systems. Similarly, high resolution X-ray spectroscopy with future \textit{Hitomi}-like missions or \textit{ATHENA} will enable comparisons of the velocity distribution.

\subsection{The effect of ICM magnetic fields on galactic hot coronal gas}
\label{sec:disc_mhd_coronae}
The results in \S~\ref{sec:mhd_stripping} and Figure~\ref{fig:gaslossratemhd}, compared to galaxies' differential gas mass loss rate in Paper I, show that the presence of weak initial magnetic fields ($\beta = 100$) does not affect the overall rate at which galaxies are stripped of their hot coronal gas. Similar results have been found in other comparable wind tunnel-like simulations of galaxy stripping in a magnetized medium. \citet{Ruszkowski14}, in simulations of disk galaxies being stripped in a magnetized ICM with magnetic fields of strength $\beta \simeq 21$, with edge-on and tilted disk configurations, find that the magnetic field has a relatively weak effect on overall mass loss. \citet{Tonnesen14}, in their stripping simulations with disk magnetic fields, find that the presence of the magnetic field does not alter the overall mass loss, although the morphology and strength of the magnetic field result in minor differences in the early stages of stripping. 

The insensitivity of overall gas mass loss rates to the presence of ICM magnetic fields occurs because gas loss is primarily driven by ram pressure in the ICM and tidal forces in the background halo; for cases where $\beta \gg 1$, these forces are significantly stronger than corresponding magnetic field effects. Galaxies experiencing ram pressure stripping develop leading surfaces whose locations are determined by approximate pressure balance between ICM ram plus thermal pressure on the one hand and the radially declining pressure profile of the galaxies' coronae on the other (Paper~I). The ICM flow compresses only the component of the magnetic field that is perpendicular to the flow direction at the leading surface of a galaxy (\citealt{Dursi08}, \citealt{Shin14}), forming a thin layer of increased field strength where  $\beta$ is still $ \gg 1$. Mass loss from galaxies during early stages is therefore similar to the case without magnetic fields.

Where magnetic fields do have an effect is in the gas loss from coronal edges and stripped tails that trail galaxies in their orbits, and this where the difference in galaxy coronal gas evolution from the pure hydrodynamic case is most evident. Magnetic field lines are stretched by shearing flows along the edges of galaxies (Figures~\ref{fig:groupdensbeta1}, ~\ref{fig:clusterdensbeta1}), amplifying the magnetic field, due to the flux freezing criterion of ideal MHD. The component of the magnetic field aligned along the direction of the flow can then suppress the formation of Kelvin-Helmholtz instabilities along the coronal gas-ICM boundary  \emph{parallel} to the flow, if the relative fluid velocity at the boundary is less than the root mean squared value of the Alfv\'en speeds in the two media (\citealt{Chandrasekhar61}). Typical values of the  Alfv\'en speed ($v_A = \sqrt{B^2 / (4 \pi \rho_{\rm gas})}$) in our simulation initially range between $\sim 50 - 100$ km s$^{-1}$ in the ICM in the isolated group and cluster, before galaxies amplify the ICM magnetic field. However, when the magnetic field is amplified due to shear at the ICM-ISM interface, the Alfv\'en speed increases up to $500 - 900$ km s$^{-1}$ in these regions. The RMS Alfv\'en speed is then comparable to the relative velocity between the ISM and ICM, allowing KH instabilities to be suppressed. These shear instabilities occur where coronae and galactic tails mix with the ICM and dissipate in the hydrodynamic simulations in Paper I. With magnetic fields aligned along these edges, stripped galactic gas that is pushed downwind does not mix with the ICM as easily. 

Stripped tails, partially supported by magnetic pressure, are narrower and less susceptible to characteristic Kelvin-Helmholtz instabilities due to shear flows at the ISM-ICM interface (Figure~\ref{fig:gal_hydroMHD}). Tails supported by magnetic pressure with field lines aligned along the direction of the tail have been reported in earlier MHD simulations of galaxy stripping (\citealt{Ruszkowski14}, \citealt{Shin14}). Some galactic tails in our simulations appear to have bifurcated structures that resemble observed double tails in galaxies undergoing stripping (e.g.\ ESO 137-002, \citealt{Zhang13}). Although galactic tails are partially supported by magnetic pressure, magnetic fields alone are not necessarily responsible for the bifurcated structure, since we see bifurcated tails in the purely hydrodynamic simulations in Paper I. However, the presence of magnetic fields makes these tails less susceptible to disruption from shear instabilities. 

\subsection{The evolution of ICM magnetic fields in the presence of orbiting galaxies}
\label{sec:disc_icm_bfield}

Orbiting, stripped, gas-rich galaxies modify the strength and configuration of ICM magnetic fields. Galaxies with an initially magnetized ISM can seed cluster magnetic fields through outflows and stripping; these effects are not investigated in this work, as our galaxies do not initially have a distinct magnetic field component. The magnetic field in our static group simulation, which in its initial configuration in the absence of galaxy motions is unrelaxed, decays to a stable configuration. Consequently, the overall magnetic field strength of the ICM decreases with time (\S~\ref{sec:isobfield}, Figure~\ref{fig:beta_nogal}). In the presence of galaxies and their orbital motions, a small-scale dynamo driven by galaxy motions amplifies the magnetic field. This amplification is sustained for $\sim 2$ Gyr in the isolated group and $\sim 1.5$ Gyr in the cluster. The morphology of the ICM magnetic field is also modified to a more tangled configuration in the presence of turbulent wakes generated by galaxies. As galaxies are further stripped of most of their gas, this process becomes less effective and the magnetic field begins to decay. 

\citet{Subramanian06} propose that turbulent motions in the ICM can amplify seed cluster magnetic fields and prevent their decay. Using analytic and numerical arguments, they argue that the exponentially fast amplification of the weak initial cluster seed magnetic field by random motions, turbulence driven by cluster major mergers, and magnetic fields generated in the turbulent wakes of infalling galaxies can sustain and amplify cluster magnetic fields. The results of the simulations in this paper are consistent with these expectations. Although the ICM in our simulations has an initial magnetic field whose strength ($\sim \muG$) is significantly higher than  cosmological seed magnetic fields arising from plasma battery effects, one can qualitatively verify that galaxy motions alone, in the absence of major mergers or any other mechanism of seeding ICM magnetic fields, can amplify the magnetic pressure by a factor of $\sim 5 - 10$ in 2 Gyr. Better resolution of the turbulent dynamo might make this factor larger.

Magnetic wakes generated by galaxies in the ICM drive turbulence while trailing behind stripped galaxies before they dissipate or are detached. The magnetic field itself becomes increasingly more tangled in addition to being amplified. The evolution of the magnetic energy density power spectrum is seen in Figures~\ref{fig:powerspectrum_b_group} and \ref{fig:powerspectrum_b_cluster} for $t = 3$ Gyr of evolution in the group and cluster. From $t = 0 - 2$ Gyr, the increase in magnetic power is driven by galaxy motions, so the increase in magnetic energy occurs primarily at scales comparable to the sizes of galaxies and their tails. The rate which magnetic energy density increases slows down after $t \sim 1$ Gyr, and turbulence decays after galaxies have mostly been stripped at $t \gtrsim 2$ Gyr. Based on these results, one can conclude that infalling stripped galaxies can drive turbulence and amplify magnetic fields for about one dynamical timescale. The importance of this process depends on galaxy infall rate and the mass distribution of infalling galaxies and subclusters.  Turbulence driven by galaxies also affects the power spectrum of kinetic energy, in addition to magnetic energy. Kinetic energy is injected at large scales as galaxies lose their potential energy, and this then cascades down to small scales.

Cosmological simulations of structure formation in general show that the process of cluster formation amplifies initial seed magnetic fields by $\sim~3$ orders of magnitude. \citet{Dolag99} and \citet{Dolag02} show that $\mbox{nG}$ seed magnetic fields are amplified to $\muG$ fields in cosmological MHD simulations, largely independent of the initial magnetic field configuration. \citet{Dubois08} using cosmological simulations of cluster formation show that magnetic fields are primarily amplified during the cluster's gravitational collapse, while additional shearing motions in the outskirts of clusters generate turbulence that further amplifies cluster magnetic fields. \citet{Vazza14} show using cosmological simulations that structure formation can generate turbulence and amplify magnetic fields in clusters. In their magnetic power spectrum analyses, they show that most of the energy injected in clusters at late times ($z \sim 0$) is at scales $ \sim 100$~kpc.  With our simulations, we do not account for magnetic field amplification during the process of $10^{14} \msun$ cluster formation itself, but by $\sim 10^{10} - 10^{12} \msun$ galaxies, and we show that these galaxies amplify ICM magnetic fields by a factor of $\sim 3$.  %Our results on magnetic field amplification by turbulence generation and shearing motions agree qualitatively with those from cosmological simulations.

Controlled experiments that simulate cluster-subcluster mergers also show that mergers amplify ICM magnetic fields. \citet{Roettiger99} show using 3D MHD simulations of merging clusters that magnetic fields become filamentary and stretched by the infalling cluster, and that magnetic energy is amplified by a factor of $3 - 20$ as a result of the merger on scales comparable to the size of the cluster core. \citet{Takizawa08} show that infalling subclusters in cluster mergers generate ordered magnetic fields in their wake, and appear as cool regions surrounded by magnetic fields. The magnetic fields in the wakes of orbiting galaxies in our simulations are also ordered and aligned with galaxies' tails. \citet{ZuHone11b}, in simulations of cluster-subcluster mergers that result in sloshing of the cluster core about the potential well, show that velocity shears associated with the cold fronts amplify magnetic fields on the surfaces of cold fronts from initial values of $0.1 - 1 \muG$  up to $\sim 10 \muG$; we see a similar shearing effect along the sides of our galaxies that results in amplifying the component of the magnetic field parallel to the ICM flow. Additionally, although our simulations are of galaxies amplifying ICM magnetic fields rather than massive subclusters, the total mass of all the galaxies in our simulations ($\sim 10 -15\%$) is comparable to the lower end of merging subcluster masses ($25\%$ in \citealt{Takizawa08}, $20\%$ in \citealt{ZuHone11b}, but $40\%$ in \citealt{Roettiger99}). Overall, the magnetic pressure increases by a factor of $5 - 10$ due to galaxies alone in our simulations (Figures~\ref{fig:beta_gal} and ~\ref{fig:beta_gal_cluster}). 

\subsection{Detecting ICM turbulence and magnetic field amplification}

The detection of ICM turbulence is an outstanding problem in cluster astrophysics. X-ray missions to date have not had sufficient spectral resolution to resolve bulk flows and gas velocity dispersions of $\sim 200$ km~s$^{-1}$. Since the loss of \textit{Hitomi} (with the exception of its early science observations of the Perseus cluster; \citealt{Hitomi16}), the most promising future mission capable of making such measurements is \textit{ATHENA}, which is scheduled to launch in 2028. Fluctuations in the thermal Sunyaev-Zeldovich (SZ) effect are also potentially a promising avenue (\citealt{Khatri16}) towards characterizing ICM turbulence, given the wealth of SZ observations from various missions including \textit{Planck}, the South Pole Telescope, MUSTANG-2, ALMA, CARMA, and more. High resolution X-ray measurements have also been used to measure density and temperature fluctuations, and infer turbulent properties of the ICM \citep{Gaspari14,Hofmann16}.

With the simulations presented in this paper, we have isolated the contribution of galaxies alone to the overall amount of turbulence in the ICM. In our idealized configuration, all the galaxies in the cluster and group have all of their mass at $t = 0$ Gyr; therefore the amount of turbulence generated by these galaxies is an upper limit to the galaxy contribution to ICM turbulence in real clusters. As seen in Figures~\ref{fig:vdisp_tot_group} and \ref{fig:vdisp_tot_cluster}, less than $1\%$ of the group volume and $10\%$  of the cluster volume have turbulent velocity dispersions higher than $300$ km s$^{-1}$ at any time. \footnote{By itself this result is consistent with the recent \textit{Hitomi} results on the Perseus cluster, since these probed only a small region in the core (\citealt{Hitomi16}). However, additional dissipation mechanisms not included in our simulations could reduce the turbulent volume fraction further.} From $t = 0$ to $3$  Gyr, the net increase in the volume fraction of the ICM in the group with turbulent velocity dispersions above $300$ km s$^{-1}$ and $700$ km s$^{-1}$ is by a factor of $2 - 3$. In the more massive cluster, where galaxies are stripped more rapidly, the net increase in these volume fractions is effectively zero from $t = 0$ to $3$ Gyr, at about $0.1 - 1\%$. We can therefore conclude that while galaxies generate turbulence via their stripped wakes as well as $g$-waves, this turbulent kinetic energy is not a significant fraction of the overall ICM energy budget. The generated turbulence does, however, play a significant role in the evolution of the magnetic field. Turbulence in the ICM can be generated by multiple sources, including cluster mergers, accretion from the surrounding large scale structure, and outflows from AGN and galaxies, as described in the introduction. Energetically, the most prominent of these sources are mergers between clusters of comparable masses, where the kinetic energy of the merger is comparable to the total potential energy of the clusters. Galaxies by comparison contribute only to $\sim 5\%$ of the mass in clusters, and are therefore not energetic enough to significantly increase the overall amount of non-thermal and turbulent pressure in the ICM. 

Although ICM magnetic fields have been observed and characterized using RM techniques, the shearing and amplification effects of individual galaxies or the alignment of magnetic fields with galaxies' stripped tails, have not yet been observed, partly due to the necessity of appropriately located background radio sources. In principle, this is possible with current and future RM observations. For instance, radio galaxies in clusters can themselves be used to probe galaxy-scale effects on the ICM magnetic field \citep[e.g.,][]{Vacca12}. Globally, the effect on the RM of multiple orbiting galaxies stirring the ICM magnetic field is to cause significant fluctuations on kpc scales. The extents of these fluctuations increase as galaxy wakes occupy larger fractions of the ICM, and decrease as galaxies are stripped. These effects can in principle be observed by mapping the RM and ICM magnetic field with high spatial resolution. Specific routines to perform these RM analyses include the PACERMAN routine \citep{Dolag05b,Vogt05} which has been used to characterize the magnetic field in the Coma cluster, showing RM and $\sigma_{\rm{RM}}$ decrease with increasing cluster centric radius and $\sigma_{\rm{RM}}$ fluctuates significantly \citep{Bonafede13}.  Other RM synthesis techniques \citep{Frick10, Frick11} that detect small-scale galaxy magnetic fields can potentially be used to characterize ICM magnetic fields on galaxy scales. 

\section{Conclusions}
\label{sec:conclusions}

We present MHD simulations of galaxies evolving in a magnetized ICM in group-scale and cluster-scale environments. Using these simulations, we have studied the qualitative and quantitative effects of ICM magnetic fields on orbiting gas-rich galaxies. We show that magnetic fields in $\beta \gg 1$ plasmas do not significantly affect the gas mass loss rate of galaxies. The thermal and ram pressure in these environments are significantly higher than the magnetic pressure, so the amount of galaxies' gas lost in the presence of ICM magnetic fields is, quantitatively, not significantly different from the gas loss in simulations without ICM magnetic fields. Therefore, the dynamical effects of magnetic fields alone do not solve the problem of the ubiquity of long-lived galactic coronae in cluster environments in the presence of various gas removal mechanisms (\citealt{Sun07}, \citealt{Jeltema08}). Magnetic fields do, however, qualitatively modify the appearance of stripped galaxies.  As galaxies move through the ICM, they are initially stripped by ram pressure and tidal effects and form the characteristic corona-tail structures seen in Paper I. ICM magnetic fields lines are stretched and draped around stripped galactic coronae. Additionally, ICM magnetic fields are dragged along and stretched in galaxy tails and the wakes of galaxies. The magnetic fields in galaxies' tails suppress Kelvin-Helmholtz shear instabilities and slow down the mixing of these tails into the ICM. Magnetically supported tails are smoother and narrower than galaxy tails in the hydrodynamic simulations in Paper I.

Orbiting massive galaxies modify the strength and morphology of ICM magnetic fields. Galaxies initially amplify the ICM field through velocity shears that stretch magnetic field lines along stripped coronae and aligned with  their tails.  The magnetic field increases  dramatically during the first $t = 1$ Gyr of evolution when galaxy stripping is most rapid. Overall, the magnetic field strength in the ICM increases from $\beta \simeq 100$ to $\beta \simeq 20$ in 2 Gyr. After galaxies have been stripped of most of their gas by $t \gtrsim 1.5 - 2$ Gyr, they do not have enough momentum to continue amplifying the field. Since the initial magnetic field configuration is not force-free, the field decays after $t \gtrsim 2$ Gyr. 

Orbiting galaxies drive turbulence into the ICM, as shown in the evolution of magnetic and kinetic energy power spectra. Most of the power injected in the magnetic energy density spectrum is on small spatial scales comparable to the  sizes of galaxies and their wakes, and during the first $\sim 1.5 - 2$ Gyr of evolution. At late times, after galaxies have been mostly stripped, there is no longer any significant driver of turbulence and the magnetic power spectrum decays. Galaxies also inject turbulent kinetic energy into the ICM. Galaxies drive turbulence in their stripped tails and wakes as well as through $g$-waves. The driving of $g$-mode turbulence is evident in the evolution of the vorticity field in regions outside of the galaxies' tails.

Using our simulations we also calculate the observational consequences of magnetic field generation and turbulent amplification. We are able to quantify the effect of galaxies alone on the ICM turbulent pressure and velocity dispersion, since the ICM in our simulations is initially in hydrostatic equilibrium and is not subject to other processes that can generate turbulence like cluster mergers and outflows. We calculate the volume fraction of the ICM where the turbulent pressure and velocity dispersion exceed typical detection limits of future X-ray missions like \textit{ATHENA}. As the turbulent kinetic energy increases from $t = 0$  to $t \simeq 1.5 - 2$ Gyr, the volume fraction of gas where the turbulent pressure and velocity dispersions lie above these limits increases by a factor of $2 - 3$. This increase in the turbulent volume fraction corresponds at the maximum to $10\%$ of the cluster ICM with turbulent velocity  dispersion above $300$ km s$^{-1}$ and $< 1\%$ above  $700$ km s$^{-1}$; these relatively low volume fractions with high turbulent velocities make sense when considering that galaxies only contribute to $\lesssim 5 - 10\%$ of the mass in clusters.  The potential observed quantitative effect of magnetic field amplification by galaxies is more subtle. High spatial resolution Faraday rotation measure mapping can reveal galaxy-scale structures in the ICM magnetic field. Azimuthally averaged radial profiles of the RM show significant fluctuations due to these galaxy-scale structures. As these structures become less prominent, the dispersion in azimuthally averaged profiles, $\sigma_{\rm{RM}}$, decreases.

Our results show that galaxies can significantly amplify existing ICM magnetic fields. Our simulations are idealized in the sense that all galaxies begin to orbit within the group and cluster simultaneously. Therefore, the total ICM magnetic field amplification is the net effect of  all orbiting cluster galaxies. Additionally, since all galaxies are stripped within $\sim 2$ Gyr, the magnetic field decays after this time. In real clusters, the constant infall of galaxies and groups of galaxies implies that their ICM magnetic fields will be constantly stirred, although the effect of any single galaxy will be much lower than the total effect of all galaxies on the ICM magnetic field. This applies to the generation of turbulent kinetic energy as well: we can place upper limits on the maximum turbulent velocity dispersion due to galaxies alone, but in real clusters, the generation of turbulence will be due to the constant infall of galaxies and subclusters. Additionally, the extent to which dynamo action can amplify magnetic fields in real clusters is dependent on spatial resolution and the initial seed magnetic field.

\section*{Acknowledgments}

This work was partially supported by the Graduate College Stutzke Dissertation Completion Fellowship at the University of Illinois at Urbana-Champaign, NASA Chandra theory award TM5-16008X, and an NSF Astronomy and Astrophysics Postdoctoral Fellowship under award AST-1501374. The simulations presented here were carried out using the NSF XSEDE Stampede system under project TG-AST040024 and the NASA Pleiades system under project SMD-15-5693. FLASH was developed largely by the DOE-supported ASC/Alliances Center for Astrophysical Thermonuclear Flashes at the University of Chicago. The figures in this paper were generated using the \texttt{yt} software package \citep{Turk11}. We are grateful to Craig Sarazin and Zhi-Yun Li for their useful insight and suggestions on this work, and to Tracy Clarke, Dongsu Ryu, Hyesung Kang, Christoph Federrath, and Jim Stone for stimulating conversations. We thank the anonymous referee for useful comments that helped improve this paper. 

\bibliography{ms}

\end{document}